\documentclass[11pt,epsf]{article}
\pdfoutput=1
\DeclareMathAlphabet{\scr}{U}{rsfs}{m}{n}
\usepackage{latexsym}
\usepackage{epsfig}
\usepackage[mathscr]{eucal}
\usepackage{amsfonts}
\usepackage{amscd}
\usepackage{array}
\usepackage{amssymb}
\usepackage{colordvi}
\usepackage[centertags]{amsmath}
\usepackage{enumerate}
\usepackage{graphicx}
\usepackage{booktabs}
\usepackage{theorem}
\usepackage[footnotesize]{caption}
\usepackage{soul}
\usepackage[merge,numbers,compress]{natbib}

\usepackage{pdfpages}

\newcommand{\sq}{\tilde{q}}
\newcommand{\su}{\tilde{u}}
\newcommand{\sd}{\tilde{d}}
\newcommand{\gl}{\tilde{g}}
\def\bom#1{{\mbox{\boldmath $#1$}}}
\newcommand\nn         {\nonumber}
\newcommand{\sul}{\tilde{u}_L}

\newcommand{\sdl}{\tilde{d}_L}

\newcommand{\sur}{\tilde{u}_R}
\newcommand{\sdr}{\tilde{d}_R}

\newcommand\sss{\mathchoice%
{\displaystyle}%
{\scriptstyle}%
{\scriptscriptstyle}%
{\scriptscriptstyle}%
}
\newcommand{\newc}{\newcommand}
\newc{\be}{\begin{equation}}
\newc{\ee}{\end{equation}}
\newc{\bi}{\begin{itemize}}
\newc{\ei}{\end{itemize}}
\newc{\benu}{\begin{enumerate}}
\newc{\eenu}{\end{enumerate}}
\newc{\bc}{\begin{center}}
\newc{\ec}{\end{center}}
\newc{\bfig}{\begin{figure}}
\newc{\efig}{\end{figure}}
\newc{\qbar}{\bar{q}}
\newc{\go}{\tilde{g}}
\newc{\PB}{\textsc{Powheg-Box}}
\newcommand\matB{{\cal B}}
\newcommand\matR{{\cal R}}
\newcommand\matV{{\cal V}}
\newcommand\matO{{\cal O}}
\newcommand\matF{{\cal F}}

\begin{document}

\title{\hfill ~\\[-30mm]
\phantom{h} \hfill\mbox{\small KA--TP--12--2013}\\[-1.1cm]
\phantom{h} \hfill\mbox{\small SFB/CPP--13--32} \\[-1.1cm]
\phantom{h} \hfill\mbox{\small PSI--PR--13--07} \\[-1.1cm]
\phantom{h} \hfill\mbox{\small TTK--13--14} 
\\[1cm]
\vspace{13mm}   \textbf{Matching Squark Pair Production at NLO with Parton Showers}}

\date{}
\author{
R.~Gavin$^{1\,}$\footnote{E-mail: \texttt{ryan.gavin@psi.ch}},
C.~Hangst$^{2\,}$\footnote{E-mail: \texttt{christian.hangst@kit.edu}},
M.~Kr\"amer$^{3\,}$\footnote{E-mail:
  \texttt{mkraemer@physik.rwth-aachen.de}},
M.~M\"uhlleitner$^{2\,}$\footnote{E-mail:
  \texttt{margarete.muehlleitner@kit.edu}}, \\
M.~Pellen$^{3\,}$\footnote{E-mail: \texttt{pellen@physik.rwth-aachen.de}},
E.~Popenda$^{1\,}$\footnote{E-mail: \texttt{eva.popenda@psi.ch}},
M.~Spira$^{1\,}$\footnote{E-mail: \texttt{michael.spira@psi.ch}}
\\[9mm]
{\small\it
$^1$Paul Scherrer Institut, CH-5232 Villigen PSI, Switzerland}\\[3mm]
{\small\it
$^2$Institute for Theoretical Physics, Karlsruhe Institute of Technology,} \\
{\small\it D-76128 Karlsruhe}\\[3mm]
{\small\it
$^3$Institute for Theoretical Particle Physics and Cosmology,}\\{\small \it RWTH Aachen University, D-52056 Aachen}\\
}

\maketitle

\begin{abstract}
\noindent  
The pair production of squarks is one of the main search channels for supersymmetry at the LHC. We present a fully differential calculation of the next-to-leading order (NLO) SUSY-QCD corrections to the on-shell production of a pair of squarks in the Minimal Supersymmetric Standard Model (MSSM), supplemented by the leading-order decay of the 
 squarks to the lightest neutralino and a quark. In addition, we use the \textsc{Powheg} method to match our NLO calculation with parton showers. To this end, the process 
 was implemented in the \PB~framework and interfaced with \textsc{Pythia6} and \textsc{Herwig++}. We study the differential scale dependence and $K$-factors, and investigate 
 the effects of the parton showers for a benchmark scenario in the constrained MSSM.
\end{abstract}
\thispagestyle{empty}
\vfill
\newpage
\setcounter{page}{1}

\tableofcontents
\newpage
\section{Introduction}
\label{ch:introduction}
Supersymmetry (SUSY) \cite{Volkov,golfand,Wess,Sohnius,Nilles,HaberKane,Gunion1,Gunion2,Gunion3} is one of the most attractive extensions of the Standard Model (SM). Besides its theoretical appeal, SUSY can provide an explanation for conceptual problems and observations which cannot be accommodated within the SM. These include, for example, the hierarchy problem and the existence of dark matter, which emerges naturally in SUSY with R-parity conservation. With the start of the LHC, the direct search for SUSY has entered a new era. It is now possible to 
discover (or exclude) SUSY particles in the TeV mass range favoured by the solution to the hierarchy problem and dark matter.
The main SUSY production processes at the LHC in  R-parity conserving SUSY models are the pair production of the strongly interacting squarks ($\sq$) and gluinos
($\go$), {\it i.e.} the processes $pp \to \sq\sq$, $\sq\overline{\sq}$, $\sq\go$ and $\go\go$.

The leading order (LO) cross section predictions for the pair production of strongly interacting SUSY particles in hadron collisions were first calculated some 
time ago \cite{squarklo1,squarklo2,squarklo3,squarklo4}. The calculation of the next-to-leading order (NLO) SUSY-QCD corrections has been performed 
in \cite{squarknlo1,squarknlo2,prospino,squarknlo3}, assuming all squarks to be degenerate in mass (except for stop pair production, 
where all squarks apart from the stop have been assumed to be degenerate). The NLO corrections have been found to be positive 
and in general large, between $5\%$ and $90\%$ depending on the process and the input parameters. The inclusion of the NLO corrections is required for quantitative phenomenological studies not only because of the large corrections, but also because the higher-order contributions reduce the dependence of the prediction on the unphysical factorization and renormalization scales from about $\pm50\%$ at LO to 
typically $\pm15\%$ at NLO. Recently, a calculation of squark 
pair production without any assumptions on the sparticle spectrum has been published \cite{hollik,hollik2}, including the subsequent decay of each squark into a quark and the lightest neutralino with NLO 
corrections in both production and decay. Furthermore, completely general NLO predictions for squark and gluino production based 
on the \textsc{MadGolem} framework have been presented and compared to resummed predictions from jet merging \cite{plehn}.
In the past years a lot of effort has been put in calculating the production processes beyond NLO, taking into account resummation and threshold effects
\cite{sqbnlo1,sqbnlo2,sqbnlo3,sqbnlo4,sqbnlo5,sqbnlo6,sqbnlo7,sqbnlo9,sqbnlo10,sqthresh1,sqthresh2,sqthresh3}. These corrections can increase the inclusive cross section
by up to $10\%$ and lead to a further reduction of the scale uncertainty. Furthermore, electroweak contributions have been considered at LO
\cite{ewlo1,ewlo2} and at NLO \cite{ewnlo1,ewnlo2,ewnlo3,ewnlo4,ewnlo5,ewnlo6,ewnlo7}. These corrections can be significant, but strongly depend on 
the model parameters and the flavour and chirality of the produced squarks.\\

The LO cross sections and NLO corrections in SUSY-QCD can be calculated with the publicly available computer program {\tt Prospino} \cite{prospino_manual}. 
Since the program is based on the calculations in \cite{prospino,squarknlo3} the NLO corrections can only be evaluated for degenerate squark masses. 
Furthermore, these corrections are implemented such that the various subchannels, characterized by different flavour and chirality combinations,
are always summed up. Results for individual subchannels can be returned, but these are obtained by scaling the exact LO cross section 
for the specific subchannel with the global $K$-factor, the ratio of the total NLO cross section and the total LO cross section, obtained for degenerate squark masses.\footnote{Note that this is only true for the second version of {\tt Prospino}, in the following denoted {\tt Prospino2}. The original version instead returns the LO and NLO results for all subchannels summed up, although it could be modified in principle such that the different channels are calculated separately.} This approach is based 
on the assumption that the $K$-factors do not vary significantly between the different subchannels. Besides the NLO corrections to the total cross section, NLO
differential distributions in transverse momentum and rapidity of the produced SUSY particles have been presented in \cite{prospino}.  
It was found that for these distributions, and for the SUSY scenarios considered, the NLO corrections mainly scale the LO distribution
by a global $K$-factor, with shape distortions of at most $\matO(10\%)$. Based on these results it has been assumed that differential $K$-factors are rather flat in general.\\

In the first part of this paper the calculation of squark pair production for squarks of the first two generations is presented at NLO in SUSY-QCD without any assumptions
on the squark masses. All subchannels are treated individually and the results are implemented in a parton-level Monte-Carlo program, 
which allows to calculate arbitrary distributions at NLO \cite{mythesis}. Our calculation for squark pair production is understood as the first step towards the calculation and
implementation of all squark and gluino production channels at NLO in a fully flexible partonic Monte-Carlo program. 
Anticipating to include SUSY-QCD corrections also in the decays of the produced particles, squark-squark production constitutes an excellent channel 
for setting up the framework for this project. Since squarks are scalar particles, no spin correlations have to be taken into account when decays of the 
squarks are added. Additionally, as illustrated in \cite{sqbnlo8,sqbnlo11}, squark pair production is the dominant channel in the higher mass region for squarks 
and gluinos, which is probed in the current and upcoming searches at the LHC. Our calculation is completely independent of the calculation of squark pair production at NLO presented in \cite{hollik}, since the methods used to treat and cancel the soft and collinear divergences in the virtual and real corrections are different: in our calculation we apply the Catani-Seymour 
subtraction formalism whereas in \cite{hollik} phase space slicing has been used. Moreover, we present a new approach to handle contributions with intermediate on-shell $\go$ and compare the results with the existing methods.\\

Besides calculating higher-order corrections in perturbation theory, it is mandatory to combine these fixed order parton level results with a parton shower to obtain more precise predictions 
for measurements at hadron colliders. The combination of a fixed order NLO calculation with the all-order effects of a parton shower is non-trivial, as
the double counting of contributions contained in both the NLO result and the parton shower has to be avoided  (see {\it e.g.} \cite{skandsQCDforcolliders}). Several methods exist 
to perform such a matching consistently, the two most widely used being \textsc{MC@NLO} (see \cite{mcatnlo}) and \textsc{Powheg} (see \cite{nason} and \cite{powheg} for a detailed 
description). We follow the \textsc{Powheg} method and use the program package \PB~\cite{powhegbox} as a framework for matching our NLO calculation for squark pair production with parton showers.\\

The paper is organized as follows: Sec.~\ref{ch:nlo} is devoted to the details of the NLO calculation. Besides the standard problems of treating ultraviolet (UV) divergences in the virtual parts and 
canceling infrared (IR) divergences in the real contributions, another type of divergences related to intermediate on-shell gluinos emerges in some channels for the real parts, and requires 
a non-trivial subtraction formalism. The implementation in the \PB~is described in Section~\ref{ch:pwg}. Section~\ref{ch:res} summarizes our main findings. In addition to the discussion of the pure NLO effects, we investigate the impact of different parton showers by interfacing our results with three shower programs: the $p_T$-ordered shower from \textsc{Pythia6}
\cite{pythia6} and both the default and the Dipole Shower of \textsc{Herwig++} \cite{herwigpp,herwigpp26,herwigdp1,herwigdp2}. Our conclusions are given in Sec.~\ref{ch:conclusion}.

\section{Squark Pair Production at NLO}
\label{ch:nlo}

\subsection{Elements of the NLO Calculation}

\begin{figure}[t]
  \centering
  \includegraphics[width=4.2cm]{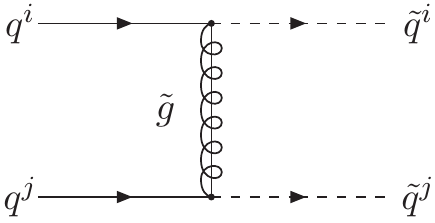}
  \hspace{1.2cm}
  \includegraphics[width=4.2cm]{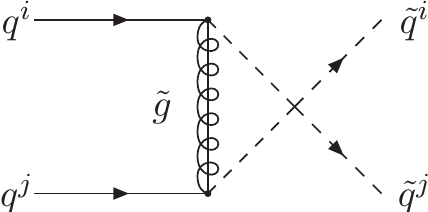}\\
  (a) \hspace{4.8cm} (b)
  \caption{Feynman diagrams contributing to LO squark pair production via {\it t}-channel (a) or {\it u}-channel (b) exchange of a gluino. The latter does not contribute to the production of 
  squarks with different flavour.}
  \label{fig:LOdiags}
\end{figure}

At LO the pair production of squarks of the first two generations proceeds through two quarks in the initial state:
\begin{equation}
 q^i + q^j \to \sq^i +  \sq^j \,,
\end{equation} 
where the indices $i,j$ characterize the flavour and chirality of the corresponding particle.
The Feynman diagrams contributing to this process are depicted in Fig.~\ref{fig:LOdiags}.
In the following we take into account only the production of squarks of the first two generations ($\su,\sd,\tilde{c},\tilde{s}$). The corresponding quarks are treated as massless. The amplitudes depend on the flavours and chiralities of the particles and can be categorized
into four different subchannels: The first two are those where the squarks have the same flavour, and the same or different chiralities. 
The $u$-channel in Fig.~\ref{fig:LOdiags} (b) only contributes to these two subchannels. The remaining two categories of subchannels are those where
the squarks have different flavour, and the same or different chiralities.\\

\begin{figure}[t]
  \centering
  \vspace{0.6cm}
  \includegraphics[width=4cm]{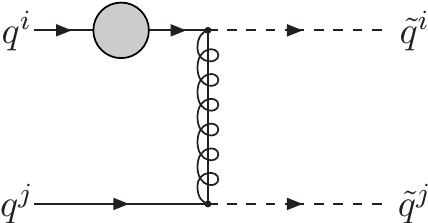}
  \hspace{1cm}
  \includegraphics[width=4cm]{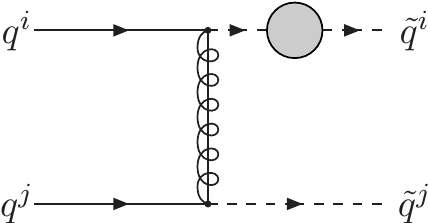}
  \hspace{1cm}
  \includegraphics[width=4cm]{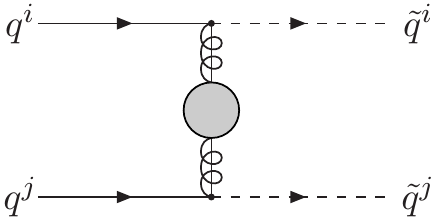}\\
  (a) \hspace{4.5cm} (b) \hspace{4.5cm} (c) \\[0.3cm]
  \includegraphics[width=4cm]{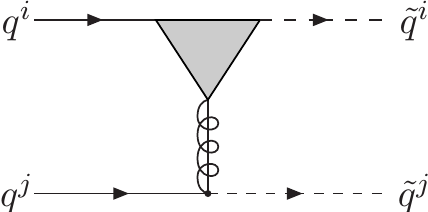}
  \hspace{1cm}
  \includegraphics[width=4cm]{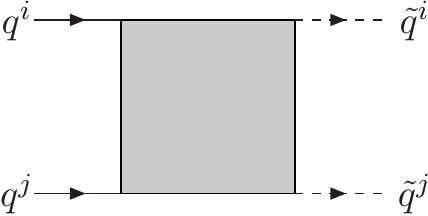}\\
  (d) \hspace{4.5cm} (e)
  \caption{Generic Feynman diagrams for virtual corrections like quark (a), squark (b), gluino (c) self-energies, vertex corrections (d) and box diagrams (e).}
  \label{fig:virtgen}
\end{figure}

\begin{figure}[h!]
  \vspace{1.5cm}
  \includegraphics[width=15.5cm]{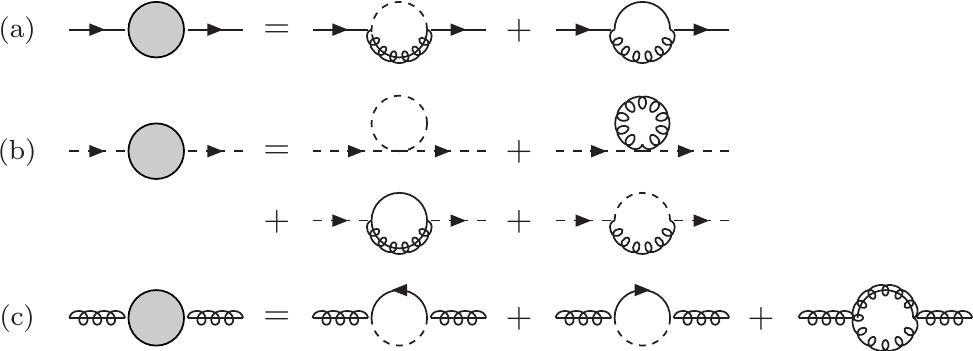}\\[1.5cm]
  \includegraphics[width=12cm]{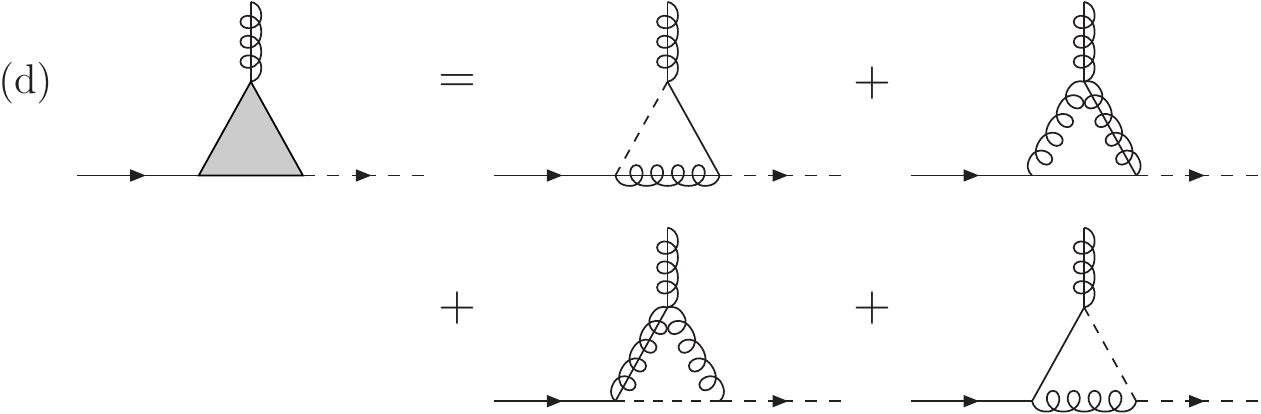}\\[1.5cm]
  \includegraphics[width=12cm]{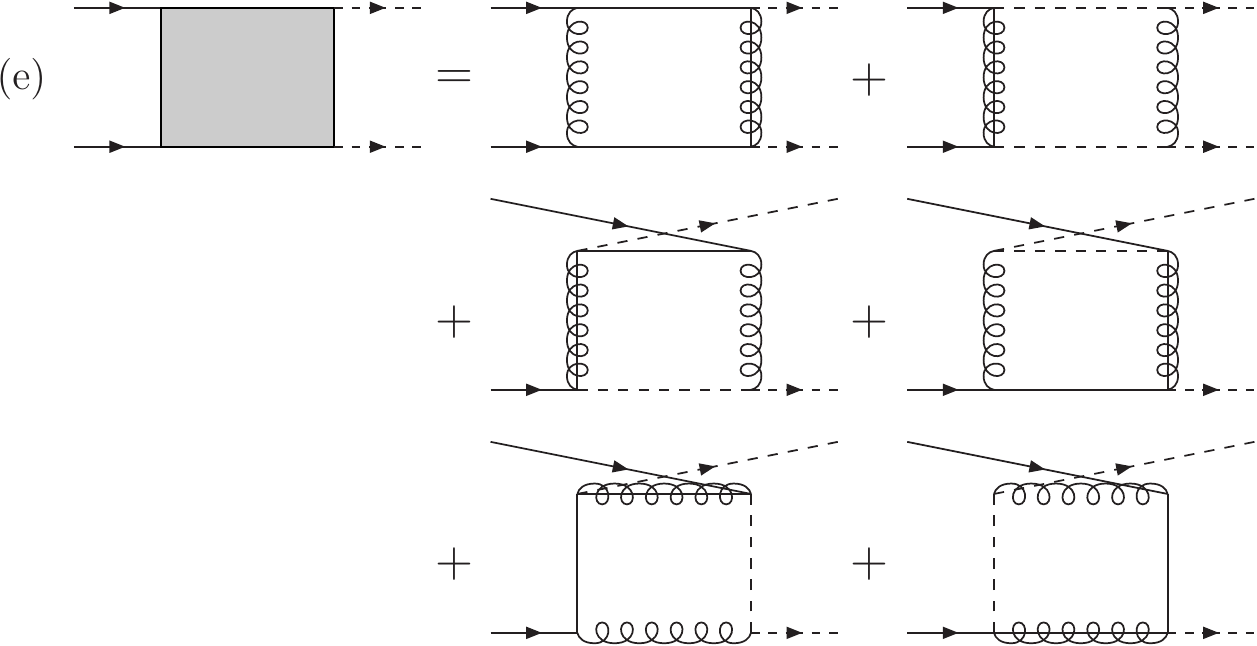}
  \caption{Feynman diagrams contributing to quark (a), squark (b) and gluino (c) self-energies, vertex corrections (d) and box diagrams (e). The box diagrams in the last line do not 
  contribute when both squarks have different flavours.}
  \label{fig:virtself}
\end{figure}

Squark pair production at NLO receives contributions from real emissions of one additional parton, a gluon or anti-quark, as well as from
virtual SUSY-QCD (SQCD) corrections. The virtual corrections to squark pair production consist of gluino, quark and squark self-energies, vertex corrections and 
box diagrams. Generic Feynman diagrams for these corrections are depicted in Fig.~\ref{fig:virtgen}. The individual Feynman diagrams
contributing to the quark, squark and gluino self-energies, to the vertex correction diagrams and to the box diagrams are listed in Fig.~\ref{fig:virtself}. 
Note that the diagrams in the last line of this figure do not contribute when both squarks have different flavours.\\

The loop diagrams of the self-energies and vertex corrections lead to UV divergencies. We use dimensional regularisation \cite{dimreg}
to handle these UV divergencies. Dimensional regularisation is a convenient regularisation scheme because it respects all gauge symmetries. However, it breaks SUSY as it introduces
a mismatch between fermionic and bosonic degrees of freedom. Invariance under SUSY transformations inquires the strong gauge coupling $g_s$ and 
the SUSY Yukawa coupling $\hat{g}_s$ be equal to all orders in perturbation theory for large scales. At one-loop level, when using dimensional regularisation, 
this relation is violated and needs to be restored by adding a finite counterterm \cite{martinvaughn},
\begin{equation} \hat{g}_s = g_s \left(1+\frac{\alpha_s}{3\pi} \right). \end{equation}
The UV divergencies can be absorbed by introducing renormalization constants for the non-vanishing squark and gluino masses, the quark, squark and 
gluino fields and the strong coupling constant. For the mass and field renormalization constants we choose the on-shell renormalization conditions. 
In case of the strong coupling constant we work in the $\overline{\textnormal{MS}}$-scheme \cite{msbar}, where only the $1/\epsilon$ UV poles 
along with some universal terms are absorbed into the counterterm $\delta g_s$ which relates the bare strong coupling $g^{(0)}_s$ and the renormalized coupling $g_s$ according to
\[g^{(0)}_s = g_s + \delta g_s \ .\] 
The counterterm $\delta g_s$ is determined from the transverse part of the gluon self energy, 
which contains contributions from SM as well as SUSY particles. The experimental value of $\alpha_s$ is given in SM QCD with
five active quark flavours at the scale of the $Z$ 
boson mass \cite{pdg}. We have decoupled the heavy squarks and gluinos as well 
as the top quark from the running of $\alpha_s$ in order to avoid 
artificial large logarithms in our calculation.
This can be accomplished by subtracting the logarithms of the masses of the heavy particles \cite{Collins}, hence
\begin{equation} 
\delta g_s=\frac{\alpha_s}{8 \pi} \left[ \beta_0 \left(-\Delta+\log \frac{\mu_R^2}{\mu^2}\right)- 2 \log \frac{m_{\gl}^2}{\mu_R^2} -\frac{2}{3} \log \frac{m_t^2}{\mu_R^2} - \sum_{i=1,12} \frac{1}{6} \log \frac{m_{\sq_i}^2}{\mu_R^2} \right]
\end{equation}
with 
\begin{equation*}
\beta_0 = \left[11-\frac{2}{3}\cdot 5 \right] + \left[-2-\frac{2}{3}-\frac{1}{6}\cdot 12 \right]
\end{equation*}
and
\[ \Delta = 1/\epsilon - \gamma + \log 4\pi \]
denoting the UV pole and the universal constants that have been absorbed in the counterterm together with a logarithm of the renormalization scale $\mu_R$ over the 't Hooft scale $\mu$. Here, $\gamma$
is the Euler-Mascheroni constant and $\beta_0$ the one-loop beta function coefficient. This definition of $\delta g_s$ assures that only the gluon and the five light quarks 
contribute to the running of $\alpha_s$.\\
 
The code for the LO amplitude and the virtual corrections has been generated with the \textsc{Mathematica} packages \textsc{FeynArts 3.5} \cite{feynarts,feynartsmssm} 
 and \textsc{FormCalc 6.1} \cite{formcalc1,formcalc2}. The one-loop integrals in the calculation are evaluated by the program package \textsc{LoopTools 2.6} \cite{formcalc1}. 
\textsc{Feyn\-Arts} provides a model file with the Feynman rules of the MSSM. In contrast to the model file of the SM, in the MSSM model file no counterterms 
are specified. These have been added according to the renormalization procedure described above. It has been checked explicitly that this procedure 
renders the calculation UV finite.
After canceling all UV divergencies by renormalization the IR divergencies remain. These will cancel against the IR divergencies of the 
real emission diagrams by applying the Catani-Seymour subtraction formalism \cite{cs,cdst}.\\

\begin{figure}[t]
  (a) \hspace{0.3cm} \includegraphics[width=3.7cm]{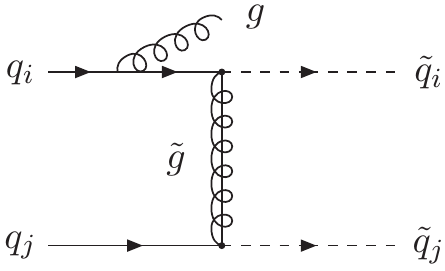} \hspace{1cm}
                   \includegraphics[width=3.7cm]{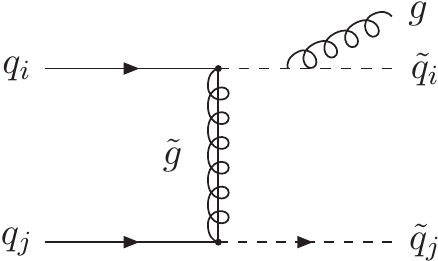}\\[1em]
  (b) \hspace{0.3cm} \includegraphics[width=3.7cm]{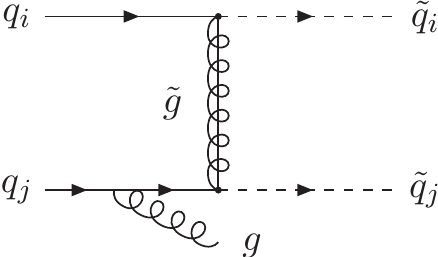} \hspace{1cm}
                   \includegraphics[width=3.7cm]{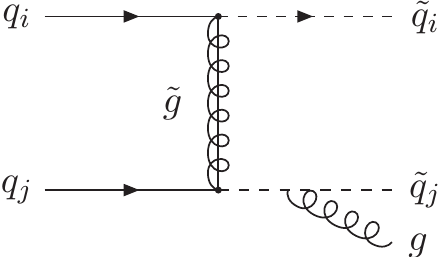}\\[1em]
  (c) \hspace{0.3cm} \includegraphics[width=3.7cm]{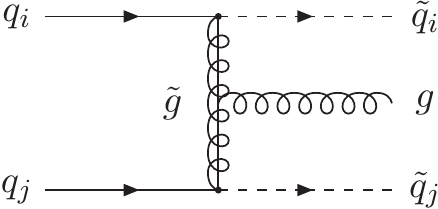}\\
  \caption{Feynman diagrams contributing to real emission matrix elements with $qq$ initial states and an emitted gluon. Diagrams which lead to soft and collinear divergencies are depicted in (a) and (b), the diagram in (c) is IR finite. }
  \label{fig:realdiags_qq}
\end{figure}
\begin{figure}[t]
  (a) \hspace{0.3cm} \includegraphics[width=3.7cm]{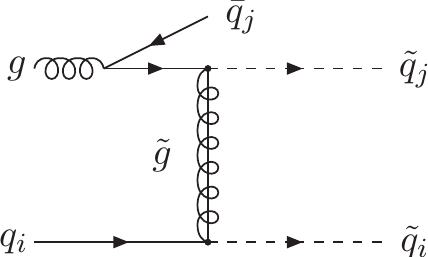}\\[1em]
  (b) \hspace{0.3cm} \includegraphics[width=3.7cm]{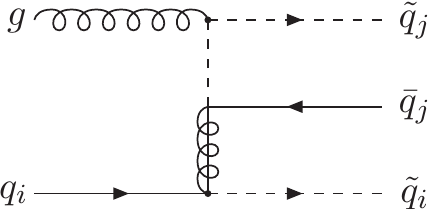}\\[1em]
  (c) \hspace{0.3cm} \includegraphics[width=14.4cm]{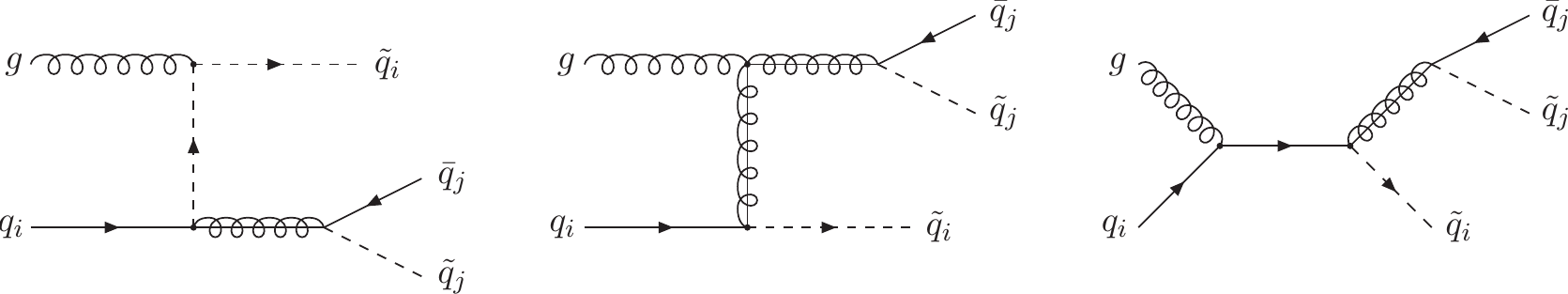} 
   \caption{Feynman diagrams contributing to real emission matrix elements with $qg$ initial states. The diagram in (a) gives rise to collinear singularities. The diagrams in (b) and (c) are IR finite. The diagrams in (c) can contribute to the production of a squark and a resonant gluino.}
  \label{fig:realdiags_qg}
\end{figure}

The matrix elements of the real emission can be classified in two different topologies. The first topology contains diagrams with two quarks in the initial state and an additionally emitted gluon: 
\begin{equation}
q_i \ q_j \to \sq_{i}\ \sq_{j} \ g\ .
\label{eq:qqreals}
\end{equation}
The $t$-channel diagrams contributing to this process are shown in Fig.~\ref{fig:realdiags_qq}. The second topology is comprised of diagrams with a quark and a gluon in the initial state and an emitted, 
massless antiquark. These diagrams are depicted in Fig.~\ref{fig:realdiags_qg}. Apart from implementing the process 
\begin{equation} 
g \ q_i \to \sq_{i} \ \sq_{j}\ \bar{q}_j 
\label{eq:qgreals1}
\end{equation}
it is important to include for $i\neq j$ also 
\begin{equation} 
g \ q_j \to \sq_{i} \ \sq_{j} \ \bar{q}_i 
\label{eq:qgreals2}
\end{equation}
in order to account for all possible initial state configurations.
Both topologies lead to IR/collinear divergencies. Diagrams with $qq$ initial states, 
which contain soft and collinear divergencies, are collected in Figs.~\ref{fig:realdiags_qq} (a) and (b). The diagrams with $qg$ initial states which emit a massless anti-quark, result in collinear divergencies only.
The corresponding diagram is shown in Fig.~\ref{fig:realdiags_qg} (a).\\

The soft and collinear divergencies are subtracted by the Catani-Seymour dipoles which have been generated using the \textsc{SuperAutoDipole 1.0} package
\cite{autodipole,superautodipole}. \textsc{SuperAutoDipole} itself provides an interface with the program \textsc{MadGraph 4.4.30} \cite{madgraph1,madgraph2}, 
which automatically produces a code for the squared matrix elements of the real emission diagrams by calling the \textsc{HELAS} subroutines based on the helicity
amplitude formalism \cite{helas}.\\
The dipoles needed to render the real emission matrix elements finite are organized in pairs of potentially collinear partons with an additional reference 
to a spectator particle. For diagrams with two quarks in the initial state this gives rise to twelve individual dipoles: The emitted gluon can be collinear 
or soft and in each case any of the other three particles in the initial or final state can serve as spectator particle.
For diagrams with a quark and a gluon in the initial state only three dipoles are necessary: The emitted antiquark can only become collinear to the initial
state gluon while the other three particles can act as the spectator particle. Hence, the counterterms $d\sigma^A$ which are subtracted from the squared real emission matrix 
elements read: 
\begin{equation}
 d\sigma^A_{qq} = \sum_{i=1}^{12} \mathcal{D}_i^{qq} \qquad \mbox{and} \qquad  d\sigma^A_{qg} = \sum_{i=1}^{3} \mathcal{D}_i^{qg} \ .
\label{eq:cts}
\end{equation}

The real emission diagrams in Fig.~\ref{fig:realdiags_qg} (c) have to be handled with care in parameter regions where the gluino is heavier than one or both squarks in the final state. In this case these dia\-grams give rise to another kind of singularity since the intermediate gluino can be produced on-shell. 
The subtraction procedure for these divergencies is described in detail in Sec. \ref{sec:onshellsub}.\\

Having subtracted the counterterm $d\sigma^A$ from the real emission matrix elements the IR divergencies in the virtual corrections are still left. 
With the choice of dipoles as published in \cite{cs,cdst} the counterterms in Eq.~(\ref{eq:cts}) can be integrated analytically over the one-parton phase space. 
This integration yields the so-called {\bom I}-terms and \bom{PK}-terms which can be evaluated in the 2-particle phase space used for the Born matrix elements
and virtual corrections. The former contain all the $1/\epsilon$ poles that are necessary to cancel the poles in the virtual contributions. The latter are the 
finite collinear remainders which are left after initial state collinear singularities have been factorized into the non-perturbative parton distribution functions (PDFs) defined in the $\overline{\textnormal{MS}}$-scheme. These \bom{PK}-terms involve an 
additional integration over $x$, which is the longitudinal momentum fraction after the splitting in the initial state.\\ 
The program \textsc{SuperAutoDipole} generates a Fortran code for the {\bom I}-terms as functions of the momenta and masses of the partons. 
It provides a flag in order to separately extract the coefficients of the $1/\epsilon^2$ and $1/\epsilon$ poles as well as the finite parts.
In principle the program \textsc{LoopTools}, which has been used to evaluate the virtual corrections, provides the same feature for the coefficients 
of the poles of the loop diagrams. By combining these two tools it is possible to compare the coefficients of the poles for every phase space point
during the numerical evaluation of the process and check whether the cancellation of the divergencies in the virtual corrections works.\\ 
However, it has to be taken into account that in the code generated by \textsc{LoopTools} the term 
\begin{equation}
\frac{(4 \pi)^{\epsilon}}{\Gamma(1-\epsilon)} = 1+\epsilon \ [\log4\pi-\gamma]+ \epsilon^2 \ \left[ \frac{1}{2} (\log^2 4\pi + \gamma^2) - \frac{\pi^2}{12} -\gamma \cdot \log4\pi \right] +\mathcal{O} (\epsilon^3)
\end{equation}  
has been factored out. In order to achieve agreement between the coefficients of the poles from the virtual corrections and the {\bom I}-terms this factor
has to be added back in by hand. This changes the coefficient $C_{-1}$ of the $1/\epsilon$ poles and the finite part $C_0$ of the virtual corrections 
calculated by \textsc{LoopTools}:
\begin{eqnarray}
  && \left( 1+\epsilon \ [\log4\pi-\gamma] + \epsilon^2 \ \left[ \frac{1}{2} (\log^2 4\pi + \gamma^2) - \frac{\pi^2}{12} -\gamma \cdot \log4\pi \right] \right)\ \left(C_{-2} \frac{1}{\epsilon^2} + C_{-1} \frac{1}{\epsilon} + C_0 \right)=  \nn \\
  && \Bigl( C_0 + C_{-1}\ [\log4\pi-\gamma] + C_{-2}\ \left[ \frac{1}{2} (\log^2 4\pi + \gamma^2) - \frac{\pi^2}{12} -\gamma \cdot \log4\pi \right] \Bigr) + \nn \\
  && \Bigl( C_{-1} + C_{-2} \ [\log4\pi-\gamma] \Bigr) \ \frac{1}{\epsilon} + C_{-2} \ \frac{1}{\epsilon^2}  \ .
\end{eqnarray}
With this modification the cancellation of the IR divergencies from the virtual corrections by subtracting the integrated Catani-Seymour dipoles can be carried
out successfully.\footnote{Furthermore, in some scalar integrals where a UV and an IR pole cancel each other, like {\it e.g.} in the $B_0 (0,0,0)$ function, the pole structure had 
to be restored by hand in \textsc{LoopTools}.}\\ 
Implementing the finite collinear remainder terms in the numerical evaluation as part of the 2-particle phase space has a drawback. 
The code slows down dramatically as for every phase space point an additional integration over the longitudinal momentum fraction $x$ has to be carried out. 
To reduce the computing time it is convenient to perform the integration over the 3-particle rather than the 2-particle phase space with an additional integration over $x$. By taking advantage of the fact that the phase space factorizes, the Born-level phase space can be mapped onto the
real emission phase space and the \bom{PK}-terms can be integrated together with the real emission matrix elements and dipoles \cite{dieter&co}. Apart from 
speeding up the numerics, this factorization of the phase space allows for consistency checks of the program, since the finite collinear cross section 
can be determined either as part of the 2-particle or as part of the 3-particle phase space. 

\newpage
\subsection{Subtraction of On-shell Intermediate Gluinos}
\label{sec:onshellsub}
 The gluon-initiated real channels ({\it cf.}~Fig.~\ref{fig:realdiags_qg}) $q_i g \rightarrow \sq_i \sq_j \qbar_j$ give rise to another type of singularity: for $m_{\sq_j}<m_{\go}$ the intermediate $\go$ in the diagrams (c) can be produced on-shell. In principle, the resulting divergence originating from the $\go$ propagator can be cured by the introduction of a finite $\go$ width $\Gamma_{\go}$,
  \be
   \label{eq:bwprop}
     \frac{1}{(p_{\sq_j}+p_{\qbar_j})^2-m_{\go}^2}\rightarrow\frac{1}{(p_{\sq_j}+p_{\qbar_j})^2-m_{\go}^2+i m_{\go}\Gamma_{\go}}\quad.
  \ee
But looking at these resonant contributions from a different point of view, they correspond to the LO production $q_i g \rightarrow \sq_i \go$, followed by the decay $\go\rightarrow \sq_j \qbar_j$. Keeping it as part of the real corrections to $\sq\sq$ production would spoil the predictivity of the NLO calculation, as for a very large region in the parameter space this resonant contribution easily exceeds the full NLO corrections. Moreover, considering all SQCD pair production channels (notably $\sq\go$ production) and their subsequent decays, these channels would be double counted. Therefore, these contributions have to be removed in a consistent way.\\

  The general structure of the $qg$ channels can be written as
  \be
  \label{eq:ossme}
    |M_{\text{tot}}|^2 = |M_{nr}|^2+2\,\text{Re}(M_{r}M_{nr}^*)+|M_r|^2,
  \ee
  where $M_{nr}$ comprises the non-resonant diagrams (denoted (a) and (b) in Fig.~\ref{fig:realdiags_qg}), and the resonant ones are combined in $M_{r}$. Note that in case of the production of same flavour $\sq$ both of them can lead to a resonant behaviour (if both $m_{\sq_1}<m_{\go}$ and $m_{\sq_2}<m_{\go}$), {\it i.e.} there are two resonant regions to consider. To simplify the following considerations, we will discuss only the case with one singular region, the result for two singular regions is obtained by taking Eq.~(\ref{eq:ossme}) into account twice, with $\sq_1\leftrightarrow\sq_2$, and adding for identical $\sq$ the additional interference terms between these two contributions.

  On-shell intermediate states which require a subtraction formalism are not a unique feature of SQCD pair production processes, but occur in other processes, too. There exist several methods to cope with them, the most relevant ones for Monte Carlo (MC) event generators being the following:
  \bi
    \item \textbf{Diagram Removal - type I (DR-I):}
      This approach was first used in the context of $tW$ production, see \cite{twmcatnlo}. It simply amounts to leave out all resonant diagrams, {\it i.e.} not only $|M_{r}|^2$ but also the interference term $2\, \text{Re}(M_{r}M_{nr}^*)$ is completely removed.

    \item \textbf{Diagram Removal - type II (DR-II):} This method was proposed in a recent calculation of the NLO corrections to $\sq\sq$ production \cite{hollik}. Here, only the $|M_{r}|^2$ part is dropped, whereas the interference term is kept. The interference terms between contributions originating from two resonant regions, which occur solely for identical $\sq$, are taken into account.
  \ei
  Both approaches are easy to implement in a MC event generator, but obviously break gauge invariance and therefore give in principle arbitrary results, as it is not guaranteed that the neglected terms are small.
  \bi
    \item \textbf{Diagram Subtraction (DS):} In this approach a \lq counterterm' is introduced which removes the resonant parts for $(p_{\sq_j}+p_{\qbar_j})^2\rightarrow m_{\go}^2$ locally, {\it i.e.} only the contributions originating from on-shell gluinos are subtracted. This method retains both the interference terms and off-shell contributions from $|M_{r}|^2$. Furthermore, by construction it allows a pointwise subtraction, and thus represents an ideal method for MC event generators. However, it respects gauge invariance only in the limit $\Gamma_{\go}\rightarrow 0$, if the width is introduced by simply replacing the resonant propagator as sketched in Eq.~(\ref{eq:bwprop}).
  \ei
  
 To obtain a fully gauge invariant result, we modified the DS method such that the (gauge dependent) matrix elements are no longer used as building blocks. Instead, we extract the poles in $(p_{\sq_j}+p_{\qbar_j})^2-m_{\go}^2\equiv s_{jg}$ analytically after choosing a specific phase space parametrisation in terms of invariants:
 \be
   |M_{\text{tot}}|^2 = \frac{f_0}{s_{jg}^2}+\frac{f_1}{s_{jg}}+f_2(s_{jg}).
 \ee
 The coefficients $f_k$ ($k=0,1,2$) are gauge invariant quantities, {\it i.e.} introducing a regulator $\Gamma_{\go}$ at this point preserves gauge invariance and we get
  \be
   \label{eq:expan}
   |M_{\text{tot}}|^2 = \frac{f_0}{s_{jg}^2+m_{\go}^2\Gamma_{\go}^2}+\frac{s_{jg}}{s_{jg}^2+m_{\go}^2\Gamma_{\go}^2} f_1 +f_2(s_{jg}).
 \ee
 
 Comparing this expression with the one obtained by introducing $\Gamma_{\go}$ at the level of matrix elements we can quantify the difference $\Delta(\Gamma_{\go},s_{jg})$ between the two methods, which gives indirectly a measure for the \lq gauge dependence' of the result:
 \be
  \label{eq:gaugedep}
   \Delta(\Gamma_{\go},s_{jg}) = \tilde{f}_2(s_{jg})\frac{m_{\go}^2\Gamma_{\go}^2}{s_{jg}^2+m_{\go}^2\Gamma_{\go}^2},
 \ee
 where $\tilde{f}_2(s_{jg})$ comprises the parts of $f_2(s_{jg})$ which originate from $2\,\text{Re}(M_{r}M_{nr}^*)+|M_r|^2$. For $\Gamma_{\go}\rightarrow 0$ the results are equivalent, but close to the resonant region the discrepancy is solely determined by the gauge dependent quantity $\tilde{f}_2(s_{jg})$.
 
 As mentioned above, the results for same flavour $\sq$ can be obtained by taking these terms with $\sq_1\leftrightarrow \sq_2$ into account twice. The additional interference terms between these two contributions for identical $\sq$ lead to terms $\propto 1/(s_{1g} s_{2g})$, again with $s_{kg}\equiv (p_{\sq_k}+p_{\qbar})^2-m_{\go}^2$. These terms arise from the interference of the two resonant parts $M_{r,1}$ and $M_{r,2}$, where the first/second $\sq$ couples to the resonant $\go$, respectively. They do not require any subtraction, however the singular structure requires again the introduction of a regularising width $\Gamma_{\go}$. To this end these terms are expanded in both $1/s_{1g}$ and $1/s_{2g}$ before this regulator is introduced:
  \begin{align}
    2\,\text{Re}(M_{r,1}M_{r,2}^*) = &\frac{s_{1g} s_{2g}+m_{\go}^2\Gamma_{\go}^2}{(s_{1g}^2+m_{\go}^2\Gamma_{\go}^2)(s_{2g}^2+m_{\go}^2\Gamma_{\go}^2)} g_0 + \frac{s_{1g}}{s_{1g}^2+m_{\go}^2\Gamma_{\go}^2}\tilde{g}_1(s_{2g})\notag\\ &+ \frac{s_{2g}}{s_{2g}^2+m_{\go}^2\Gamma_{\go}^2}\tilde{g}_2(s_{1g}) 
    + \tilde{g}_3(s_{1g},s_{2g}).
  \end{align}
  Correspondingly, the interference terms between the non-resonant and the resonant terms are expanded in either $1/s_{1g}$ or $1/s_{2g}$, depending on the type of the singular structure. Together with the interference terms from non-resonant contributions the expansion coefficients obtained in this way render the expressions $\propto \tilde{g}_{1,2,3}$ gauge invariant. 
  
  Considering again the difference between this expanded gauge invariant expression and the one obtained by performing the replacement Eq.~(\ref{eq:bwprop}) directly in the matrix elements this contribution yields additional terms. Qualitatively the effect of these terms is the same as in Eq.~(\ref{eq:gaugedep}): For $\Gamma_{\go}\rightarrow 0$ they vanish as expected, but close to the resonant region the difference is determined solely by the gauge dependent coefficients $\tilde{g}_{1,2,3}$.
 
 This modified DS method (in the following denoted as $\text{DS}^*$) is in principle equivalent to the method used originally in the implementation of NLO corrections to SQCD pair production processes in \textsc{Prospino} \cite{prospino}. However, the actual implementation of the DS$(^*)$ scheme in a MC generator is quite involved. In the following we will make some remarks on the different building blocks required for both the original and the modified DS scheme. Note that the actually subtracted quantity is in both schemes identical, as appropriate for an unambiguous subtraction scheme. For more details on the implementation of the (original) DS scheme see \cite{twmcatnlo}.\\

 The general form of the subtraction term for the DS method can be written as follows\footnote{We discuss again only the case of one singular region, the generalization to two singular configurations as needed in the same flavour case is straightforward.}:
   \be
      d\sigma_{\text{sub}} = \Theta(\sqrt{\hat s}-m_{\go}-m_{\sq_i})\,\Theta(m_{\go}-m_{\sq_j})\,|M_r(\tilde{\varPhi}_3)|^2\,\frac{m_{\go}^2\Gamma_{\go}^2}{(m_{\sq_j \qbar_j}^2-m_{\go}^2)^2+m_{\go}^2\Gamma_{\go}^2}\, d\tilde{\varPhi}_3
   \ee
   with the invariant mass of the $\go$ defined as $m_{\sq_j \qbar_j}^2 = (p_{\sq_j}+p_{\qbar_j})^2$.
 Correspondingly we obtain for the DS$^*$ scheme
    \be
      d\sigma_{\text{sub}} = \Theta(\sqrt{\hat s}-m_{\go}-m_{\sq_i})\,\Theta(m_{\go}-m_{\sq_j})\,\frac{f_0(\tilde{\varPhi}_3)}{(m_{\sq_j \qbar_j}^2-m_{\go}^2)^2+m_{\go}^2\Gamma_{\go}^2}\, d\tilde{\varPhi}_3.
   \ee
 
   The different elements guarantee the following properties:
   \bi
      \item The case of an on-shell intermediate $\go$ can only occur if the energy in the partonic center-of-mass system is sufficient to generate both an on-shell $\go$ and the $\sq_i$ not originating from the \lq $\go$ decay'. This is ensured by the first step-function, $\Theta(\sqrt{\hat s}-m_{\go}-m_{\sq_i})$.
      \item Only the case $m_{\go}>m_{\sq_j}$ requires subtraction, which is ensured by the factor $\Theta(m_{\go}-m_{\sq_j})$. This is a non-trivial restriction only in the case of same flavour $\sq$ with different chiralities for a hierarchy like $m_{\sq_1}<m_{\go}<m_{\sq_2}$. In all other cases there is either only one type of $\sq$ involved, or flavour conservation dictates which $\sq$ can originate from the on-shell $\go$.
      \item The choice $d\sigma_{\text{sub}}\propto  |M_r|^2$ ensures the exact cancellation of the $\sq_i \go$ contribution in the limit $m_{\sq_j \qbar_j} \rightarrow m_{\go}$. In this limit this term reproduces the term $\propto f_0$ in the analytical expansion, see Eq.~(\ref{eq:expan}), {\it i.e.} the subtraction term in both approaches is indeed identical. Moreover, using the full amplitude squared retains spin correlations.
      \item The subtraction term is supposed to remove only contributions with $ m_{\sq_j \qbar_j} =m_{\go}$. An arbitrary phase space point in the 3-particle phase space $\varPhi_3$ will usually not fulfil this criterion. Therefore the kinematics has to be adapted appropriately by a mapping $\varPhi_3\rightarrow\tilde{\varPhi}_3$. Besides putting the $\go$ on its mass-shell, this momentum reshuffling has to respect energy-momentum conservation. Furthermore, it should preserve the on-shell conditions for the final state squarks and become an identity transformation for $ m_{\sq_j \qbar_j} =m_{\go}$.  This situation is similar to the construction of the transformed kinematics in the Catani-Seymour formalism. Therefore we adopted the formulae for the case where both the spectator and the emitter are final state massive particles from \cite{cdst} to construct the momenta of the $\sq_i$, $\tilde{p}_{\sq_i}$, and of the intermediate $\go$, $\tilde{p}_{\go}$. The momenta of $\sq_j$ and $\qbar_j$ are then obtained 
      by performing the \lq decay' of the $\go$ in its rest frame, preserving the original direction of $\sq_j$, and boosting the result along $\tilde{p}_{\go}$.
      \item In the limit $\Gamma_{\go}\rightarrow 0$, the subtracted term has to reduce to
      \be
          \hat{\sigma}_{\sq\go} \,  BR(\go\rightarrow \sq\qbar),
      \ee
      which requires the Breit-Wigner form of the (squared) $\go$ propagator, as
      \be
         \frac{m_{\go} \Gamma_{\go}}{(m_{\sq_j \qbar_j}^2-m_{\go}^2)^2+m_{\go}^2\, \Gamma_{\go}^2}\stackrel{\Gamma_{\go}\rightarrow 0}{\longrightarrow} \pi \delta\left(m_{\sq_j \qbar_j}^2-m_{\go}^2\right)
      \ee
      leads to $m_{\sq_j \qbar_j}^2=m_{\go}^2$ upon integration over $m_{\sq_j \qbar_j}^2$\footnote{Note that this holds strictly speaking only if the range of integration for $m_{\sq_j \qbar_j}^2$ comprises the complete real axis. The physical phase space boundaries for $m_{\sq_j \qbar_j}^2$, however, are finite. For a discussion of the size of these (usually small) effects see appendix D of \cite{hollik}.}.  
      The reshuffling procedure obviously destroys this form in $|M_r|^2$, hence it has to be restored in the DS scheme explicitly. As actual value of $\Gamma_{\go}$ we do not use the physical width, but consider this parameter as a pure regularisation parameter, which is chosen such that the result is independent of its value. Note that we introduce a non-vanishing $\Gamma_{\go}$ solely where it is necessary, {\it i.e.} in $M_r$, but not in $M_{nr}$, as this would change the IR behaviour and invalidate the cancellation of these divergencies against the Catani-Seymour subtraction terms. Moreover, terms linear in $\Gamma_{\go}$ which appear in the interference term $M_r\, M_{nr}^*$ are discarded, as we only aim to reproduce the first term in an expansion in $\Gamma_{\go}/m_{\go}$, {\it i.e.} we consider the limit $\Gamma_{\go}\rightarrow 0$. As the separation of the different terms and the correct treatment of the $\go$ width (especially in the interference 
term) in an implementation based completely on \textsc{MadGraph} routines is quite involved, we calculated the real amplitudes squared analytically with the help of \textsc{FeynCalc} \cite{feyncalc}. In the DS$^*$ scheme the terms containing $\Gamma_{\go}$ are unambiguously determined by the construction of the expansion.
    
      \item The last subtlety in the implementation is the form of the Jacobian for the MC integration over the 3-particle phase space. While the applied formalism for the reshuffling of the final-state kinematics guarantees that the transformed momenta lie within a \lq restricted' phase space, {\it i.e.} they fulfil $(p_{\sq_j}+p_{\qbar_j})^2=m_{\go}^2$ by construction, a naive implementation of the subtraction term in the integral over the whole phase space would not only remove on-shell contributions, but also off-shell terms, if the integration limits are not adapted appropriately. To clarify this point, consider a specific parametrisation of the 3-particle phase space with 2 invariants (chosen as $s_2=(p_{\sq_j}+p_{\qbar_j})^2$ and $t_1=(p_g-p_{\sq_i})^2$) and 2 angles which describe the $\go$ decay, see \cite{byckling}. In terms of these integration variables the phase space element has the form
      \be
        \frac{d\varPhi_3}{d\Omega}\propto\int_{s_2^-}^{s_2^+}  ds_2 \int_{t_1^-(s_2)}^{t_1^+(s_2)}dt_1 \frac{s_2-m_{\sq_i}^2}{s_2}= \int_{s_2^-}^{s_2^+} ds_2 \int_{0}^{1}dx \frac{s_2-m_{\sq_i}^2}{s_2} \lambda^{1/2}(\hat{s},s_2,m_{\sq_i}^2)
        \label{eq:3pps}
      \ee
      where the integration over $t_1$ has been mapped on the interval $[0,1]$ as needed for a MC integration\footnote{Considering the Breit-Wigner form of the integrand, the integration over $s_2$ should be mapped such that the resonant region is probed efficiently. A convenient way to achieve this is the Breit-Wigner-mapping: \\
      $s_2= m_{\go}^2+m_{\go} \Gamma_{\go} \tan(y)$ with $y=\left[\tan^{-1}\left(\frac{s_2^+-m_{\go}^2}{m_{\go} \Gamma_{\go}}  \right)-\tan^{-1}\left(\frac{s_2^--m_{\go}^2}{m_{\go} \Gamma_{\go}}  \right)\right]\, x + \tan^{-1}\left(\frac{s_2^--m_{\go}^2}{m_{\go} \Gamma_{\go}}  \right),\, x\in[0,1]$.}, {\it i.e.} $t_1(s_2)=(t_1^+(s_2)-t_1^-(s_2))\, x + t_1^-(s_2)$ with $(t_1^+(s_2)-t_1^-(s_2))=\lambda^{1/2}(\hat{s},s_2,m_{\sq_i}^2)$ and $\lambda(x,y,z)=x^2+y^2+z^2-2xy-2xz-2yz$. Using the same parametrisation for the phase space integration of the subtraction term with its reshuffled kinematics $\tilde{\varPhi}_3$, one has to take into account that in the \lq restricted' phase space with $s_2=m_{\go}$ the Jacobian has to be rescaled according to the replacement $s_2\rightarrow m_{\go}$ in Eq.~(\ref{eq:3pps}):
     \be
       \label{eq:jaco}
       d\tilde{\varPhi}_3 = d\varPhi_3 \frac{\lambda^{1/2}(\hat{s},m_{\go}^2,m_{\sq_i}^2)}{\lambda^{1/2}(\hat{s},s_2,m_{\sq_i}^2)} \frac{(m_{\go}^2-m_{\sq_i}^2)\, s_2}{(s_2-m_{\sq_i}^2)\, m_{\go}^2}.
     \ee
   \ei

In the following, we will discuss some results obtained with the different schemes. Furthermore, the (in)dependence of the predictions on the actual value of the regularising $\go$ width is analyzed\footnote{For the DR-I method the obtained results are by construction independent of $\Gamma_{\go}$, for the DR-II method we set $\Gamma_{\go}\neq 0$ only in $M_r$, as described in the discussion of the DS scheme. However, as already mentioned earlier, both DR methods and the DS method (for finite $\Gamma_{\go}$) are gauge dependent. In our calculation we use a lightcone gauge for the external gluons.}.
\begin{figure}[t]
  \begin{minipage}{0.49\textwidth}
 \includegraphics[scale=0.3]{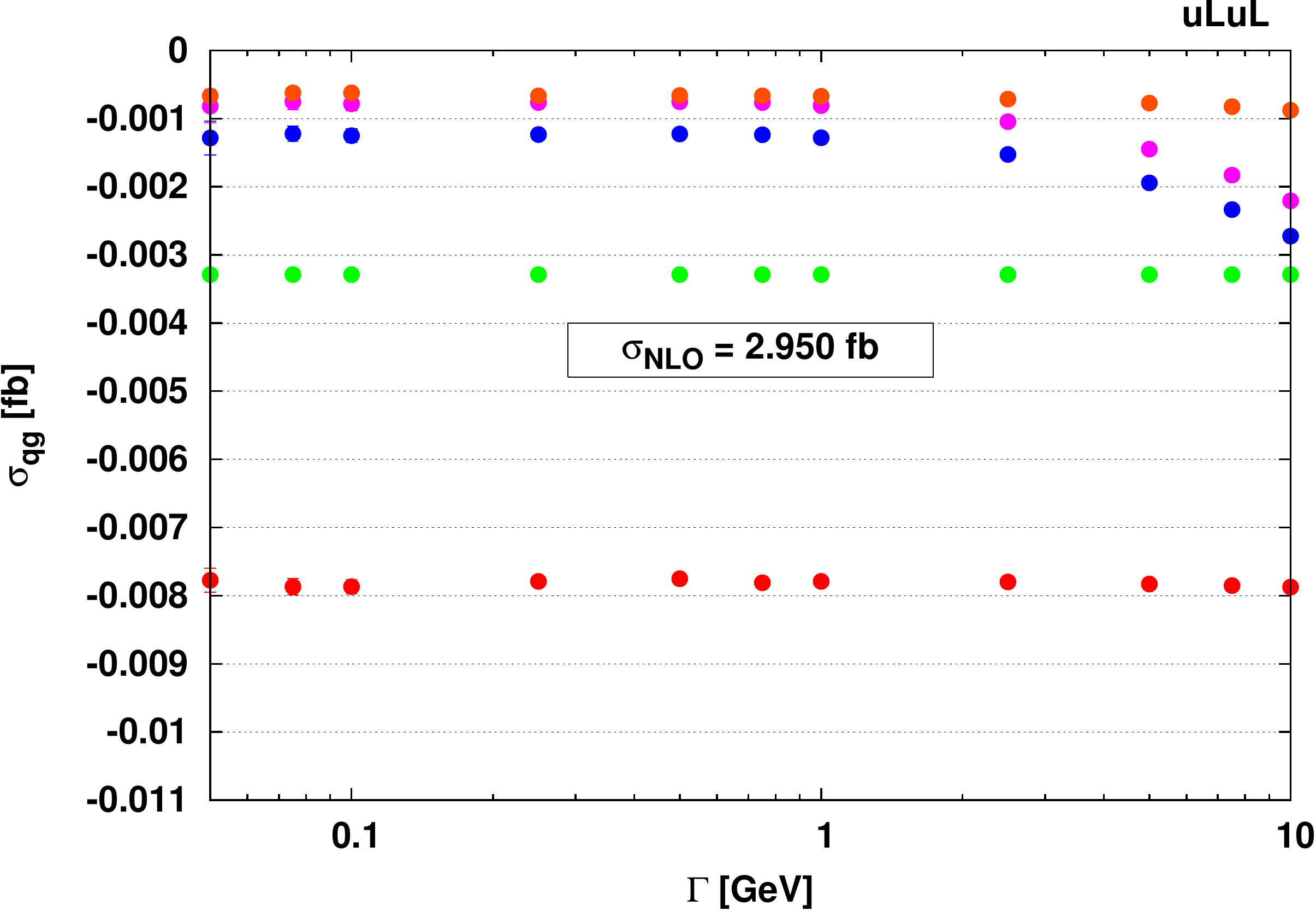}
 \newline
 \includegraphics[scale=0.3]{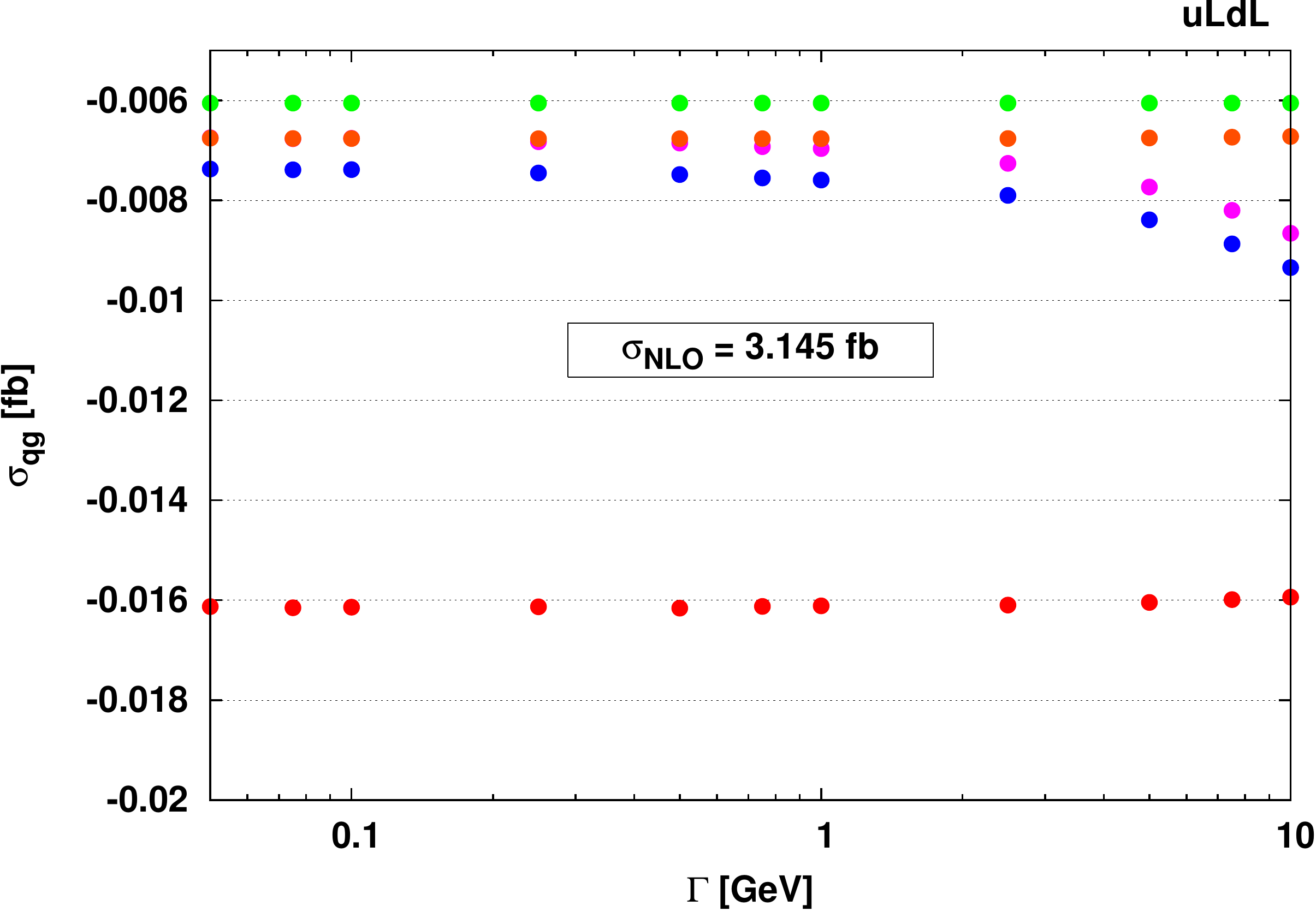}
\end{minipage}
\begin{minipage}{0.49\textwidth}
 \includegraphics[scale=0.3]{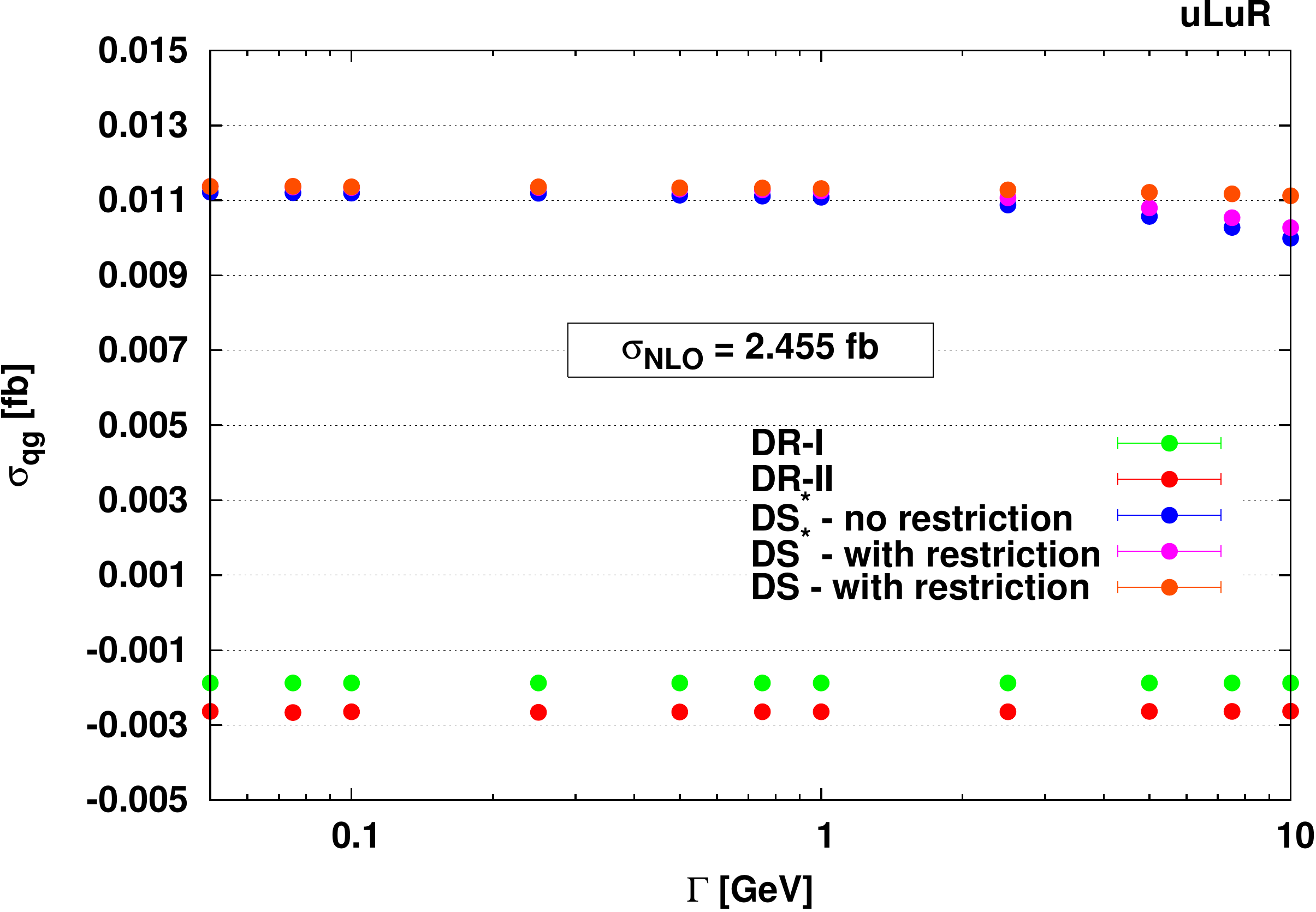}
 \newline
 \includegraphics[scale=0.3]{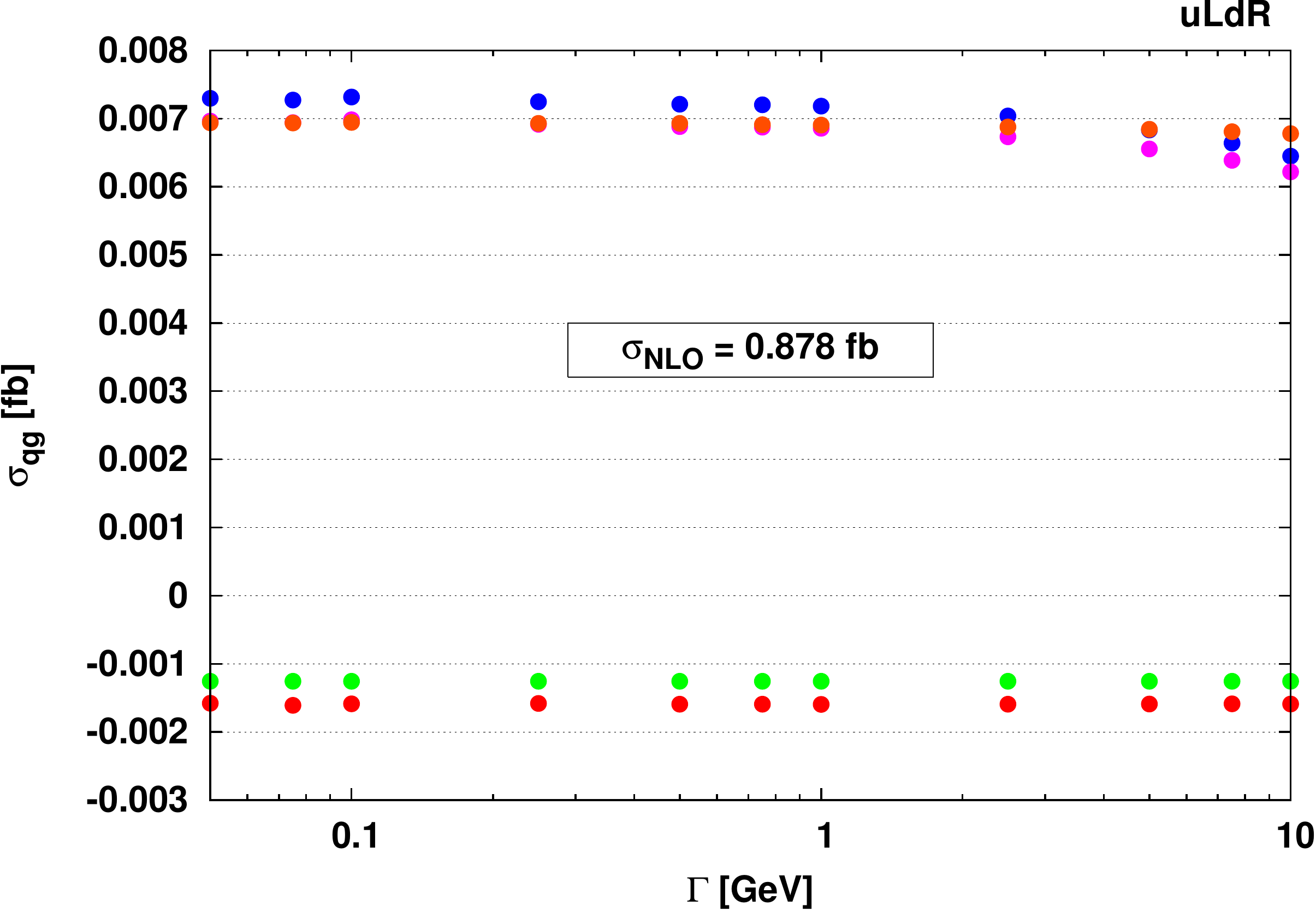}
\end{minipage}
\caption {The $qg$ contributions as obtained by using the different subtraction schemes for the four representative $\sq\sq$ channels with different choices for the regularisation parameter $\Gamma_{\go}$. For \lq DS$^*$-with restriction' the Jacobian has been modified according to Eq.~(\ref{eq:jaco}), while \lq DS$^*$-no restriction' shows the (incorrect) results without applying this factor. To illustrate the differences between the (not gauge invariant) DS and the DS$^*$ scheme we plot the results obtained with the DS method (with restriction), too. Also given is the full NLO cross section for the respective channels as obtained with the DS$^*$ scheme with the corrected Jacobian.}
\label{fig:oss}
\end{figure}

To this end we consider a scenario within the CMSSM with $m_{\sq}<m_{\go}$ (for the actual values of the masses, see Tab.~\ref{tab:masses} in Sec.~\ref{ch:res}). We take into account only the first generation and show in Fig.~\ref{fig:oss} for a representative set of the four possible channels (same/different flavour with same/different chirality) the contribution of the $qg$ initiated channels $\sigma_{qg}$, as a function of the regularising width $\Gamma_{\go}$. This comprises only the $2\rightarrow 3$ parts of the respective processes, {\it i.e.} the real amplitudes squared and the counterterms for the subtraction of the IR divergencies. Thus the differences visible in this quantity (which is not a physical observable) directly indicate the influence of the neglected terms in the on-shell subtraction methods. We note that for both DR methods and the DS scheme the result is rather stable with respect to the value used for $\Gamma_{\go}$ and thus insensitive to this parameter. (This holds of course only for 
the
considered range. If $\Gamma_{\go}$ is increased further its influence becomes visible, while for smaller values the result becomes numerically unstable.) Applying the DS$^*$ method we see in all channels more or less pronounced effects of this regulator for $\Gamma_{\go}\gtrsim 1\ {\rm GeV}$. Moreover, we note that the magnitude of the terms neglected in both DR schemes can be sizable. The two curves shown for the DS$^*$ scheme have been obtained by including/excluding the correction factor for the Jacobian, see Eq.~(\ref{eq:jaco}). While not being as drastic as in case of \lq DR vs. DS$^*$', the influence of this term is nevertheless non-negligible. Comparing the results for the gauge invariant method $DS^*$ with the corresponding DS results we observe a good agreement for small values of $\Gamma_{\go}$, while the differences for larger values are sizeable. This behaviour was to some extent expected, as the deviations between these methods are controlled by the width, see Eq.~(\ref{eq:gaugedep}). 

However, these observations have to be interpreted with a grain of salt: comparing the absolute size of the considered quantity $\sigma_{qg}$ to the full NLO cross section for the different channels, we note that it amounts only to a sub percent effect in the scenario considered here\footnote{This statement holds only for the production of $\sq$ of the first generation. If second generation $\sq$ are involved, the discrepancies in the results for the total cross sections of the different subchannels 
obtained with the different methods can become rather large. Comparing {\it e.g.} the DS$^*$ and the DR-II method for the benchmark point CMSSM 10.1.5, we observe deviations up to $\matO(20\%)$ for channels including second generation squarks. These large effects can be explained by the fact that in these cases the $qg$ contributions gain in relative importance due to larger PDF factors $f$ ({\it e.g.} $f_u f_g > f_u f_c$). Nevertheless, the impact of these channels on the total cross section after summing all subchannels is very small, of $\matO(1\%)$.}.
The actual numbers of $\sigma_{\rm NLO}$ for the DS$^*$ scheme with the corrected Jacobian are depicted in the plots. The effect on distributions is in general small. We will show some examples in Sec.~\ref{sec:pwgonshellsub}.

\subsection{Tests and Comparisons}

The complete NLO calculation, as described in the previous sections, has been implemented in a Fortran program in order to perform the phase space integration
and the convolution with the PDFs numerically by means of statistical Monte Carlo methods. The integration routine used for this purpose is \textsc{MONACO},
which is a modified version of the Fortran subroutine \textsc{VEGAS} \cite{vegas} and is part of the Monte Carlo program \textsc{VBFNLO} \cite{vbfnlo1,vbfnlo2,vbfnlo3}.\\ 
In order to check the various parts of the implementation of the calculation and in order to exclude possible error sources, numerous internal tests have 
been performed. Among these are the check whether the Catani-Seymour dipoles cancel the real emission contributions in the singular regions, the check 
whether the {\bom I} terms of the integrated dipoles render the correct coefficients of the $1/\epsilon$ and $1/\epsilon^2$ terms and the check whether 
the cross section of the finite collinear remainder coincides in the implementations as part of the 2-particle and as part of the 3-particle phase space.
We have checked carefully that the recalculated matrix elements for the gluon-initiated real contributions lead to the same results as the corresponding
\textsc{MadGraph} routines.
To further validate the code these tests have been supplemented, as far as possible, by a comparison of the results for the LO and NLO
cross section to results obtained with the program \textsc{Prospino2} \cite{prospino_manual}.\\ 
The program \textsc{Prospino2} computes NLO cross sections efficiently for the production of SUSY particles at hadron colliders based on the calculations 
accomplished in \cite{prospino}. However, some simplifications have been made which have to be taken into account for a consistent comparison of the results. 
While the LO cross section for squark pair production is calculated correctly, {\it i.e.} taking the individual masses into account, and separately for the various 
flavour and chirality combinations, the NLO corrections are always summed over the subchannels assuming a common mass for all squarks. The $K$-factor, {\it i.e.}~the ratio 
between the NLO and LO cross section
\begin{equation}
 K = \frac{\sigma_{NLO}}{\sigma_{LO}}\ ,
\end{equation}  
is determined for the total cross section, with all subchannels summed up. Results for the NLO cross sections of different subchannels can be returned but 
have been obtained by scaling the LO cross sections with the $K$-factor obtained from the total cross section at LO and NLO. Thus, it is assumed that the 
$K$-factor does not change for different flavour and chirality combinations. Since \textsc{Prospino2} reads SUSY Les Houches Accord (LHA) \cite{susylha} spectrum files but calculates an 
average squark mass for the evaluation of the NLO corrections, it is most sensible to compare results for a scenario with degenerate squark masses. For that 
purpose all squark masses have been set to
\[m_{\sq} = 1800 \ {\rm GeV}\ ,\]
the gluino mass is chosen to be 
\[m_{\gl} = 1600 \ {\rm GeV}\ .\]
Additionally, \textsc{Prospino2} uses CTEQ6 PDFs throughout, {\it i.e.} the CTEQ6L1 set for the LO and the CTEQ6M set for the NLO cross section. For the strong coupling $\alpha_s$ the 1-loop (2-loop) RGEs are used for the LO (NLO) results.
With these choices results for a cross-check against \textsc{Prospino2} at a center-of-mass energy of $8$ TeV have been produced. In case of degenerate squark masses several of the 
36 subchannels yield the same result. For example $\sul \sul$ and $\sur \sur$ have the same cross section and so have $\sul \sdr$ and $\sur \sdl$. 
As a consequence, only 20 out of the 36 possible channels have cross 
sections that differ from each other. Several of these 20 cross sections differ just due to PDFs, {\it i.e.}~different flavours in the initial state, and thus all subchannels can 
be summarized in the 4 categories $\sul \sul, \sul \sdl, \sul \sur$ and $\sul \sdr$. The sum of the LO and NLO cross sections for all subchannels contributing 
to these categories with the corresponding $K$-factors in comparison to the ones obtained with \textsc{Prospino2} are listed in Table~\ref{table:proscheck}. As everywhere else in this 
work the charge conjugated processes are included for every subchannel. In the last line of Table~\ref{table:proscheck} the sum of all subchannels is stated.
\begin{table}[t]
\vspace*{0.3cm}
\begin{center}
$
\renewcommand{\arraystretch}{1.6}
\begin{array}{|c||c|c|c||c|c|c|} \hline
 {\rm channel}& \sigma_{\rm LO} {\rm\ [fb]} & \sigma_{\rm NLO} {\rm\ [fb]}& K & \sigma_{\rm LO}^{\rm Prospino} {\rm\ [fb]}& \sigma_{\rm NLO}^{\rm Prospino} {\rm\ [fb]}& K^{\rm Prospino} \\ \hline \hline
 \sul \sul    & 1.29 \cdot 10^{-1} & 1.43\cdot 10^{-1} & 1.11 & 1.29\cdot 10^{-1} & 1.50\cdot 10^{-1} & 1.16 \\ \hline
 \sul \sdl    & 8.00 \cdot 10^{-2} & 9.92\cdot 10^{-2} & 1.23 & 8.00\cdot 10^{-2} & 9.28\cdot 10^{-2} & 1.16 \\ \hline
 \sul \sur    & 3.40 \cdot 10^{-2} & 4.00\cdot 10^{-2} & 1.18 & 3.40\cdot 10^{-2} & 3.95\cdot 10^{-2} & 1.16 \\ \hline
 \sul \sdr    & 1.39\cdot 10^{-2}  & 1.74\cdot 10^{-2} & 1.26 & 1.39\cdot 10^{-2} & 1.62\cdot 10^{-2} & 1.16 \\ \hline\hline\hline
 {\rm Sum}    & 2.57 \cdot 10^{-1} & 3.00 \cdot 10^{-1}& 1.16 & 2.57\cdot 10^{-1}  & 2.99\cdot 10^{-1}  & 1.16 \\ \hline
\end{array}
$
\caption{\label{table:proscheck} LO and NLO cross sections and $K$-factors for individual subchannels and the sum of all $36$ subchannels in comparison
to \textsc{Prospino2}. Charge conjugated processes are included. The values have been obtained for a common squark mass $m_{\sq} = 1800 \ {\rm GeV}$, 
a gluino mass of $m_{\gl} = 1600 \ {\rm GeV}$ and a center-of-mass energy of 8 TeV.} 
\end{center}
\vspace*{-0.2cm}
\end{table}
As can be inferred from the table the LO cross sections are in perfect agreement. The NLO total cross sections agree within their errors and consequently 
the total $K$-factors are the same. While \textsc{Prospino2} assumes that this total $K$-factor is constant in the various subchannels, calculating the NLO 
cross sections for the subchannels individually shows that this approximation is not entirely satisfactory. The $K$-factors of the individual subchannels vary in the range between $1.11 - 1.26$. 
Therefore, an independent treatment of subchannels seems appropriate, as in general squarks of different chiralities and thus different channels have different masses, decay widths and 
kinematic distributions. We have also verified that for a scenario with $m_{\sq} < m_{\gl}$, {\it i.e.} $m_{\sq} = 1600 \ {\rm GeV}$ and $m_{\gl} = 1800 \ {\rm GeV}$, which corresponds to the case where a gluino is resonantly produced,
the LO cross sections in all subchannels and the total NLO cross section perfectly agree with the results obtained with \textsc{Prospino2}.\\
In order to check our calculation also for non-degenerate squark masses, we have compared LO and NLO cross sections to all combinations of subchannels given in Table 6 of \cite{hollik}, 
where the benchmark point CMSSM 10.1.5 from \cite{susybench} was used. 
We find perfect agreement at LO but deviations of about $1\%$ to $20\%$ depending on the subchannel at NLO which can be attributed to the implementation of the DR-II type on-shell 
subtraction scheme instead of the DS subtraction scheme, which was chosen in our case. Using the DR-II scheme in our calculation we 
find very good agreement with \cite{hollik} at NLO. Furthermore, we have cross checked our results against those presented in Table II in \cite{plehn} for the benchmark points
CMSSM 10.2.2 and CMSSM 40.3.2 (defined again according to \cite{susybench}). In the former, the gluino is heavier than the squarks of the first two generations, in the latter the gluino is lighter.
We find agreement on the sub-percent level in all subchannels independent of the scenario chosen and therefore independent of whether subtraction of on-shell intermediate 
gluinos has to be performed or not.


\section{Matching $\sq\sq$ Production with Parton Showers Using the\\ \textsc{Powheg} Method}
\label{ch:pwg}
To obtain realistic predictions for measurements at the LHC, a combination of the fixed order NLO results described in the last chapter with parton shower programs is mandatory. The \textsc{Powheg}~method \cite{nason,powheg} is one option to perform this matching consistently and will be used in the following.
The basic idea of the \textsc{Powheg}~method consists of generating the hardest emission first, maintaining full NLO accuracy, and adding subsequent radiation with a $p_T$-vetoed shower program. If the ordering variable in the parton shower is different from $p_T$ one has to add a truncated shower to obtain a complete description. Formally, the \textsc{Powheg} cross section for $n$ particles in the final state derived with this procedure has the following form:
\begin{eqnarray}
  d\sigma_{\sss PWG}&=&\overline{\matB}(\varPhi_n)\,d \varPhi_n \left[
\Delta(\varPhi_n, p_{T}^{min})+ \Delta(\varPhi_n, k_T)
 \frac{\matR_s(\varPhi_{n},\varPhi_{rad})}{\matB(\varPhi_n)} \theta(k_T-p_T^{min}) d\varPhi_{rad}\right] \nonumber\\
 &+&(\matR-\matR_s)d\varPhi_{n+1},
 \label{pwgmaster}
\end{eqnarray}
 with $\varPhi_n$ representing the underlying Born phase space. The phase space for the real emission is constructed from $\varPhi_n$ and the radiation variables, denoted $\varPhi_{rad}$ here, thus $\varPhi_{n+1}=\{\varPhi_n,\varPhi_{rad}\}$. $\matR$ corresponds to the full real amplitude squared, whereas $\matR_s$ is chosen such that in the limit of a soft/collinear emission $\matR_s\rightarrow \matR$.\footnote{This guarantees that the \textsc{Powheg} Sudakov form factor has the same leading-log accuracy as a shower MC program.} Choosing $\matR=\matR_s$ obviously simplifies the expression, but in some cases a different choice is more appropriate, see the discussion below. The scale $p_{T}^{min}$ determines the lower limit for the $p_T$ of the radiated parton. It is of the order of a typical hadronic scale, $p_{T}^{min}=\matO(1 \, \text{GeV})$. The two main ingredients in Eq.~(\ref{pwgmaster}) are the $\overline{\matB}$ function which ensures the NLO accuracy of the method and comprises the typical 
elements 
of a NLO calculation, namely the born ($\matB$), virtual ($\matV$) and real ($\matR_{s}$) terms,
 \be
  \overline{\matB}(\varPhi_n) =  \Bigl[ \matB(\varPhi_n) + \matV(\varPhi_n)+
 \int  \matR_s(\varPhi_{n},\varPhi_{rad})  d \varPhi_{rad} \Bigr],
 \ee
 and the \textsc{Powheg} Sudakov form factor
 \be
\Delta(\varPhi_n,p_T)=\exp\left[-\int d\varPhi_{rad}'\frac{\matR_s(\varPhi_{n},\varPhi_{rad}')}{\matB(\varPhi_n)}
\theta(k_T(\varPhi_n, \varPhi'_{rad})-p_T) \right].
 \ee
 Note that for $\matR\neq\matR_s$, only $\matR_s$ affects the generation of the first emission, while the contributions of the remnant term $(\matR-\matR_s)$ are \lq regular', {\it i.e.} they do not contain any soft/collinear divergent terms and can thus be generated with usual MC methods.
 
 
The main steps of the method as the actual generation of the first emission or the subtraction of the IR divergencies in the real terms are process-independent and have been automatised in the \PB~framework (see \cite{powhegbox} for details).

\subsection{Implementation in the \PB}
\subsubsection{SQCD Processes in the \PB}
So far only SM processes have been implemented in this program package (the only exceptions being slepton pair production \cite{dislepton} and $tH^-$ production \cite{toph}, however in both processes the created BSM particles do not interact strongly and are therefore not affected by the radiation generation). That is why as a first step towards the implementation of our pure (S)QCD process we had to make sure that all steps in the existing code are suited for dealing correctly with this type of processes.
To this end, the following aspects implemented in the \PB~ had to be considered:
\benu
\item The automatised version of the FKS method \cite{fks} used in \textsc{Powheg} for the IR divergencies in the real contributions might be affected. In the first step of the implemented algorithm, all singular regions for the flavour structures of the process under consideration are identified. Here only pairs of massless partons are relevant. Therefore the occurrence of massive colour-charged sparticles does not spoil this procedure.

The subtraction terms used in this method consist of the eikonal factors for the soft singularities and the factorization formulae for the collinear singularities (see appendices A and B in \cite{powhegbox} for details). For $\sq\sq$ production, collinear singularities can only appear in initial state (IS) radiation. Hence the corresponding formulae are unchanged. Soft gluons can be radiated off final state (FS) squarks, but as the eikonal factors are independent of the spin of the emitter, only minor changes in the code were necessary. To be more specific, the routine {\tt softalr} in {\tt sigsoftcoll.f} was modified in a way that the sums over massive coloured particles comprise also the PDG codes of SQCD particles. Moreover, we had to ensure that the correct $SU(3)$ Casimir factors for squarks ($C_F$) and gluinos ($C_A$) are used.

Correspondingly, the implemented formulae for the soft-virtual cross section had to be adapted. Again, only in the parts concerning massive coloured particles the occurring sums had to be extended to SQCD particles with the correct Casimir factors. These changes affect solely the subroutine {\tt btildevirt} in {\tt sigsoftvirt.f}.

\item The generation of the first emission according to the \lq \textsc{Powheg} master-formula' in Eq.~(\ref{pwgmaster}) as implemented in the \PB~is not affected by the presence of coloured sparticles.
\item Moreover, the \PB~provides several \lq utility routines' for the calculation of $\alpha_s$, calling PDF libraries, writing out LesHouchesEvent (LHE) files, performing simple analyses etc. Besides some minor changes in the output to LHE files, the only possible source of problems are the implemented formulae for $\alpha_s$. Here the $\overline{\textnormal{MS}}$ scheme with 5 active flavours is used. As we decoupled all heavy (s)particles from the running of $\alpha_s$ in our calculation, no changes were necessary at this point.
\eenu

 \subsubsection{Process-dependent Ingredients}
 Apart from these changes in the main part of the program, the usual process-dependent parts for the implementation of a process in the \PB~had to be provided. These consist of

 \bi
   \item the list of all independent flavour structures for the relevant Born and for the real channels,
   \item the Born phase space, here for a $2\rightarrow 2$ process with massive particles,
   \item the Born and the colour/spin-correlated\footnote{As there are no gluons present at tree-level and the external quarks are treated massless, the spin-correlated Born amplitudes squared vanish for $\sq\sq$ production.} Born amplitude squared,
   \item the finite part of the virtual contributions, calculated as described in Sec.~\ref{ch:nlo},
   \item the real contributions squared for all subchannels,
   \item the colour flows for the Born configurations in the large-$N_c$ limit.
 \ei

 As we do not impose any assumption on the masses of the produced squarks, we have in principle 36 configurations of same/different flavour/chirality squarks with different masses in the final state, which have to be treated in separate runs of the code and are combined afterwards.\footnote{Note that the charge conjugate processes are included, but not discussed separately here.} To reduce the computation time of our code, subchannels with final state squarks of the same mass are combined by using the {\tt smartsig} option of the \PB.



  \subsubsection{Implementation of the On-shell Subtraction}
  \label{sec:imposs}
  Implementing the subtraction of contributions with an on-shell intermediate $\go$ as described in Sec.~\ref{ch:nlo} is quite involved. The occurring problems are mostly related to the way the phase space for the real radiation is built up in the \PB: being implemented in a process-independent way it is tailored to the generation of the hardest emission, {\it i.e.}~starting from a point in the phase space for the $2\rightarrow n$ Born-like configuration, the integration over the 1-particle phase space of the radiated parton is performed using its rescaled energy and two angles relative to the emitting particle. Comparing this situation to the way the subtraction is built up in our stand-alone NLO program it is obvious that

  \benu
   \item it is not possible to perform a Breit-Wigner-mapping (BW-mapping) for the integration over the invariant $\go$ mass as discussed in footnote 5. Thus the usage of a $\Gamma_{\go}\ll m_{\go}$ would worsen the convergence of the integration (if the result converges at all) and
   \item a restriction of the phase space on the on-shell configurations as described above is not straightforward.
  \eenu


  Furthermore, all different schemes except for the simplest DR-I scheme lead to real contributions which are no longer positive definite. This has two consequences: first of all the fraction of events with negative weights is increased, as the $\overline{\matB}$ function in Eq.~(\ref{pwgmaster}) is no longer guaranteed to be positive. Second the mechanism for the actual generation of the hardest emission is based on the assumption that the ratio $\matR_s/\matB$ in the \textsc{Powheg} Sudakov form factor is positive. Both problems were discussed in the context of the implementation of $tW$ production in the \PB~\cite{twpwg}. While the fraction of events with negative weights can be reduced by applying a \lq folded' integration over the radiation variables, a feature that is implemented in the \PB~and described in \cite{powhegbox}, the second problem cannot be solved directly. The proposition in \cite{twpwg} adapted to our process consists in introducing a cut on the invariant mass of the intermediate $\go$ 
close to the resonant region:
  \be
  \label{eq:cutmgo}
  \matR_s\rightarrow \matR_s\, \Theta(|m_{\go}-m_{\text{inv}}|-\Delta)
  \ee
  with $\Delta = \matO(\Gamma_{\go})$. The motivation for this procedure was based on the observation that the situation $\matR_s/\matB<0$ occurs most often close to the resonant region. We have checked that this holds for our process, too. Nevertheless, in view of all these problems we have opted against a \lq direct' implementation of the DS scheme.

  Instead we implemented the subtraction mechanism such that for the actual \textsc{Powheg} generation of an event with $q g$ in the initial state only parts of the real amplitudes squared from Eq.~(\ref{eq:ossme}) are used as $\matR_s$ in Eq.~(\ref{pwgmaster})\footnote{To simplify the notation, we consider similar to the considerations in Sec.~\ref{sec:onshellsub} only the case of one singular region. The generalization to the same flavour case, where two singular regions may occur, is straightforward, except for the case of identical $\sq$, where additional interference terms between these two contributions occur. We attribute these terms always to $\matR_s$.}. The remaining terms, which include the parts with potentially resonant intermediate $\go$, are then treated as regular remnants, which are integrated separately, using a phase space tailored to the resonant structure. The subtraction term for these on-shell contributions is part of these terms, too, {\textit i.e.} the phase space can be easily restricted. As this splitting has to preserve the leading-log accuracy of the whole \textsc{Powheg} formalism these remnant terms must not comprise any of the IR divergent parts. There are several possibilities to perform this splitting:
  
  \bi
    \item $\matR_s=|M_{\text{tot}}|^2$: This choice corresponds to the original DS method, {\it i.e.} the case without any splitting. As already discussed in Sec.~\ref{sec:onshellsub}, this quantity is only gauge independent if an analytical expansion in the poles is performed. We call this option in the following DS$^*$-I.
    \item $\matR_s=|M_{nr}|^2$: In this case the interference term and the resonant amplitude squared ($2\,\text{Re}(M_{r}M_{nr}^*)+|M_r|^2$ in Eq.~(\ref{eq:ossme})) with the corresponding subtraction term for the on-shell intermediate $\go$ are treated as regular remnants. These terms are not IR divergent, thus no FKS subtraction is necessary and the leading-log accuracy of the \textsc{Powheg} formalism is not spoiled. A BW-mapping is possible and there is no need for an artificial cut as defined in Eq.~(\ref{eq:cutmgo}) as $\matR_s>0$. However, in this case it is not possible to restore gauge invariance by replacing the different terms by the expanded result, as both $2\,\text{Re}(M_{r}M_{nr}^*)$ and $|M_r|^2$ provide terms of $\matO((1/s_{jg})^0)$, which would have to be combined with $|M_{nr}|^2$ to obtain a gauge invariant result\footnote{As these terms are treated differently in the event generation (see Eq.~(\ref{pwgmaster})), a residual gauge dependence is left even for $\Gamma_{\go}\rightarrow 0$.}. This option is called DS-II in the following.
    \item $\matR_s=\frac{f_1}{s_{jg}}+f_2(s_{jg})$ (see Eq.~(\ref{eq:expan}) for the definition of $f_1,f_2$): This approach is gauge invariant by construction. The IR divergent parts are contained in $f_2$. However, this choice leads again to negative values for $\matR_s$, which require the introduction of the artificial cut on the invariant mass of the resonant $\go$ described above. The points with $\matR_s<0$ occur again most often close to the region where $m_{\text{inv}}\approx m_{\go}$. We call this option in the following DS$^*$-III.
  \ei
  
%

  
  A drawback of all solutions with $\matR_s\neq |M_{\text{tot}}|^2$ is the usually quite high negative weight fraction for these remnant terms. In the original code, regular remnants are supposed to be positive, as they comprise the full matrix elements squared for subchannels which do not have any IR divergencies. Therefore the parts of the code concerning these contributions had to be adapted. 

   In Tab.~\ref{tab:ossmeth} we summarize the advantages/disadvantages of all the aforementioned methods. Comparing the different implementations in the \PB~we conclude that there is no optimal choice regarding speed, numerical stability and conceptual correctness: while the simple but incomplete DR-I method is the fastest and most stable one, the more involved solutions based on the DS($^*$) scheme either require the introduction of an artificial cut for the radiation generation or are not gauge invariant. We will compare these methods for a specific benchmark point in the next section.
  \begin{table}
\bc
\begin{tabular}{|c |c | c | c | c |   }
  \hline
  Method & BW-mapping  &  phase space restr.   &  cut for radiation  & gauge invariance\\\hline\hline
  DR-I     &      unnecessary  &     unnecessary           &       unnecessary   & violated \\
  DR-II  &      unnecessary  &     unnecessary         &       yes        &  violated \\\hline
  DS$^*$-I   & not possible &  not possible       &       yes    &     preserved  \\
  DS-II  &    possible  &    possible         &       unnecessary     &  violated \\
  DS$^*$-III  &    possible  &    possible     &       yes     &  preserved \\\hline
\end{tabular}
  \caption{  \label{tab:ossmeth}Summary of the advantages and disadvantages of the different subtraction methods. Both Diagram Removal (DR) methods are discussed in Sec.~\ref{sec:onshellsub}. The Diagram Subtraction (DS) methods listed here are only distinct w.r.t. the actual implementation: for DS$^*$-I the subtraction is performed on the complete real amplitude squared ({\it i.e.} $\matR_s=\matR$ in Eq.~(\ref{pwgmaster})). For DS-II and DS$^*$-III the real contributions are split as described in the text.}
\ec
\vspace*{-0.2cm}
\end{table}
\subsubsection{Checks}
To test the validity of our \textsc{Powheg} implementation, several tests have been performed. A first important check concerning the correct implementation of the Born and the real contributions (and in our case the changes in the routine {\tt softalr}) is the cancellation of the IR singularities in the real contributions against the corresponding FKS terms. This behaviour is checked in the \PB~by comparing the real matrix elements squared in the soft/collinear limit with the known approximations, which depend only on the Born amplitudes.

Moreover, the \PB~allows the user to produce as a by-product arbitrary LO/NLO differential distributions. We compared these with our independent NLO implementation and found full agreement for all considered observables. Note that this also validates the FKS subtraction method as implemented in the \PB~and the dipole subtraction used for our NLO calculation.

Besides these basic tests, a valid implementation should fulfil several requirements inherent in the \textsc{Powheg} method itself. First of all, the generated events should guarantee NLO accuracy for inclusive quantities. Moreover the generated hardest emission should reproduce the predictions given by the NLO calculation\footnote{This prediction is governed by the real part of the calculation, thus this quantity is in fact a LO prediction.} for large $p_T$, up to higher order corrections. To check if these requirements are fulfilled we have compared several NLO predictions for suitable differential distributions with the corresponding \textsc{Powheg} predictions after generation of the first hard emission, {\it i.e.} at the level of events written out to an LHE file. Some of these results are shown in the next section.


\section{Results}
\label{ch:res}
This section summarizes our main findings. As there are no hints for SUSY at the LHC to date, we chose for illustration two mSUGRA scenarios, one with $m_{\sq}>m_{\go}$ and the other one with $m_{\sq}<m_{\go}$ that are not yet excluded by data, see {\it e.g.} \cite{atlasexcl,cmssusy}. For the SM-parameters, we used \cite{pdg}
  \begin{align}
  m_Z = 91.1876\, \text{GeV}, \qquad G_F=1.16637\cdot 10^{-5}\, \text{GeV}^{-2}, \qquad \alpha_s(m_Z)=0.118, \notag\\
  m_b^{\overline{\textnormal{MS}}}(m_b)=4.25\, \text{GeV}, \qquad m_t=174.3\, \text{GeV}, \qquad m_{\tau} = 1.777\, \text{GeV}.
  \end{align}
Our scenarios are based on the CMSSM-points $10.3.6^*$ and $10.4.5$\footnote{We have modified $m_0$ for the point $10.3.6$ to get a mass spectrum consistent with the latest exclusion bounds.} proposed in \cite{susybench}, the input parameters are summarized in Tab.~\ref{tab:msugra}.
\begin{table}[t]
\bc
\begin{tabular}{|c |c | c | c | c | c|}
  \hline
  CMSSM-point & $m_{1/2}$  &  $m_0$  & $A_0$   &  $tan(\beta)$  & $sgn(\mu)$\\\hline\hline
  $10.3.6^*$  & $550 \, \text{GeV}$  &  $825\, \text{GeV}$  & $0\, \text{GeV}$ &    $10$        &    $+1$       \\
  $10.4.5$  & $690\, \text{GeV}$  &  $1150\, \text{GeV}$  & $0\, \text{GeV}$ &    $10$        &    $+1$       \\\hline
\end{tabular}
  \caption{\label{tab:msugra}The input parameters for the considered scenarios.}
\ec
\vspace*{-0.2cm}
\end{table}
To generate the resulting mass spectra we used \textsc{Softsusy 3.3.4} \cite{softsusy}. The thus obtained on-shell masses are then used as input variables for our calculation. As \textsc{Softsusy} implements non-vanishing Yukawa corrections, the masses of the second generation squarks are slightly different from the corresponding ones of the first generation. To simplify the analysis and save computing time, we replaced these values by taking the mean of the mass pairs, {\it i.e.}~we set $m_{\tilde{u}_L}=m_{\tilde{c}_L}=(m_{\tilde{u}_L}+m_{\tilde{c}_L})/2$ etc. The relevant mass values are listed in Tab.~\ref{tab:masses}.
\begin{table}[h]
\bc
\begin{tabular}{| c |c | c | c | c | c| c|}
  \hline
  CMSSM-point & $m_{\tilde{u}_L}$ & $m_{\tilde{u}_R}$ & $m_{\tilde{d}_L}$ & $m_{\tilde{d}_R}$ & $m_{\tilde{g}}$ & $m_{\tilde{\chi}^0_1}$\\\hline\hline

  $10.3.6^*$  & $1799.53$ & $1760.21$  & $1801.08$ & $1756.40$ & $1602.96$ &  $290.83$\\

  $10.4.5$  & $1746.63$ & $1684.31$  & $1748.25$ & $1677.82$ & $1840.58$ &  $347.71$\\\hline
\end{tabular}
\caption{\label{tab:masses}The masses obtained with the parameters from Tab.~\ref{tab:msugra} after averaging the $\sq$ masses of the first two generations as described in the text.}
\ec
\vspace*{-0.2cm}
\end{table}

Despite concentrating on the production process, we will show some distributions including the decays of the produced squarks. To this end, we consider the decay channel with the shortest \lq cascade', $\sq\rightarrow q \tilde{\chi}^0_1$. While having the largest or at least second largest branching ratio (BR) for $\sq_R$, the BR for $\sq_L$ is quite small. This is mainly due to the fact that in both scenarios the $\tilde{\chi}^0_1$ is mostly bino-like, moreover in case of point $10.3.6^*$ the channel $\sq\rightarrow \go q$ opens up. As we do not intend to perform a complete analysis for all possible cascades, we nevertheless consider only this channel.  For the calculation of the LO branching ratios we used the program \textsc{Sdecay 1.3} \cite{sdecay}. The results are listed in Tab.~\ref{tab:brs}.

\begin{table}
\bc
\begin{tabular}{| c |c | c | c | c |}
  \hline
  \rule{0pt}{14pt}
  CMSSM-point & $\text{BR}(\tilde{u}_L\rightarrow u \tilde{\chi}^0_1)$ & $\text{BR}(\tilde{u}_R\rightarrow u \tilde{\chi}^0_1)$ & $\text{BR}(\tilde{d}_L\rightarrow d \tilde{\chi}^0_1)$ & $\text{BR}(\tilde{d}_R\rightarrow d \tilde{\chi}^0_1)$ \\\hline\hline

  $10.3.6^*$  & $0.0098$ & $0.566$  & $0.0121$ & $0.254$ \\

  $10.4.5$  & $0.0137$ & $0.998$  & $0.0160$ & $0.998$ \\\hline

\end{tabular}
  \caption{\label{tab:brs}The branching ratios for the decay $\sq\rightarrow \tilde{\chi}^0_1 q$ for the two scenarios considered here.}
\ec
\vspace*{-0.2cm}
\end{table}

The renormalization $(\mu_R)$ and factorization $(\mu_F)$ scale were chosen as $\mu_R=\mu_F=\overline{m}_{\sq}$, the bar indicating the average over all $\sq$ masses of the first two generations.

For the PDFs we used the LO set \textsc{CTEQ6L1} and for NLO results \textsc{CT10NLO} with $\alpha_s(m_Z)=0.118$ \cite{cteq}. Both sets are taken from the \textsc{LHAPDF}-package \cite{lhapdf}. In the LO calculation, $\alpha_s$ was computed using the 1-loop RGEs, while for the NLO calculation the
2-loop results were used. All results shown in the following have been obtained for the LHC with a center-of-mass energy of $\sqrt{s}=14\,\text{TeV}$.
In the results including the decays of the $\sq$ or parton shower effects more than one parton occurs in the final state. These are then clustered into jets using \textsc{Fastjet 3.0.3} \cite{fastjet}. We use the anti-$k_T$ algorithm \cite{antikt} with $R=0.4$. If not stated otherwise we require the transverse momentum and the pseudorapidity of the jets to fulfil
\be
   p_T^j>20\,\text{GeV},\quad |\eta^j|<2.8.
\ee
We include the error bars in all distributions, if not indicated differently.

\subsection{NLO Results}
\begin{figure}[t]
  \centering
 \includegraphics[width=7cm,angle=270]{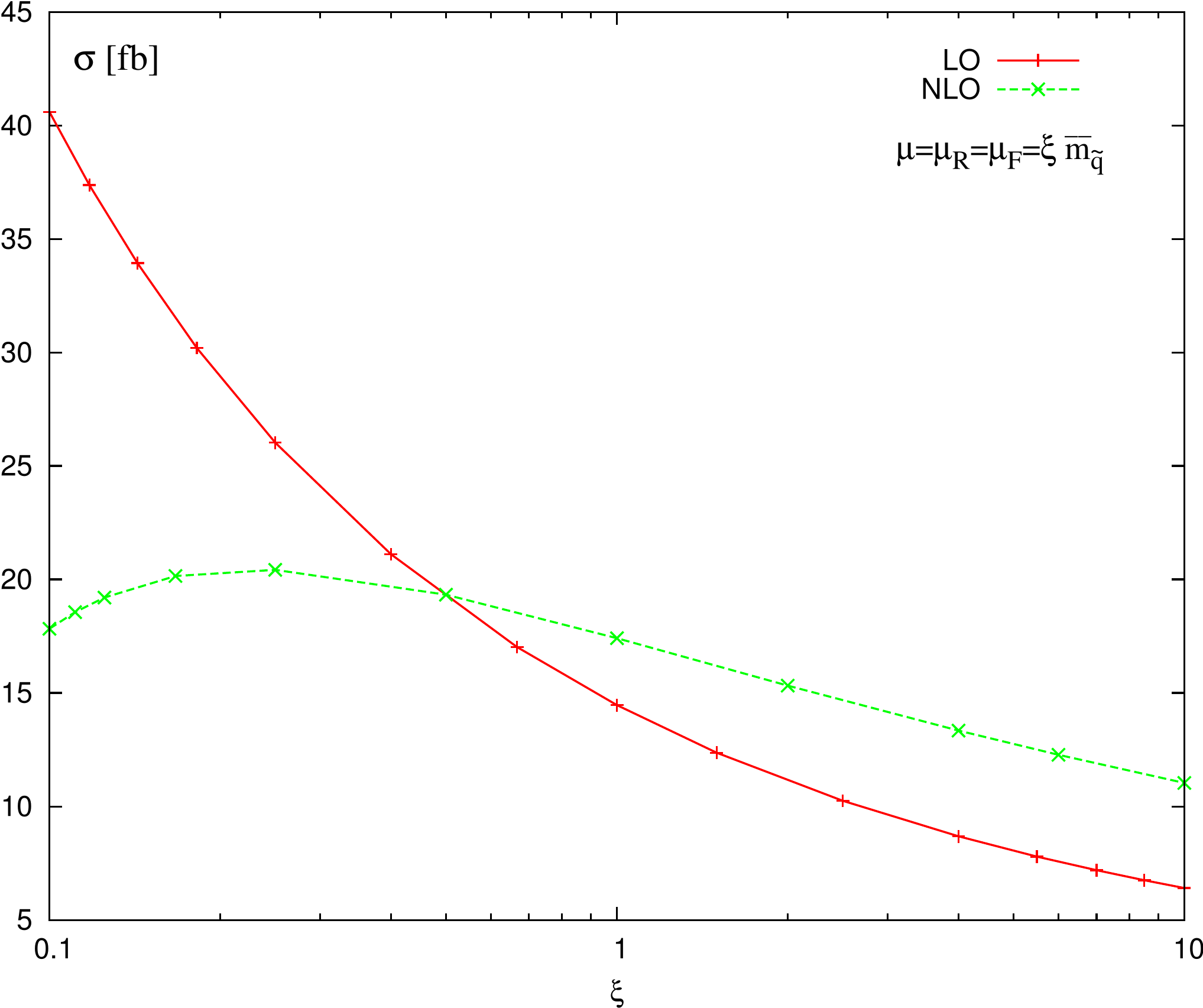}
  \caption{Scale dependence of the LO and NLO total cross section at a center-of-mass energy of $\sqrt{s}=14\,\text{TeV}$ for the CMSSM point $10.3.6^*$.}
  \label{fig:scalevar}
\end{figure}

Before investigating the effects of the NLO corrections on
differential distributions, we will analyze the scale dependence of the total cross sections. The variation of the unphysical factorization and 
renormalization scales in the LO and NLO cross sections can provide a rough estimate on the remaining theoretical uncertainties due to higher order corrections. 
Figure~\ref{fig:scalevar} shows the scale dependence of the LO and NLO cross sections of squark pair production calculated with input 
parameters according to the CMSSM point $10.3.6^*$ of Table~\ref{tab:msugra}. The renormalization and factorization scale have been set to a common value, which is varied by a
factor of $10$ in both directions around the central value given by the average squark mass $\overline{m}_{\sq}$. The NLO cross section exhibits clearly a much flatter scale dependence than 
the LO cross section. Varying the latter by a factor of two around the central value results in a dependence of about $\pm40\%$. In the NLO cross section the scale dependence in the same range
reduces to $\pm10\%$. The dependence on the factorization scale is very weak and the residual scale dependence is dominated by the renormalization scale dependence of $\alpha_s$. 
The cross sections 
at the central scale amount to
\begin{equation}
 \sigma^{LO} = 14.47\ {\rm fb} \quad \text{and} \quad \sigma^{NLO}=17.40\ {\rm fb},
\end{equation}  
implying a $K$-factor of
\begin{equation}
 K = 1.20
\end{equation}
and thus an enhancement of the LO cross section due to the NLO corrections by $20\%$.

In the rest of this section the effects of the NLO corrections on differential distributions shall be presented. These effects are exemplified based on two observables:
\begin{itemize}
 \item The invariant mass of the squark pair in the final state, 
  $m^{\sq \sq} = \sqrt{\left( p_{\sq_i} + p_{\sq_j} \right)^2}$.
 \item The transverse momentum of each squark in the final state, 
  $p_T^{\sq} = \sqrt{p^2_{\sq_{i},x} + p^2_{\sq_{i}y}}$.
\end{itemize}
\begin{figure}[t]
 \centering
 \includegraphics[width=5.5cm,angle=270]{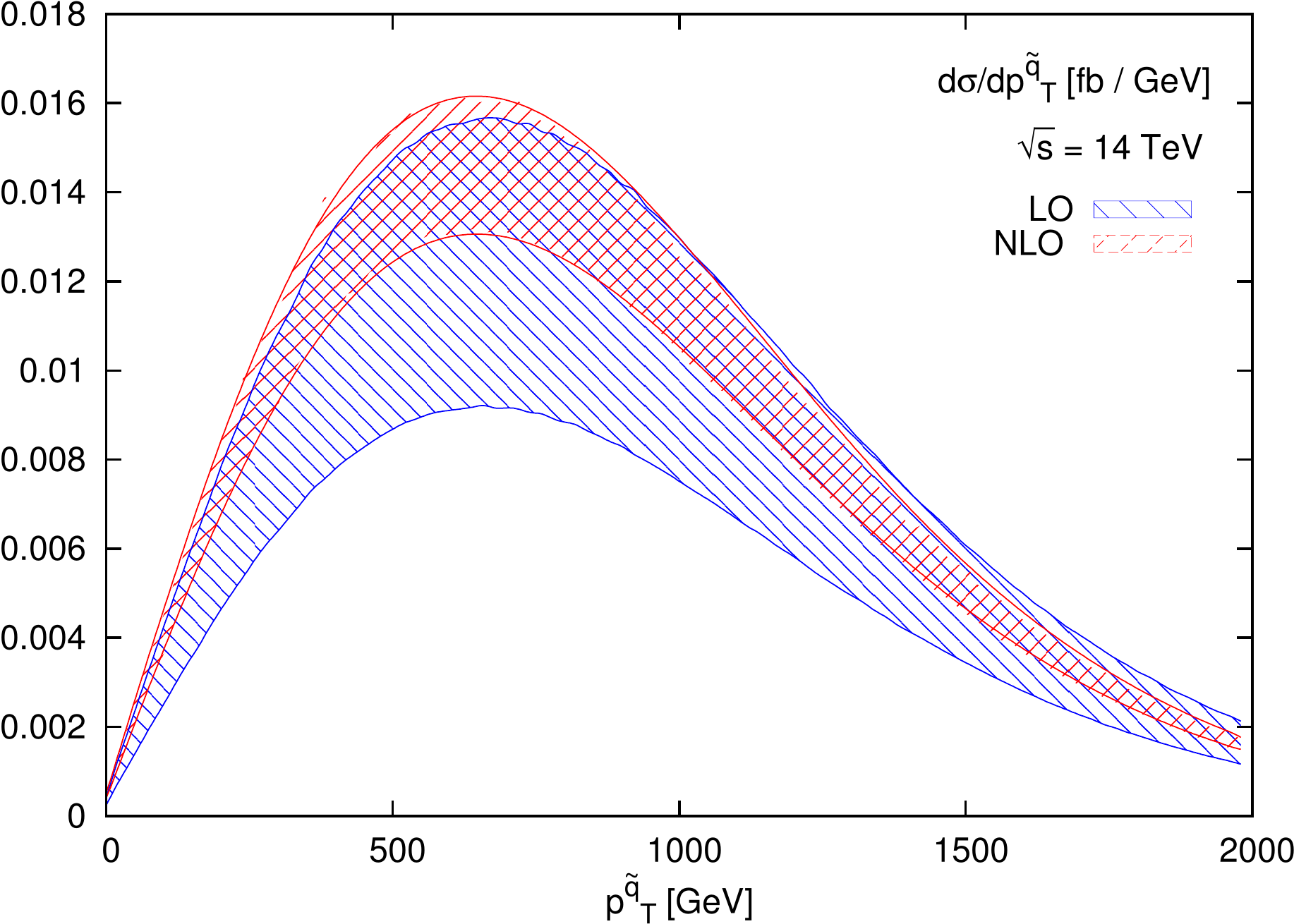}
 \hspace{0.2cm}
 \includegraphics[width=5.5cm,angle=270]{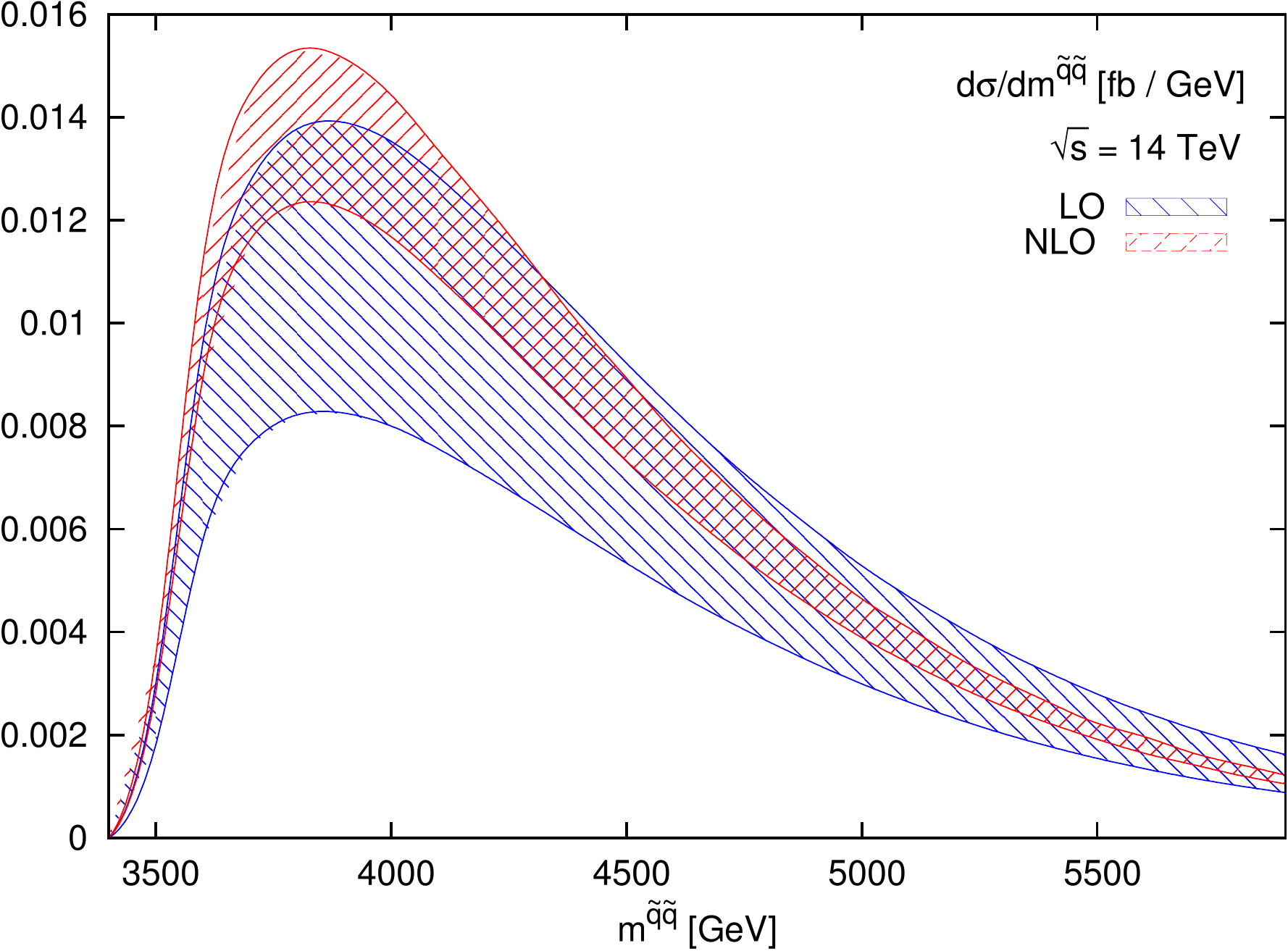}\\
 \vspace{0.3cm} (a) \hspace{7.5cm} (b)
  \caption{LO and NLO transverse momentum $p_T^{\sq}$ (a) and invariant mass $m^{\sq \sq}$ (b) distributions for the CMSSM point $10.3.6^*$ at a center-of-mass energy of $14\ {\rm TeV}$. 
  The displayed bands originate from varying the factorization and renormalization scale by a factor of 2 up and down.}
  \label{fig:dist_0_errorband}
\end{figure}
Figure~\ref{fig:dist_0_errorband} displays the comparison of the scale dependence in these distributions at LO and NLO for the CMSSM point $10.3.6^*$. 
The bands in these plots have been obtained by varying the factorization and renormalization scale by a factor of 2 up and down. As for the total cross section the scale dependence is reduced 
in both distributions at NLO. Note that the NLO bands overlap significantly with the estimated uncertainty range obtained for the LO predictions.\\

The effects of the NLO corrections on the shapes of the distributions can be visualized by normalizing the distributions to unity, {\it i.e.}~by dividing the LO distributions by the LO cross section 
and the NLO distributions by the NLO cross section. If the $K$-factor were flat, which means that the NLO distributions coincide with the LO distribution scaled by the $K$-factor of the total 
cross section, the normalized LO and NLO distributions would match exactly.\\
In \cite{prospino} it was found that the normalized $p^T$ and rapidity distributions are hardly ({\textit i.e.} within $\sim 10\%$) affected by the transition from LO to NLO. These results have been obtained with a common squark mass 
of $m_{\sq}=600\ {\rm GeV}$, a gluino mass of $m_{\gl} = 500 \ {\rm GeV}$ and a top quark mass of $m_t = 175\ {\rm GeV}$ for the factorization and renormalization scale $\mu_R=\mu_F=m_{\sq}$ at 
a center-of-mass energy of $\sqrt{s}=14\ {\rm TeV}$. Adopting these parameters in the present calculation the distributions of \cite{prospino} have been reproduced. 
For the scenario analyzed here, the normalized distributions are shown in Fig.~\ref{fig:dist_0_normalized}. The $p^T$ distribution exhibits similarly moderate effects as already found for this 
distribution in \cite{prospino}. The shape of the invariant mass distribution is affected  more by the NLO corrections. These effects can be quantified by determining the differential 
$K$-factor, defined as the NLO differential cross section divided by the LO differential cross section. The differential $K$-factor for the $p^T$ and invariant mass distributions is also 
depicted in Fig.~\ref{fig:dist_0_normalized}, lower panel. For the $p^T$ distribution the $K$-factor varies in a range of $\pm10\ \%$, while in the case of the invariant mass distribution 
the variation comprises a range of almost $\pm20\ \%$. For comparison the figures with the differential $K$-factor also include the constant $K$-factor of the total cross sections, 
depicted by the dashed line. In both cases rescaling the LO distributions by the global $K$-factor, as has been a common procedure so far, would overestimate the tail of the distributions and 
underestimate the threshold regions. Besides using the (fixed) average of the $\sq$ masses for $\mu_R$ and $\mu_F$ we have performed the same analysis with a dynamical scale, the average of the transverse masses of the $\sq$, defined as 
\[\overline{m}_T=\left(\sqrt{m_{\sq_1}^2+p_{T,\sq_1}^2 } + \sqrt{m_{\sq_2}^2+p_{T,\sq_2}^2 }\right)/2\ .\]
The total $K$-factor for this scale choice is a bit larger than before ($K=1.24$). The $p_T^{\sq}$ distribution is in this case better described by rescaling the LO result with the global $K$-factor (the values of the differential $K$-factor range from $1.3$ to $1.15$, compared to $1.3$ to $1.05$ in case of the fixed scale). For the $m_{\sq\sq}$ distribution, however, the differences between the two scale choices are smaller (with the dynamical scale we find a differential $K$-factor ranging from $K=1.4$ to $K=1.05$, compared to the range $K=1.4$ to $K=1.0$ for the fixed scale). The shapes of the differential $K$-factors are not affected by the different scale choice.\\
 \bfig[t]
  \begin{minipage}{0.5\textwidth}
 \includegraphics[width=\textwidth,height=48mm]{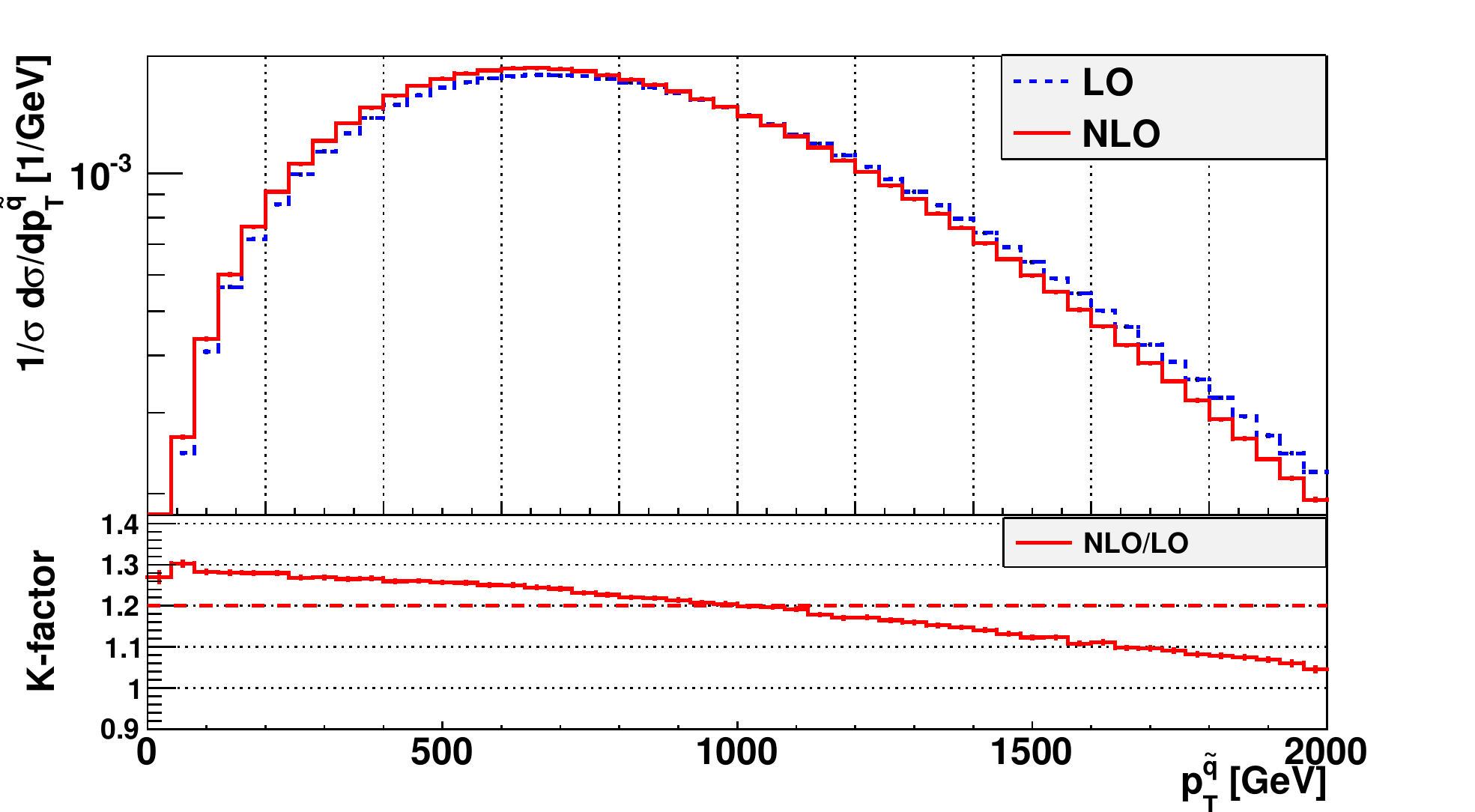}
\end{minipage}
\begin{minipage}{0.5\textwidth}
 \includegraphics[width=\textwidth,height=48mm]{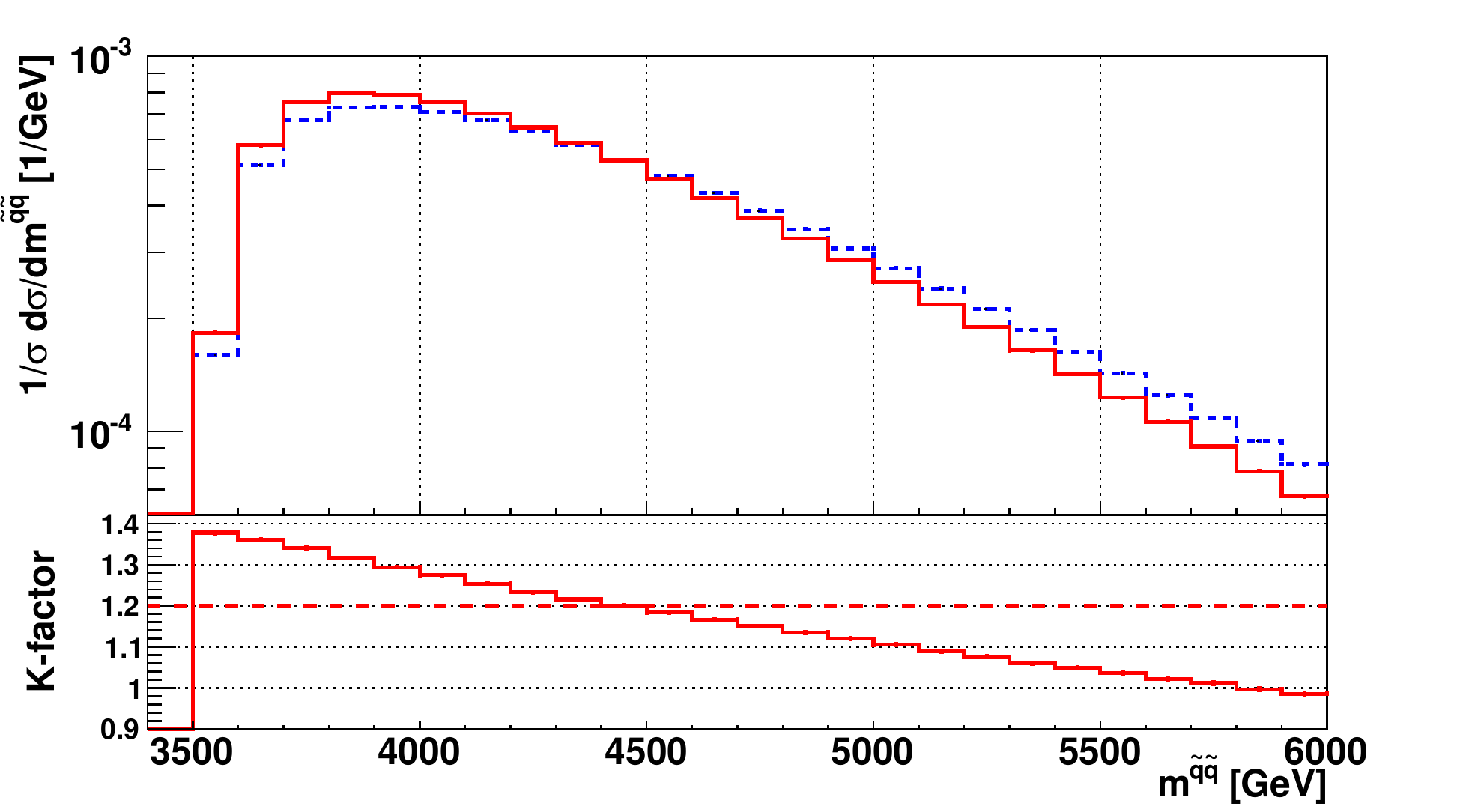}
\end{minipage}
  \caption{Normalized $p_T^{\sq}$ and $m^{\sq \sq}$ distributions and corresponding differential (full) and global (dashed) $K$-factors for the CMSSM point $10.3.6^*$
  and a center-of-mass energy of $14\ {\rm TeV}$. For the $p_T^{\sq}$ distribution the contributions of both $\sq$ have been summed.}
  \label{fig:dist_0_normalized}
\efig

 \bfig
  \begin{minipage}{0.5\textwidth}
 \includegraphics[width=\textwidth,height=48mm]{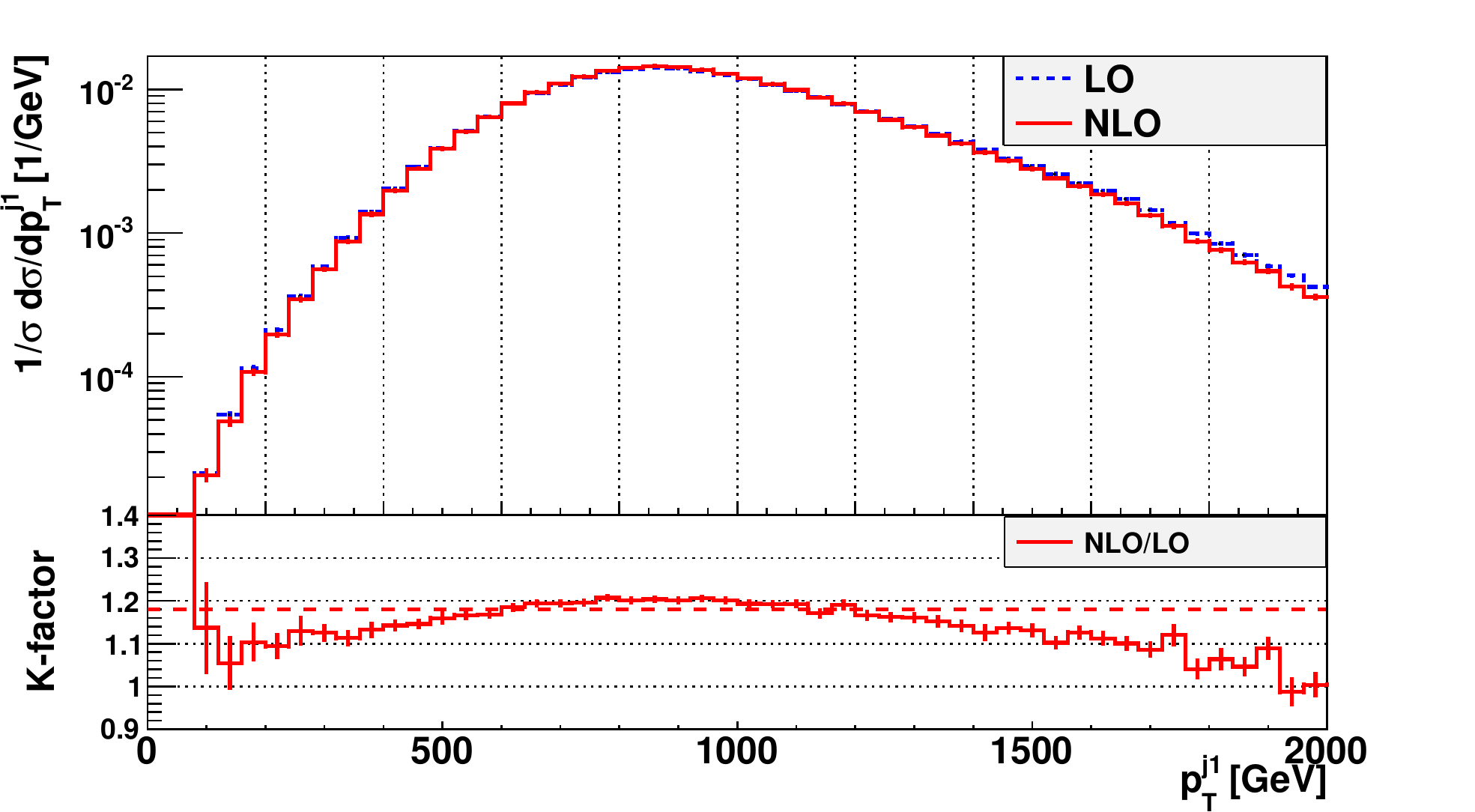}
\end{minipage}
\begin{minipage}{0.5\textwidth}
 \includegraphics[width=\textwidth,height=48mm]{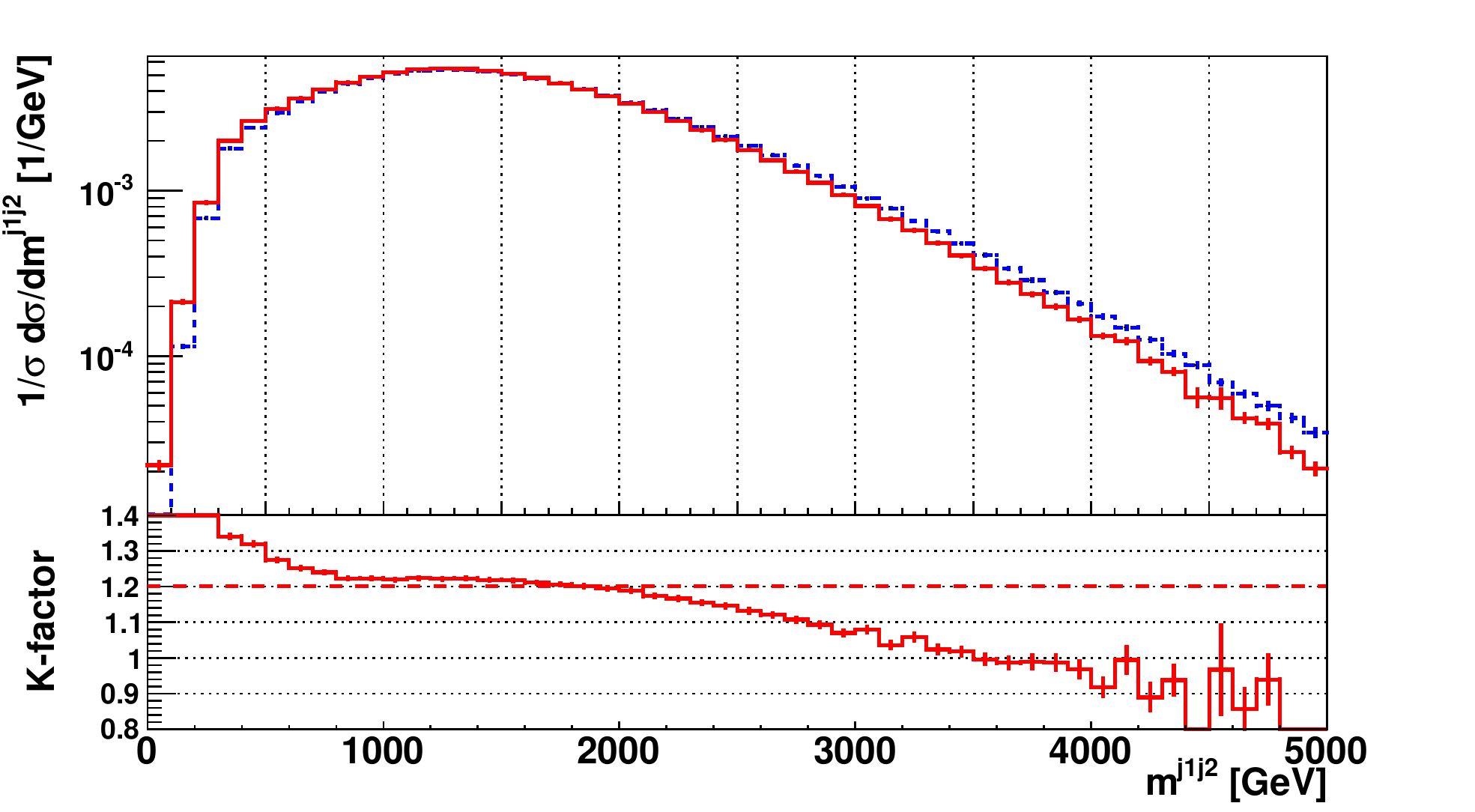}
\end{minipage}
  \caption{Normalized $p^T$ distribution of the hardest jet  and the invariant mass of the two hardest jets with corresponding differential (full) and global (dashed) $K$-factors for the CMSSM point $10.3.6^*$
  and a center-of-mass energy of $14\ {\rm TeV}$.}
  \label{fig:dist_1_normalized}
\efig

Certainly, the investigation at production level with unstable particles in the final state is only a first step towards a realistic analysis of the effects of NLO corrections on 
differential distributions. Nevertheless, it already gives a first hint that for squark pair production at the LHC the leading order distributions cannot be simply multiplied by an 
overall $K$-factor to obtain proper NLO distributions and that fully differential distributions should be used for phenomenological studies, in particular for investigations of particle properties.

In order to obtain more realistic predictions we have also added the LO decay $\sq\rightarrow q \tilde{\chi}^0_1$ for the produced squarks. The quarks originating from this decay 
and the gluon from the real corrections are clustered into jets which are ordered in $p^T$. Therefore, we show in the following the transverse momentum distributions $p^T_{j1}$ of the hardest jet
and the invariant mass distribution $m^{j_1 j_2}$ of the two hardest jets. Figure \ref{fig:dist_1_normalized} displays these distributions, again normalized with the appropriate cross 
sections, and the corresponding differential $K$-factors. While the differential $K$-factor of the $p^T$ distribution does not exhibit a strong variation, the differential $K$-factor 
of the $m^{j_1 j_2}$ distribution inherits the visible phase space dependence already observed in the $m^{\sq\sq}$ result. Using $\mu_R=\mu_F=\overline{m}_T$ does not modify these observations significantly. These distributions can be considered as examples for the fact that the observed variation of 
the $K$-factor at production level can still have a visible impact after adding decays. Similar results have been found
in \cite{hollik} where differential $K$-factors have been studied for squark pair production and decay with NLO corrections in both stages.

\subsection{Powheg Results}
For the investigation of the \textsc{Powheg} results (and the influence of different parton showers in Sec.~\ref{sec:diffPS}), we generated event samples with 5M events using our \PB~implementation. We neglected events with negative weights for the CMSSM-point $10.3.6^*$ ($m_{\sq}>m_{\go}$) by setting the flag {\tt withnegweights} to 0, which is justified by the fact that their total fraction amounts to less than one per mille. For the CMSSM-point $10.4.5$ ($m_{\sq}<m_{\go}$) we kept the events with negative weights, as they are more frequent due to the subtraction of on-shell gluinos, as discussed in Sec.~\ref{sec:imposs}.

As in the discussion of the NLO-results, we first consider results with undecayed $\sq$. Except for demanding that the emitted parton fulfils $p_{T}^j>1\,\text{GeV}$ we do not impose any cuts here. In Fig.~\ref{fig:inclu} we present several distributions of inclusive quantities for the benchmark point $10.3.6^*$: the invariant mass of the produced squarks $m^{\sq\sq}$, the transverse momentum $p_{T}^{\sq}$, the rapidity $y^{\sq}$ and the pseudorapidity $\eta^{\sq}$, where $p_T$, $y$ and $\eta$ are obtained by summing the individual distributions of both $\sq$. Shown are the NLO predictions compared to the distributions at the level of the generated \textsc{Powheg} events (which are by default written into an LHE file, thus denoted LHE in the following) after the first radiation but without further parton shower.

\bfig
  \begin{minipage}{0.5\textwidth}
 \includegraphics[width=\textwidth,height=50mm]{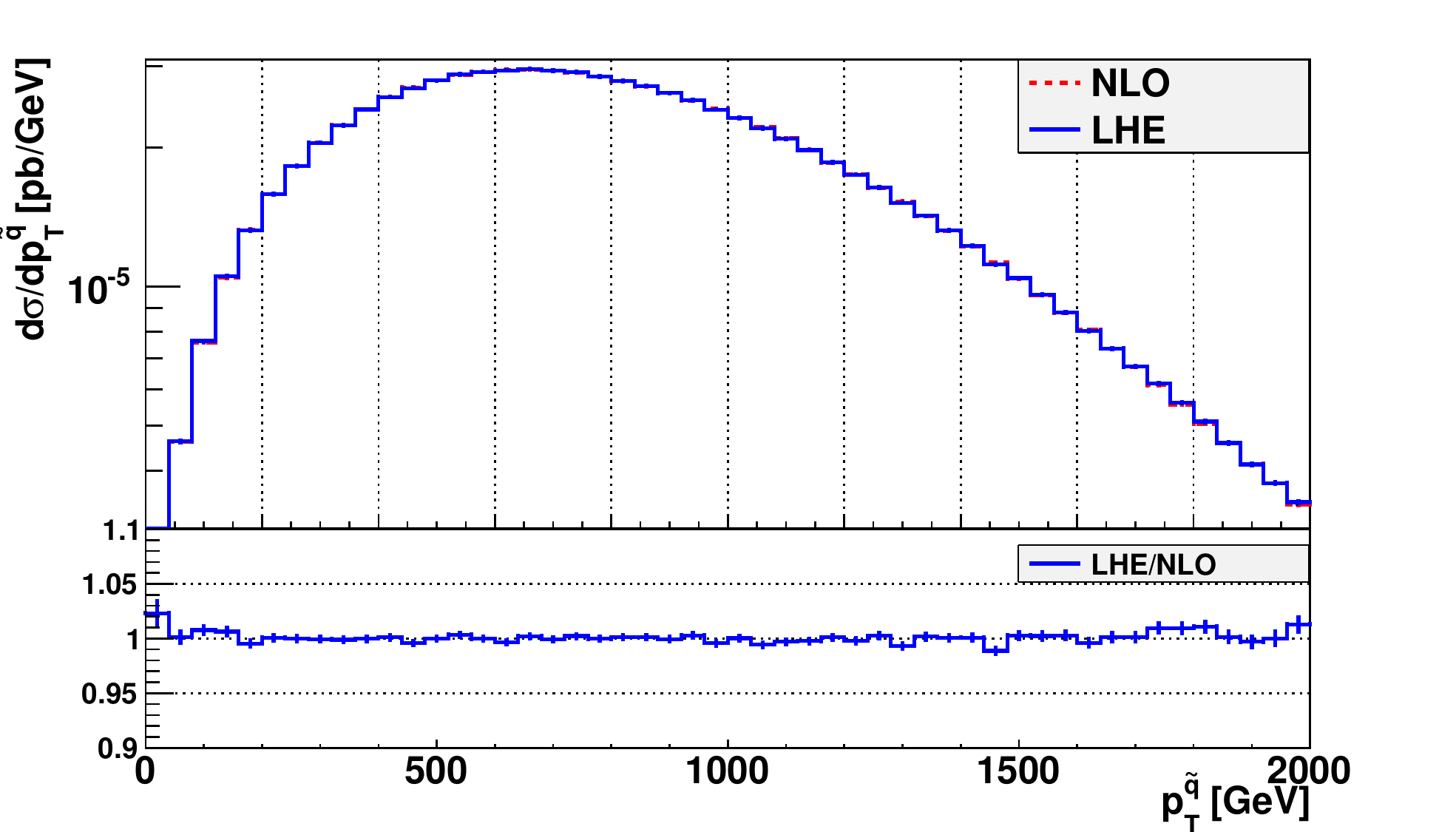}
 \newline
 \includegraphics[width=\textwidth,height=50mm]{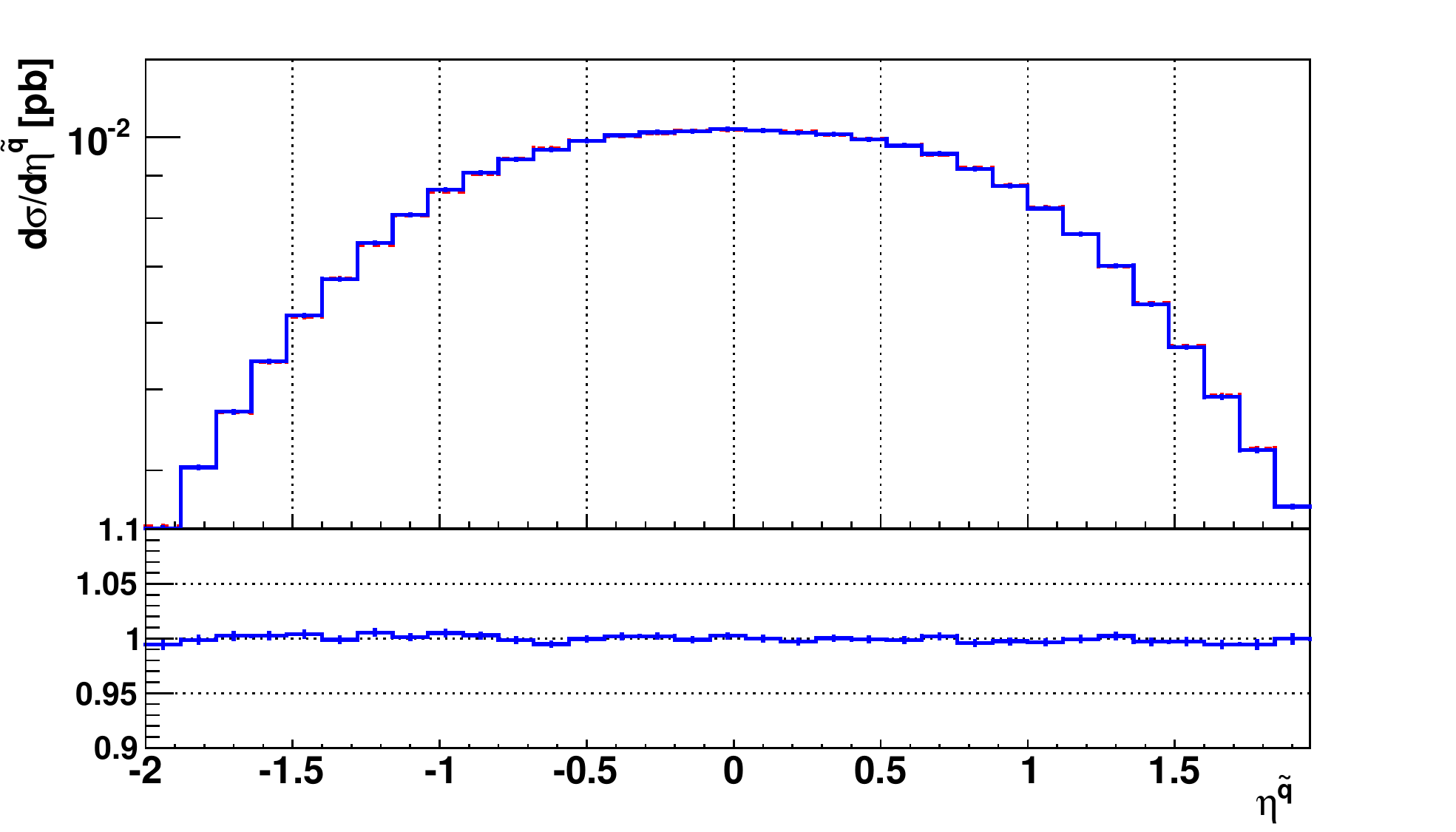}
\end{minipage}
\begin{minipage}{0.5\textwidth}
 \includegraphics[width=\textwidth,height=50mm]{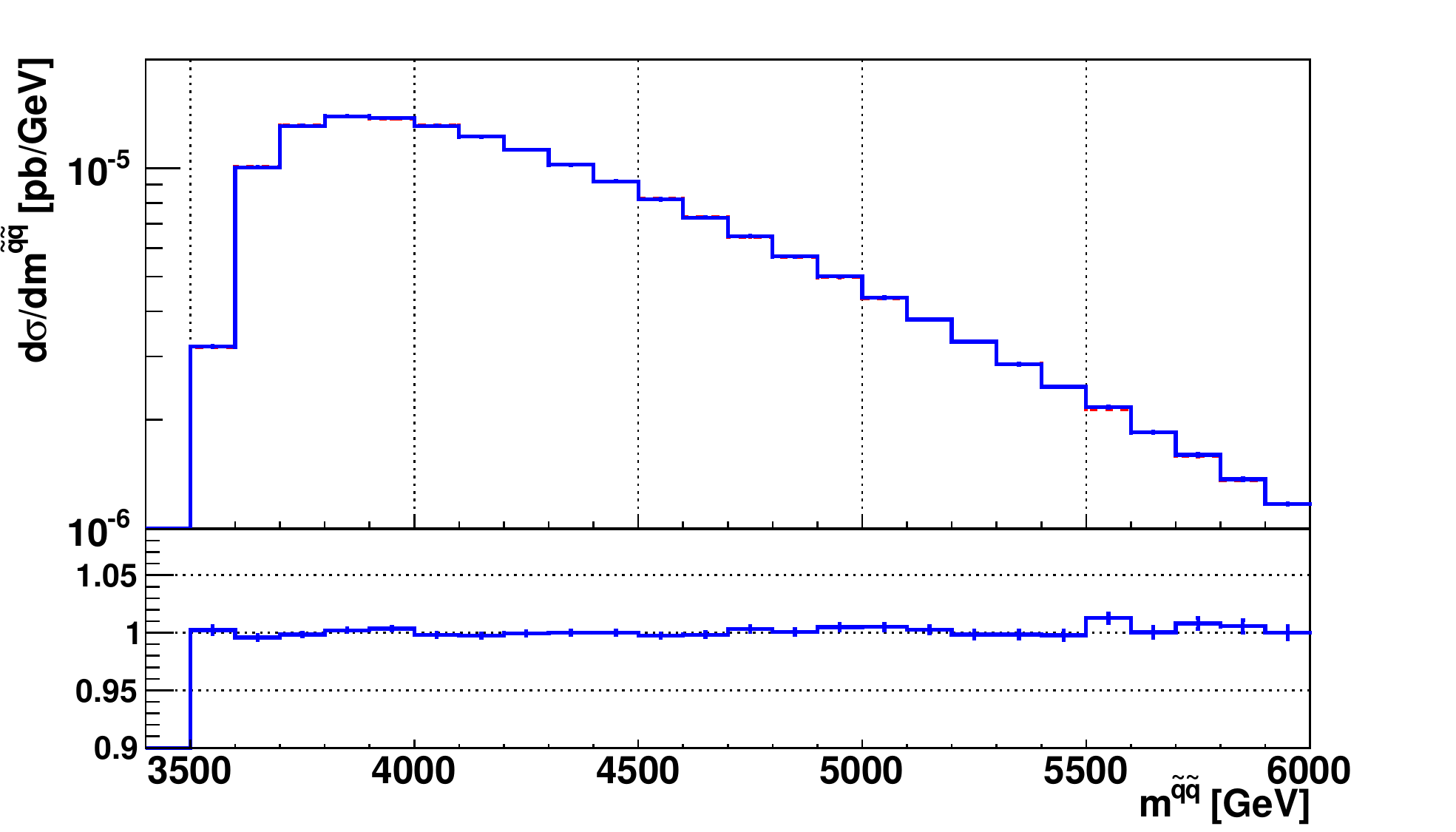}
 \newline
 \includegraphics[width=\textwidth,height=50mm]{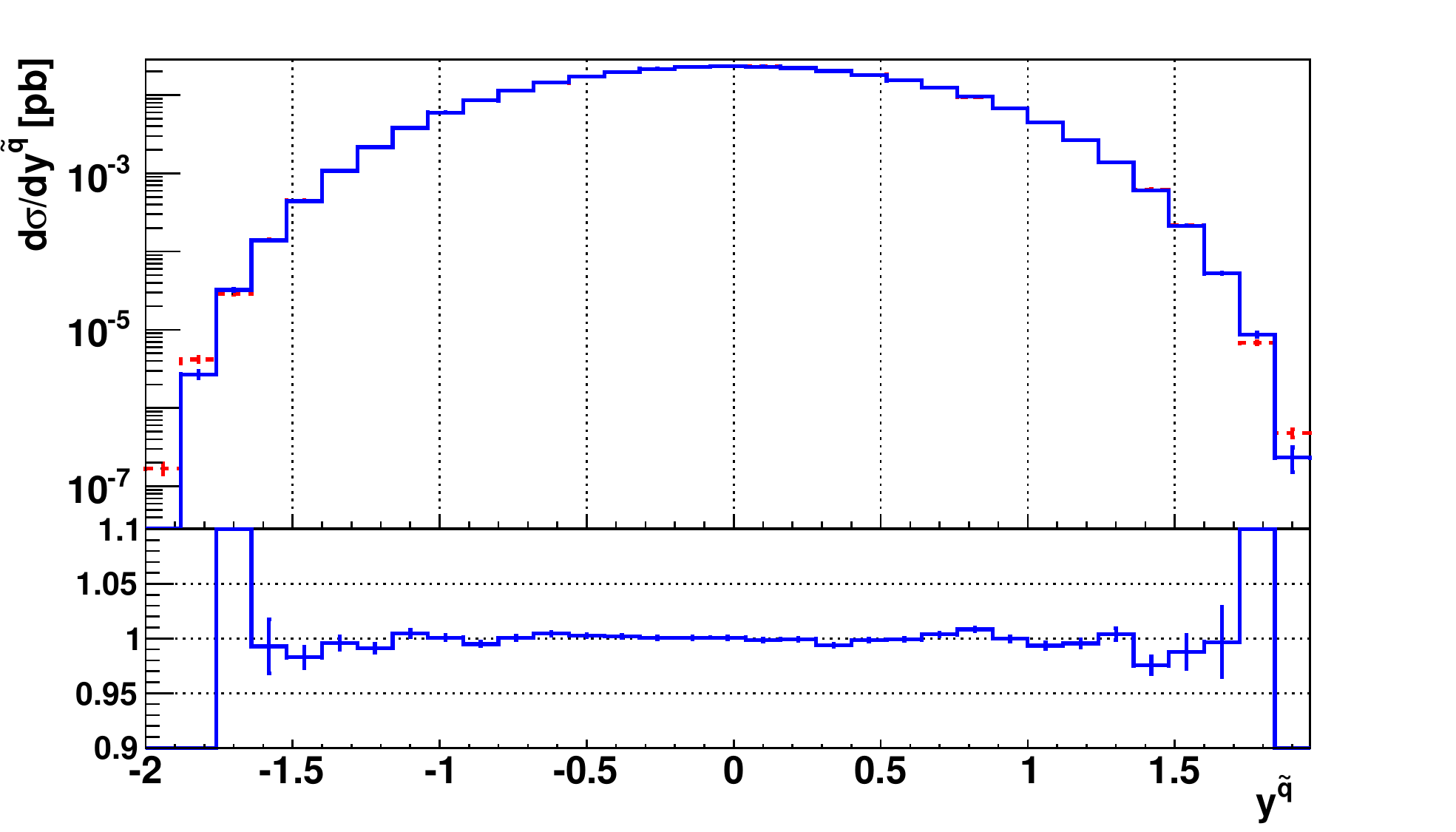}
\end{minipage}
\caption {Comparison of NLO results with the results after the first radiation generated according to the \textsc{Powheg} method (LHE) using the CMSSM point $10.3.6^*$ for several inclusive observables: the invariant mass of the two $\sq$, $m^{\sq\sq}$, and the sum of the transverse momentum, $p_{T}^{\sq}$, the rapidity, $y^{\sq}$, and the pseudorapidity, $\eta^{\sq}$, distributions for both $\sq$. The lower part of each plot shows the ratio LHE/NLO. Note that the curves are essentially identical and thus not distinguishable.}
\label{fig:inclu}
\efig

The differences between the NLO and the LHE curves are at most in the percent range. Hence the \textsc{Powheg}-events reproduce the NLO results, as expected for inclusive observables. 
A similar behaviour is observed when considering the second benchmark point, $10.4.5$, with the DS-scheme applied. 

 Next we turn to exclusive variables, which are expected to show some sensitivity to the additional emission of partons. The results shown in the following are again obtained for the benchmark point $10.3.6^*$. In Fig.~\ref{fig:ptsqsq} the $p_T^{\sq\sq}$ distribution of the $\sq\sq$-system is shown, which corresponds at NLO to the $p_T$ distribution of the emitted parton. 
 
 \bfig
  \begin{minipage}{0.5\textwidth}
 \includegraphics[width=\textwidth,height=48mm]{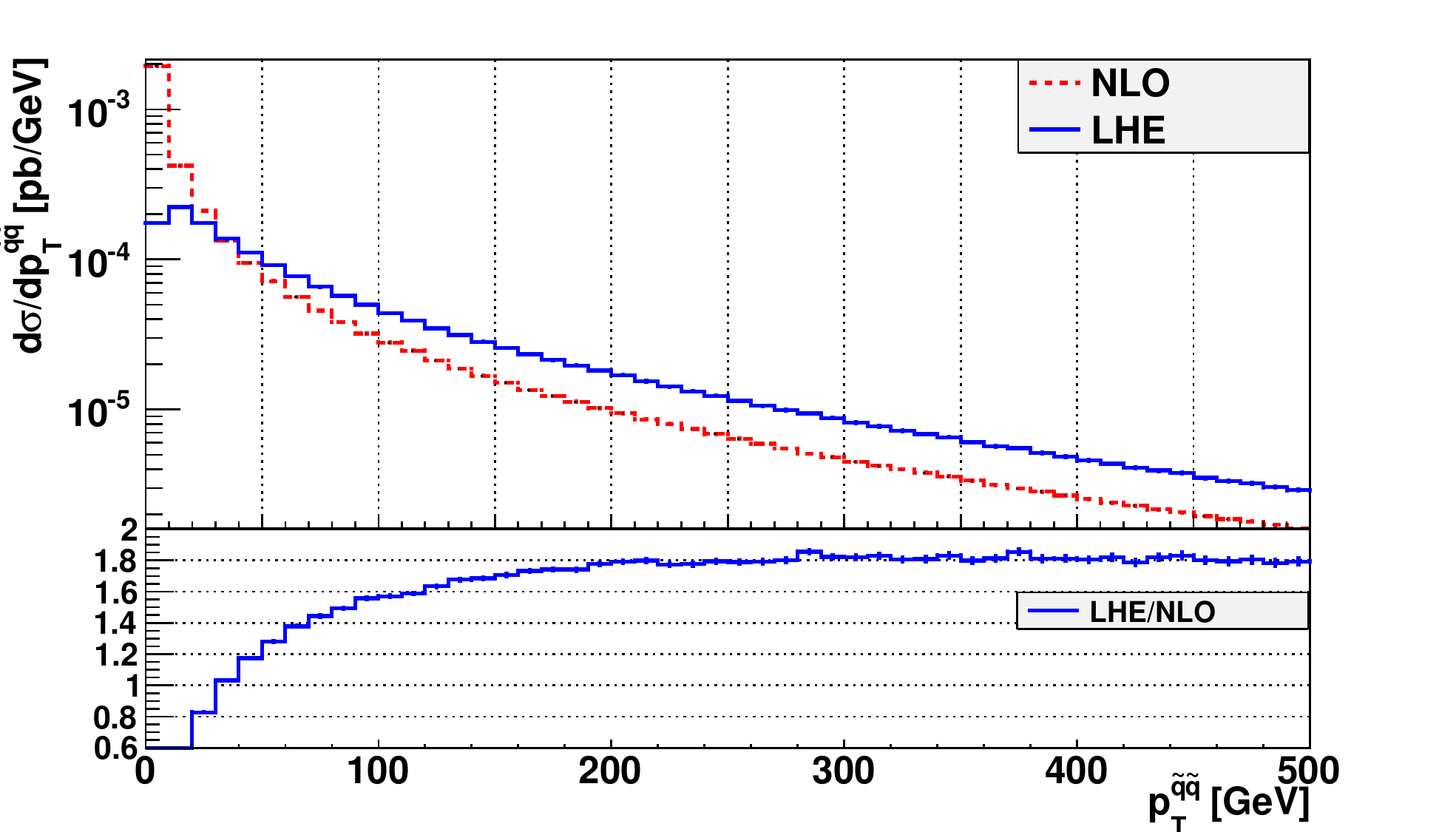}
\end{minipage}
\begin{minipage}{0.5\textwidth}
 \includegraphics[width=\textwidth,height=48mm]{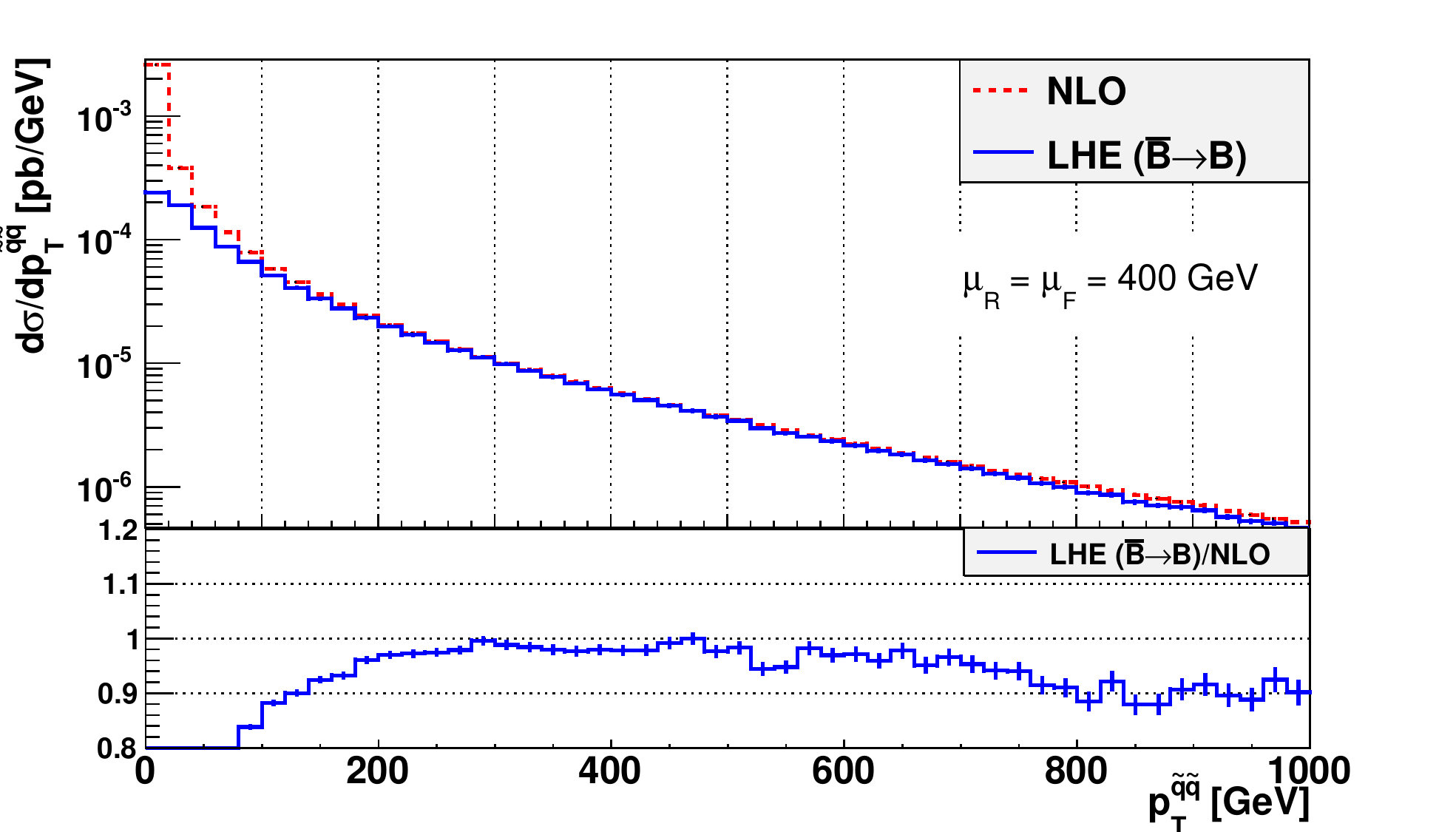}
\end{minipage}
\caption {Comparison of the NLO results with the LHE outcome for $p_T^{\sq\sq}$ with our default scale choice and a full \textsc{Powheg} simulation (left) and a lower scale choice $\mu_R=\mu_F=400\,\text{GeV}$ with the replacement $\overline{\matB}\rightarrow\matB$ in the \textsc{Powheg} simulation (right). Again the CMSSM point $10.3.6^*$ is used.}
\label{fig:ptsqsq}
\efig
 
 Comparing the NLO with the LHE curve (left plot in Fig.~\ref{fig:ptsqsq}), we note large discrepancies over the whole range. For low values of $p_T^{\sq\sq}$, the NLO result is divergent, whereas the behaviour of the LHE output reflects the Sudakov damping inherent in the way the \textsc{Powheg} events are generated according to Eq.~(\ref{pwgmaster}). For high $p_T^{\sq\sq}$-values, where the two curves should coincide again, the ratio LHE/NLO amounts to about $1.8$, {\it i.e.}~the LHE result is enhanced by this factor with respect to the outcome of an NLO simulation. A similar behaviour was already observed in other \textsc{Powheg}-implementations, {\it e.g.}~Higgs production in $gg$ fusion \cite{ggH} and vector boson pair production \cite{VV}. As discussed in these references, this enhancement
can, on the one hand, be traced back to the sizable NLO $K$-factors of the investigated process, as for large $p_T$ of the radiated parton the \textsc{Powheg} master-formula, Eq.~(\ref{pwgmaster}), behaves as
 \be
    d\sigma_{\sss PWG} \rightarrow \left(\frac{\overline{\matB}}{\matB} \matR_s +(\matR-\matR_s)\right)d\varPhi_{n+1} = \left[(1+\matO(\alpha_s))\matR_s + (\matR-\matR_s)\right]d\varPhi_{n+1}\ ,
 \ee
 {\it i.e.}~the ratio $\overline{\matB}/\matB$ enhances the (N)LO-prediction, which is described by $\matR_s$. On the other hand, this enhancement is also induced by the usage of different scales in the NLO calculation (where a fixed scale $\mu_R=\mu_F=\overline{m}_{\sq}$ is used) and the \textsc{Powheg} event generation (here the relevant scale is related to the $p_T$ of the radiated parton with respect to its emitter). The authors of \cite{ggH} proposed a simple test for this explanation: the whole event-generation is performed with $\overline{\matB}\rightarrow\matB$ in Eq.~(\ref{pwgmaster}), thus the enhancement-factor should drop out. To eliminate the effect of the different scales, we used for the comparison a lower scale of $\mu_R = \mu_F = 400\,\text{GeV}$, thus we expect to see an agreement of the (N)LO prediction and the LHE outcome with $\overline{\matB}\rightarrow\matB$ in the region $p_T^{\sq\sq}\approx 400\,\text{GeV}$. The results depicted in the right panel of Fig.~\ref{fig:ptsqsq} indeed 
show the expected behaviour.

To reduce this effect, a simple procedure was proposed in \cite{ggH}. In essence, the generalized \textsc{Powheg} master-formula with $\matR\neq \matR_s$ is used, with $\matR_s = \matF \matR $. The function $\matF$ has to fulfil $\matF\rightarrow 1$ in the limit of soft/collinear radiation and should vanish for harder radiation. This behaviour can be achieved {\it e.g.} with the following form (see \cite{ggH}):
\be
  \matF=\frac{h^2}{p_T^2+h^2}\ .
\ee
Here, $h$ is a parameter which controls the \lq damping' of the $\overline{\matB}/\matB$-enhancement (larger $h$ corresponds to a softer damping, {\it i.e.}~the (N)LO-behaviour is restored for larger values of $p_T$). This choice is also implemented in the \PB~and therefore used in the following.

In Fig.~\ref{fig:ptsqsqdamp} we show again the $p_T^{\sq\sq}$-distribution, now with different values of $h$. As expected, the actual value of $h$ determines the value of $p_T^{\sq\sq}$ where the NLO behaviour is restored. At first glance, the  ad-hoc introduction of this additional parameter seems to introduce a certain amount of arbitrariness in the prediction obtained with a \textsc{Powheg} simulation. But we recall here that all results are determined up to higher-order effects. Moreover, we have checked that the distributions of inclusive observables are not affected by the actual value of $h$, as expected.

 \bfig
\centering
 \includegraphics[width=0.8\textwidth,height=70mm]{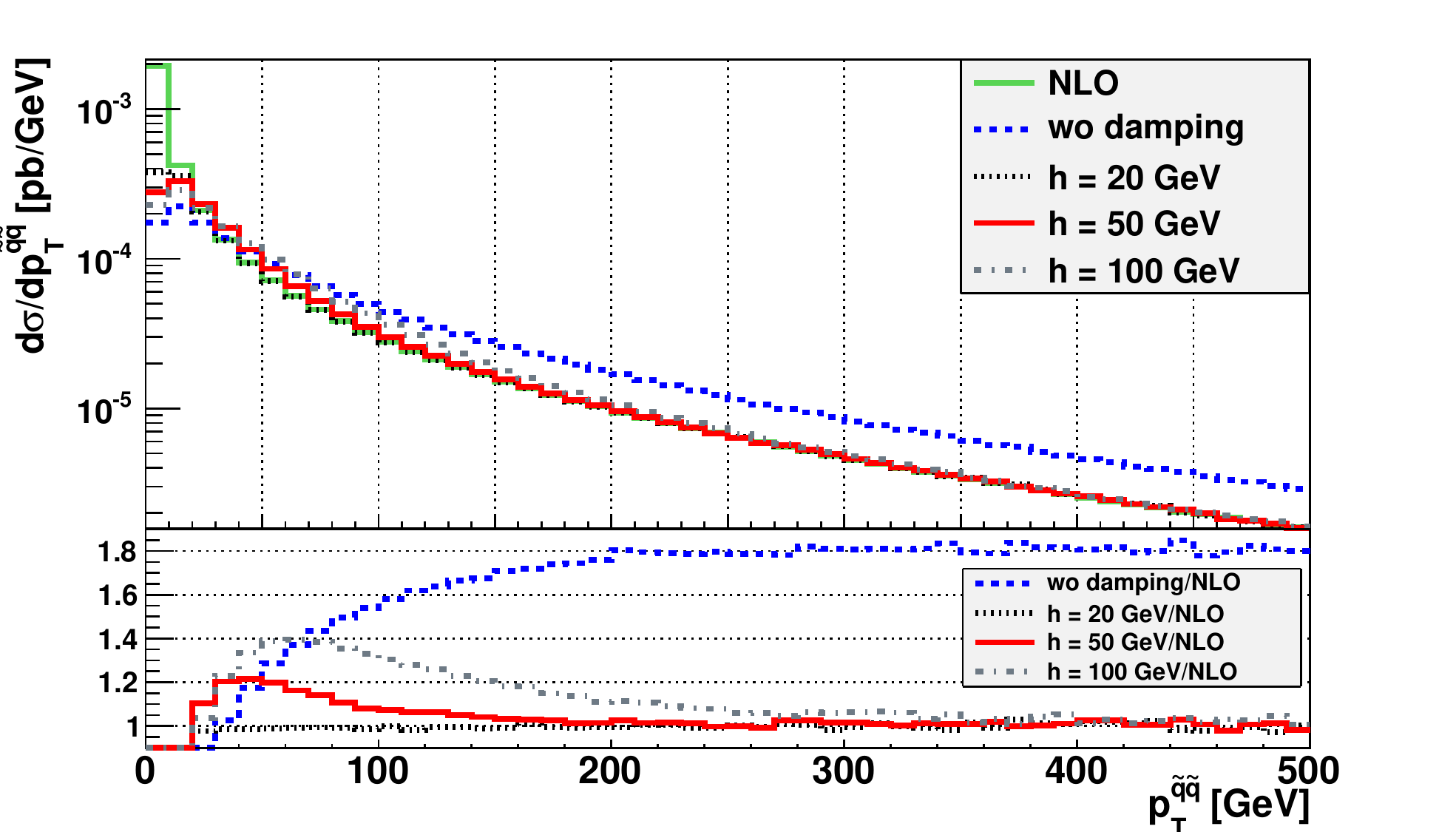}
\caption {Comparison of the $p_T^{\sq\sq}$ distribution obtained with the CMSSM point $10.3.6^*$ for the NLO case, the \textsc{Powheg} simulation without any damping of non-singular regions in the \textsc{Powheg} event generation (formally $h\rightarrow\infty$) and the results obtained with different choices of the damping parameter $h$. The error bars are not shown.}
\label{fig:ptsqsqdamp}
\efig

In the following, we will use $h=50\,\text{GeV}$, which ensures that the $p_T^{\sq\sq}$-distribution at NLO and after the generation of the \textsc{Powheg} radiation coincide for $p_T^{\sq\sq}>200\,\text{GeV}$. Of course, the agreement between these two results is not limited to this specific distribution, but can be observed in other distributions which are sensitive to the emission of an additional parton, too. As an example the rapidity distributions for the radiated parton, $y^j$, and for the $\sq\sq$-system, $y^{\sq\sq}$, (with a cut $p_T^{\sq\sq}>200\,\text{GeV}$) are shown in Fig.~\ref{fig:yjptsqsqcut}.

 \bfig
  \begin{minipage}{0.5\textwidth}
 \includegraphics[width=\textwidth,height=48mm]{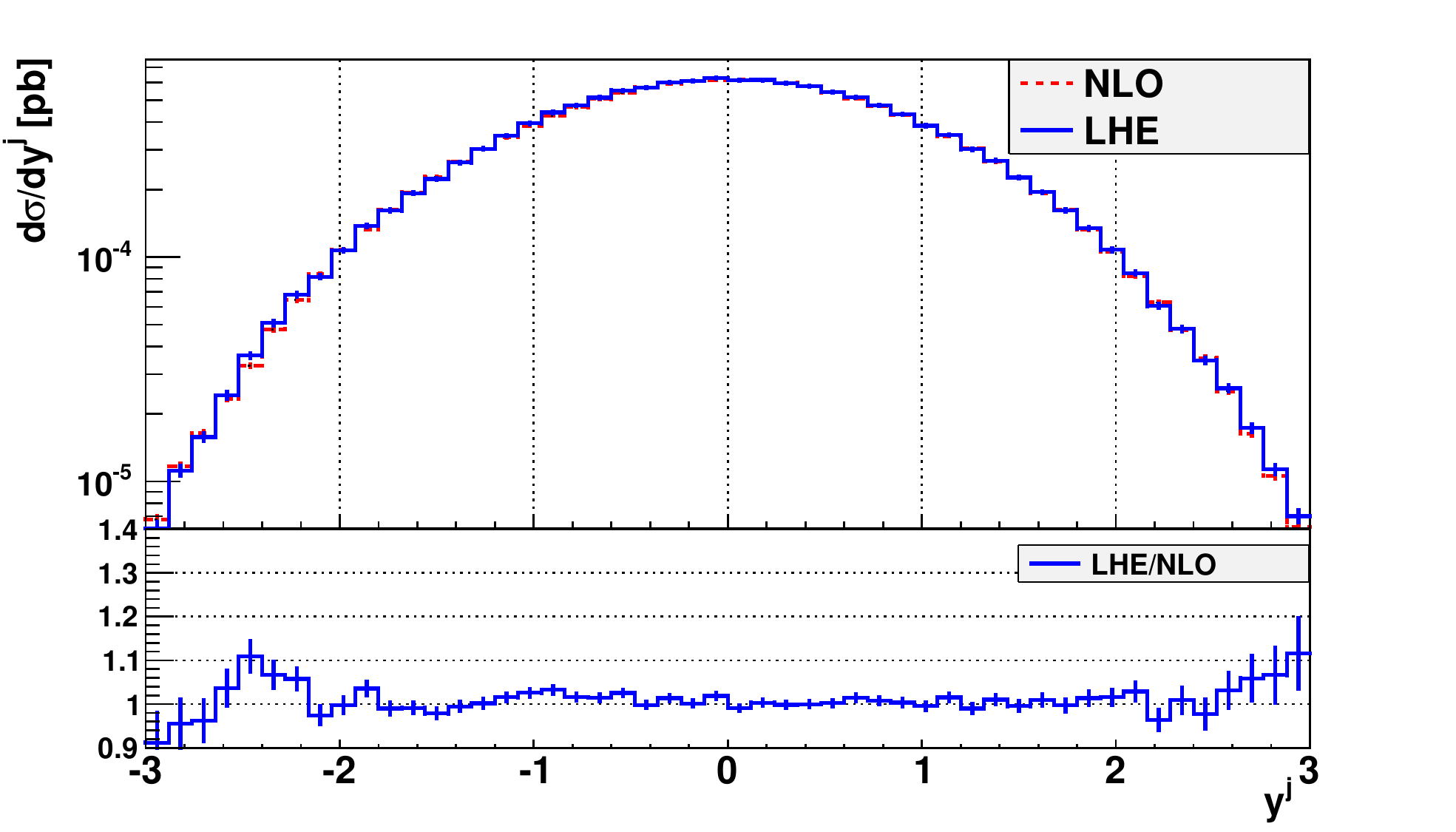}
\end{minipage}
\begin{minipage}{0.5\textwidth}
 \includegraphics[width=\textwidth,height=48mm]{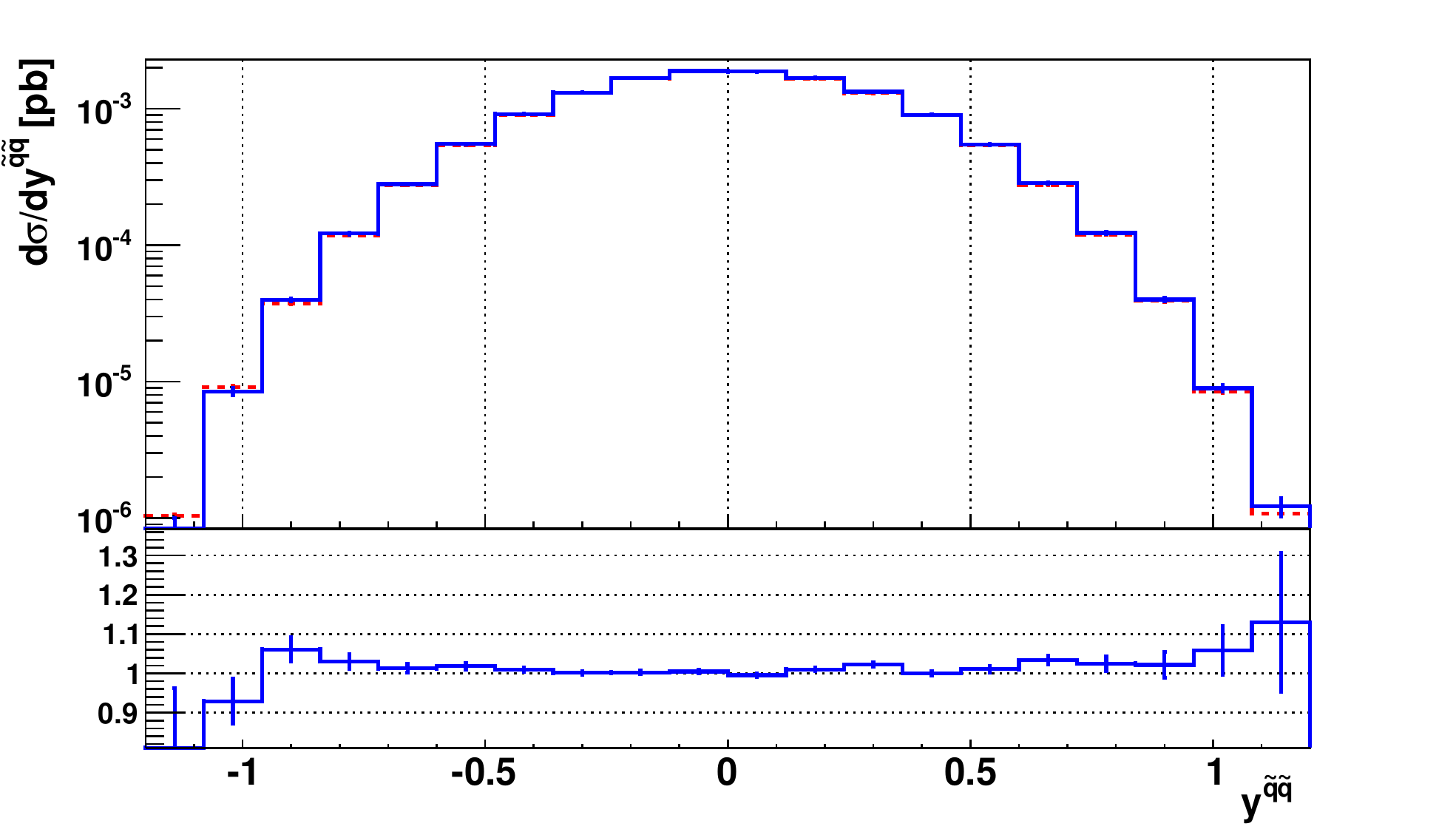}
\end{minipage}
\caption {The rapidity distributions of the emitted parton ($y^j$) and of the $\sq\sq$-system ($y^{\sq\sq}$) with a cut $p_T^{\sq\sq}>200\,\text{GeV}$ applied for the CMSSM point $10.3.6^*$.}
\label{fig:yjptsqsqcut}
\efig

 The choice $\matR\neq \matR_s$ increases the fraction of negative weights to around $5\%$, as it essentially selects the IR-divergent regions. This fraction was completely negligible for $h\rightarrow \infty$ and scenarios with $m_{\sq}>m_{\go}$. To eliminate this effect, we used the folding-option as implemented in the \PB~(see \cite{powhegbox} for details). After applying this procedure with $f_{\xi}=5,f_{y}=2, f_{\phi}=1$\footnote{The parameters $f$ correspond to the number of foldings, {\it i.e.}~the number of phase-space points considered for each radiation variable while keeping the underlying Born kinematics fixed.} for the integration over the radiation variables $\xi$, $y$ and $\phi$, respectively, with the choice $h=50\,\text{GeV}$, the fraction of events with negative weights is below one per mille and thus again completely irrelevant. 

 As in the case of the inclusive observables discussed earlier, these observations hold equally well for the other considered benchmark point, $10.4.5$, with the DS-scheme applied.

\subsection{Subtraction of Contributions with On-shell Intermediate $\go$ in Powheg}
\label{sec:pwgonshellsub}
\bfig[t]
  \begin{minipage}{0.5\textwidth}
 \includegraphics[width=\textwidth,height=50mm]{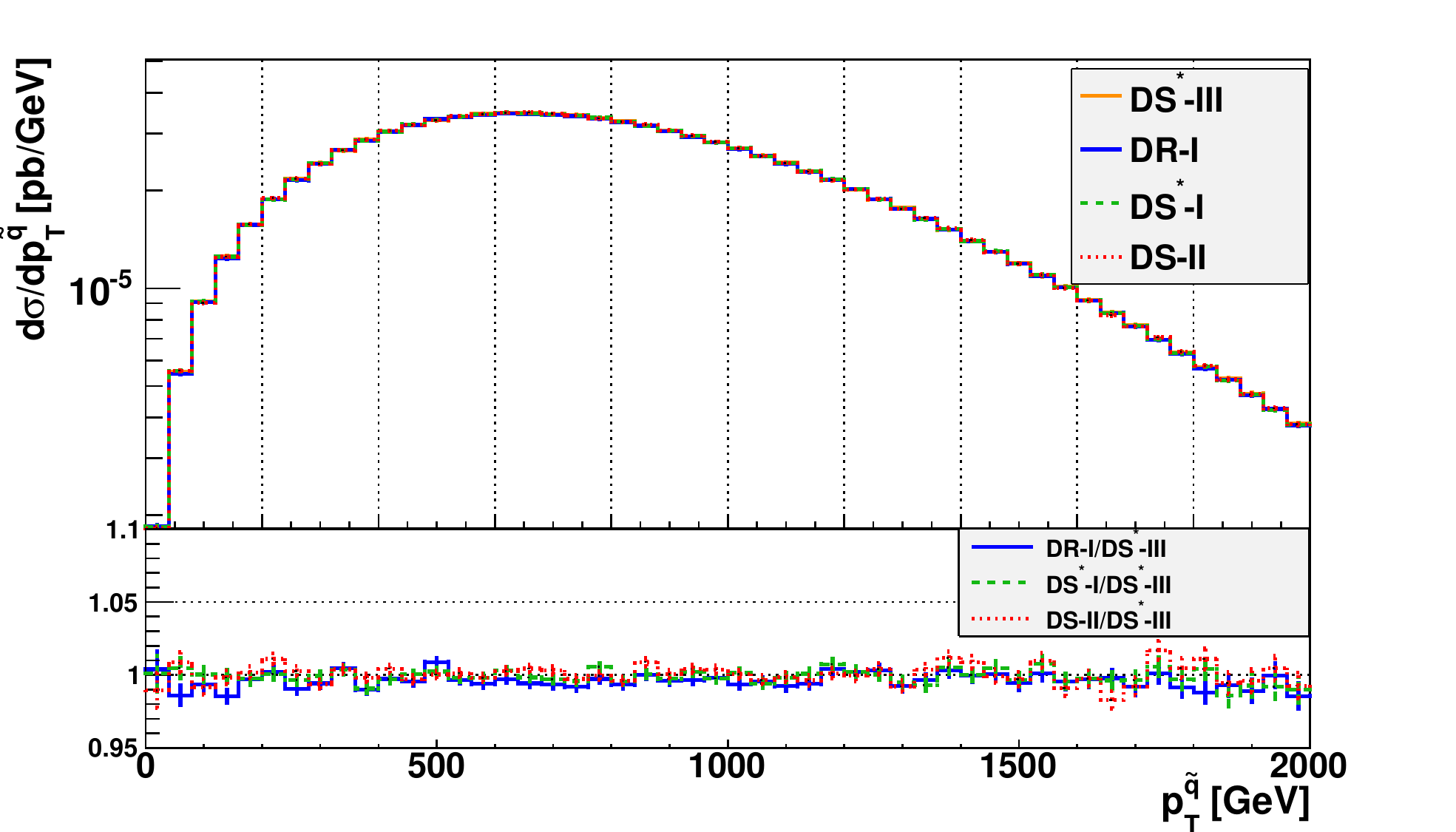}
 \newline
 \includegraphics[width=\textwidth,height=50mm]{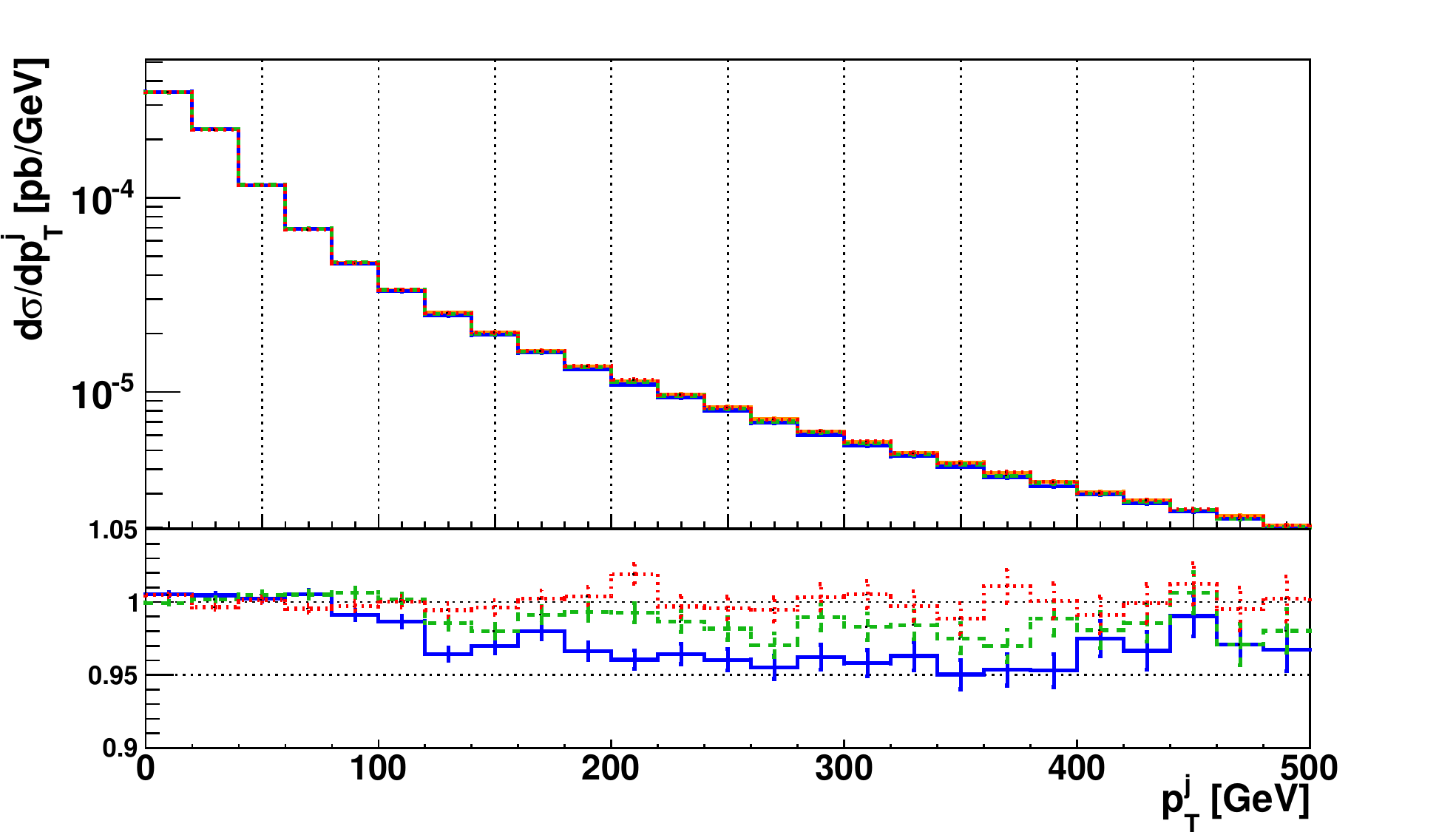}
\end{minipage}
\begin{minipage}{0.5\textwidth}
 \includegraphics[width=\textwidth,height=50mm]{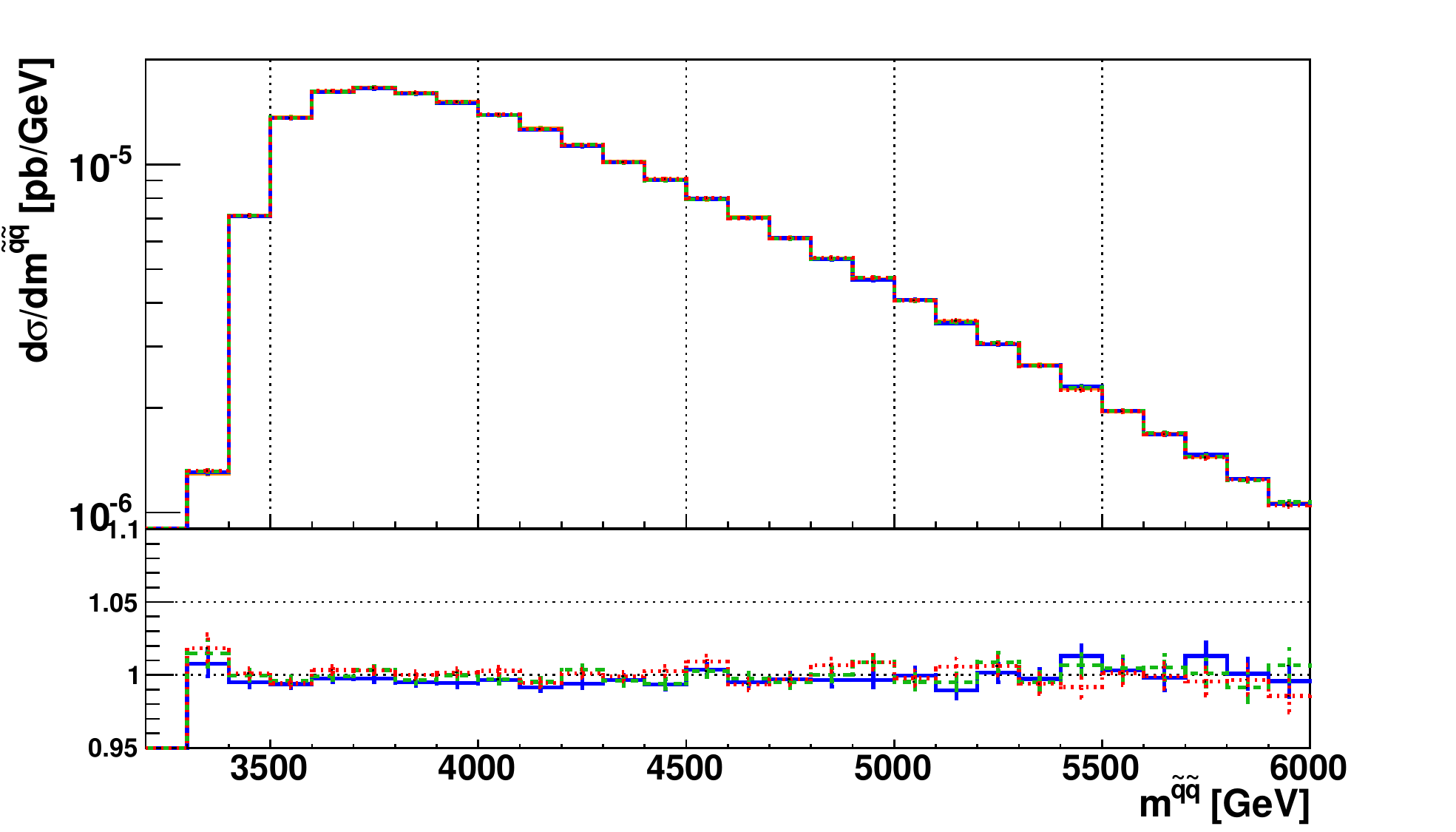}
 \newline
 \includegraphics[width=\textwidth,height=50mm]{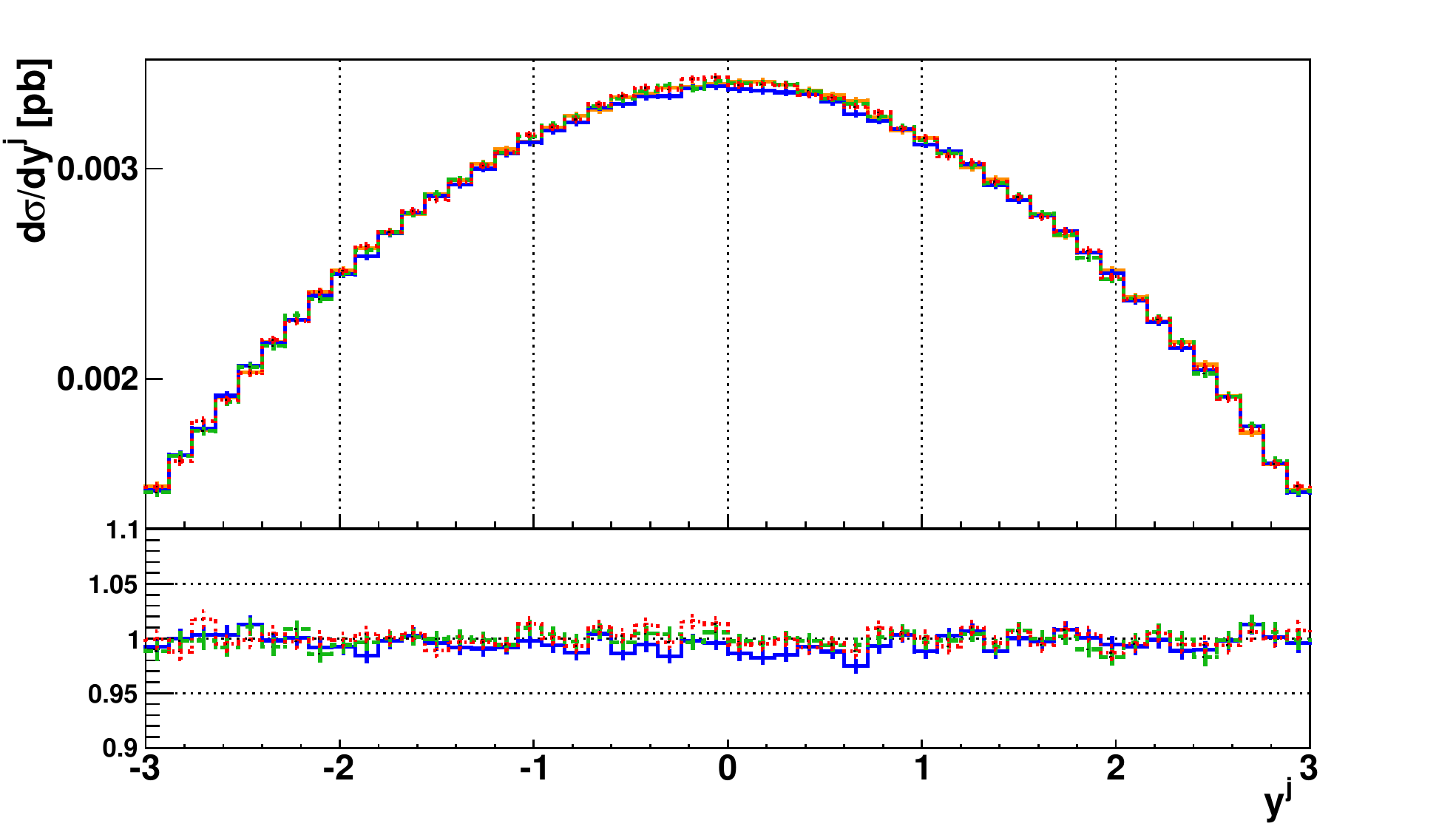}
\end{minipage}
\caption {Comparison of different subtraction methods for the contributions with on-shell intermediate $\go$. Shown are the results obtained with the CMSSM point $10.4.5$ for $p_T^{\sq}$, $m^{\sq\sq}$, $p_T^j$ and $y^j$ after the \textsc{Powheg} event generation ({\it i.e.}~at the LHE file level) for the simplest Diagram Removal scheme (DR-I) and for the three implementations of the Diagram Subtraction (DS) method for a regularizing $\go$ width $\Gamma_{\go} = 1\,\text{GeV}$. The lower part of each plot shows the ratios of the DR-I and the DS$^*$-I and DS-II results to the DS$^*$-III prediction.}
\label{fig:OSSmethods}
\efig
As discussed in Sec.~\ref{sec:imposs} there exist several possible ways to implement a pointwise subtraction scheme for the contributions with intermediate on-shell $\go$ in the \PB. In this section we will show some results obtained with the different methods discussed earlier. To this end we consider the event samples generated for the benchmark point $10.4.5$ with the settings specified in the sections above. The results presented in the following are all based on the \textsc{Powheg} LHE output, the $\sq$ are again left undecayed and no cuts are applied.

In Fig.~\ref{fig:OSSmethods} the different subtraction methods are compared. The Diagram Removal methods DR-I and DR-II (which is not displayed in the plot) have been defined in Sec.~\ref{sec:onshellsub}. The Diagram Subtraction methods DS$^*$-I, DS-II and DS$^*$-III are distinct with respect to the actual implementation and the way the regulator $\Gamma_{\go}$ is introduced: for DS$^*$-I, the event generation is performed with the complete real amplitudes squared after subtracting the on-shell contributions. The matrix elements squared are expanded according to Eq.~(\ref{eq:expan}), but as discussed in Sec.~\ref{sec:imposs} it is not possible to modify the Jacobian of the subtraction terms correctly. The DS-II results are obtained such that the resonant parts with the respective subtraction terms are treated as regular remnants. This allows for this modification, however an expansion of the matrix elements (which is required to preserve gauge invariance) is not possible. In the DS$^*$-III methods both the 
expansion of the matrix elements and the modification of the Jacobian for the subtraction terms is taken into account.  All methods except the DR-I scheme require the introduction of a regularizing $\go$ width. We use $\Gamma_{\go} = 1\,\text{GeV}$ here. Moreover, the DS$^*$-I and the DS$^*$-III method require the introduction of a cut on the invariant mass of the resonant $\go$ in the radiation generation of the \textsc{Powheg} event as defined in Eq.~(\ref{eq:cutmgo}). We used $\Delta=10\,\text{GeV}$.

Comparing the distributions for the $p_T$ and the invariant mass of the $\sq$ (upper row) we note that the differences between the methods are smaller than $2\%$ over the whole considered range and mostly dominated by statistical fluctuations. The same conclusion holds for the rapidity of the radiated parton, $y^j$. Larger discrepancies occur in the $p_T$-distribution of the radiated parton, $p_T^j$, for high $p_T$-values where the distribution is essentially dominated by the actual form of the real amplitudes squared and thus becomes more sensitive to the applied subtraction method. Here the DR-I scheme gives slightly ($\matO(3-5\%)$) smaller predictions than the DS($^*$) methods. Considering the DR-II method the observations are essentially the same. The discrepancies in the $p_T^j$ distribution are in this case even larger than those obtained with the DR-I method and amount to $\matO(5-8\%)$.

Another important point in the context of the Diagram Subtraction scheme is the independence of the result of the numerical value for the regulator width $\Gamma_{\go}$, see Sec.~\ref{sec:onshellsub}. As already stated there, the contribution of the $qg$ channels to the total cross section is independent of this value if we consider $\Gamma_{\go}\lesssim1\,\text{GeV}$. The effect of $\Gamma_{\go}$ on distributions obtained after the \textsc{Powheg} simulation can be estimated from the results depicted in Fig.~\ref{fig:OSSDSgamma}, where the DS$^*$-III method was applied for $\Gamma_{\go} = 0.1\,\text{GeV},\,1\,\text{GeV},\,10\,\text{GeV}$. As can be concluded from the plots, the results are essentially independent of the actual value of $\Gamma_{\go}$ over the whole range considered here, even if we use the a value $\Gamma_{\go}>1\,\text{GeV}$. This is a consequence of the fact that the qg-contributions are tiny in comparison to the total NLO cross section, see Sec.~\ref{sec:onshellsub}. 
\bfig[t]
  \begin{minipage}{0.5\textwidth}
 \includegraphics[width=\textwidth,height=50mm]{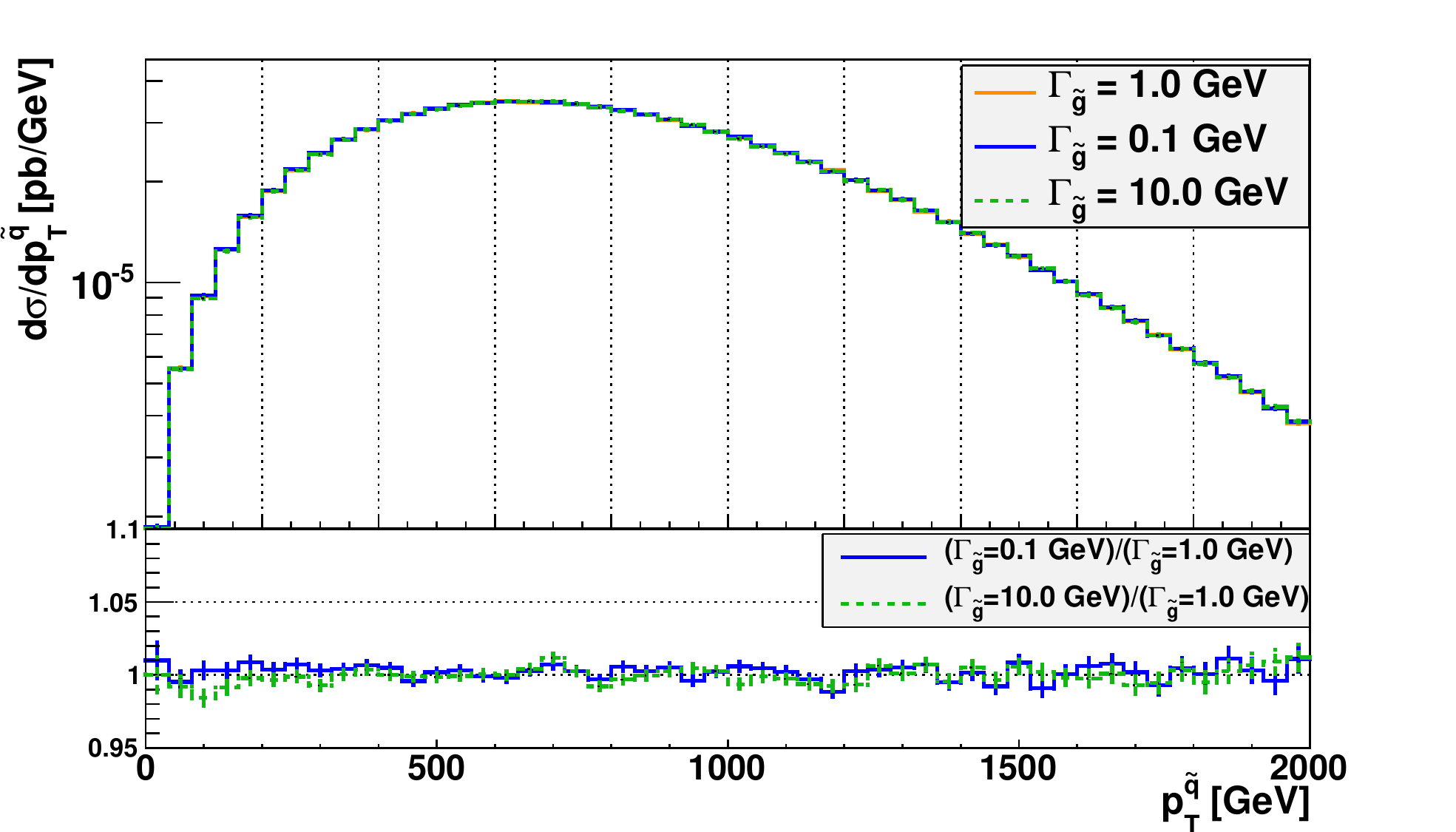}
 \newline
 \includegraphics[width=\textwidth,height=50mm]{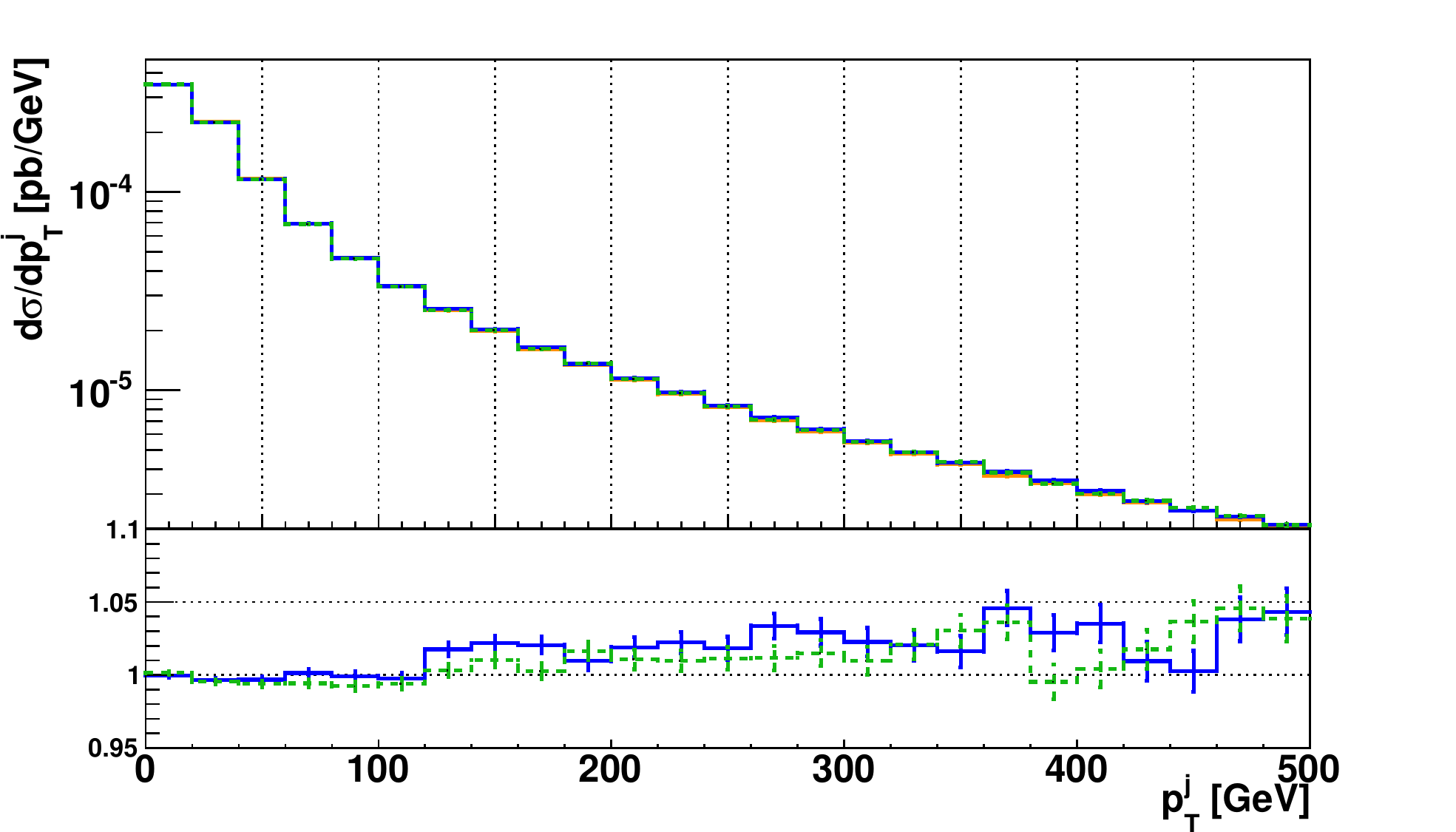}
\end{minipage}
\begin{minipage}{0.5\textwidth}
 \includegraphics[width=\textwidth,height=50mm]{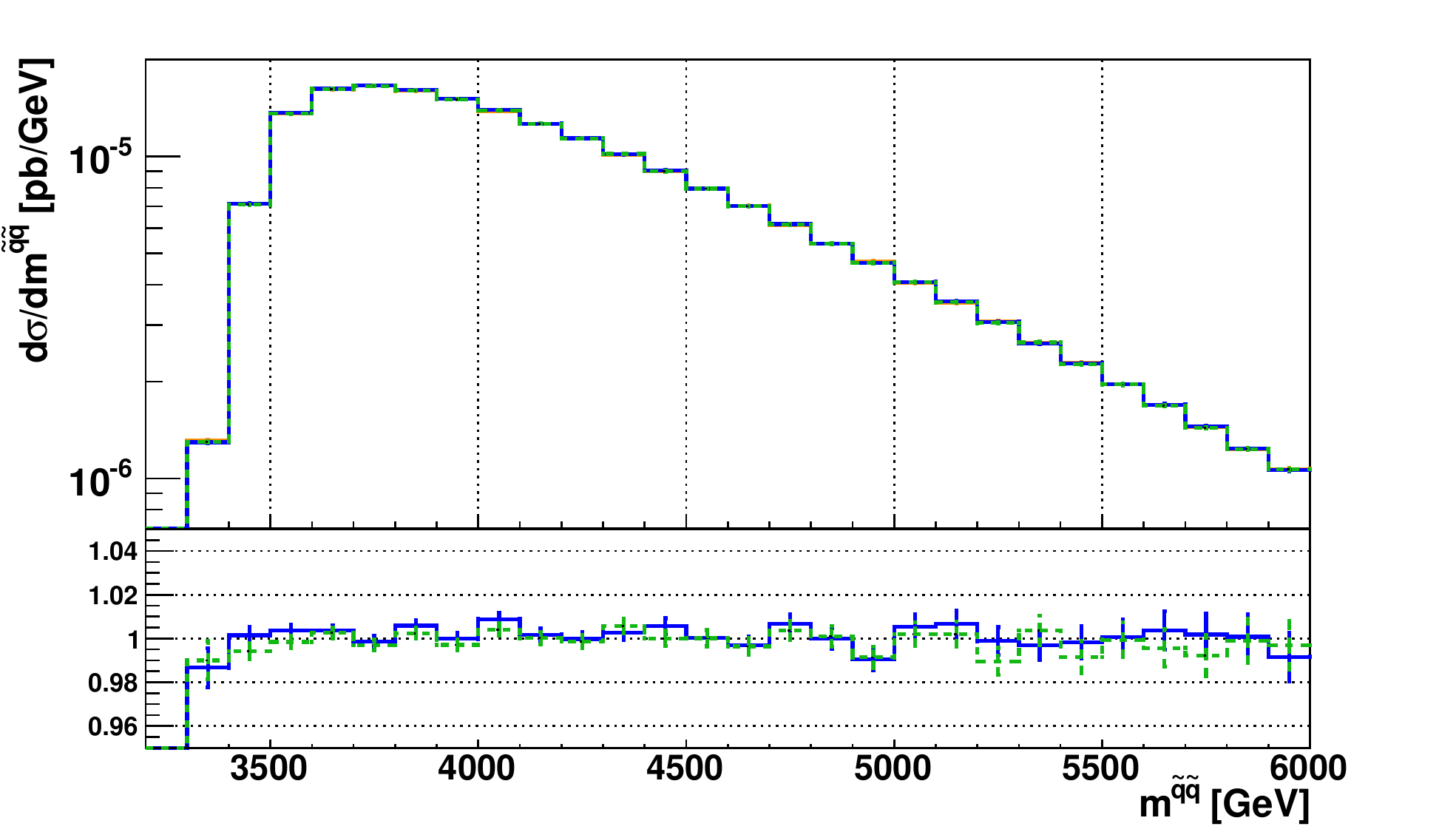}
 \newline
 \includegraphics[width=\textwidth,height=50mm]{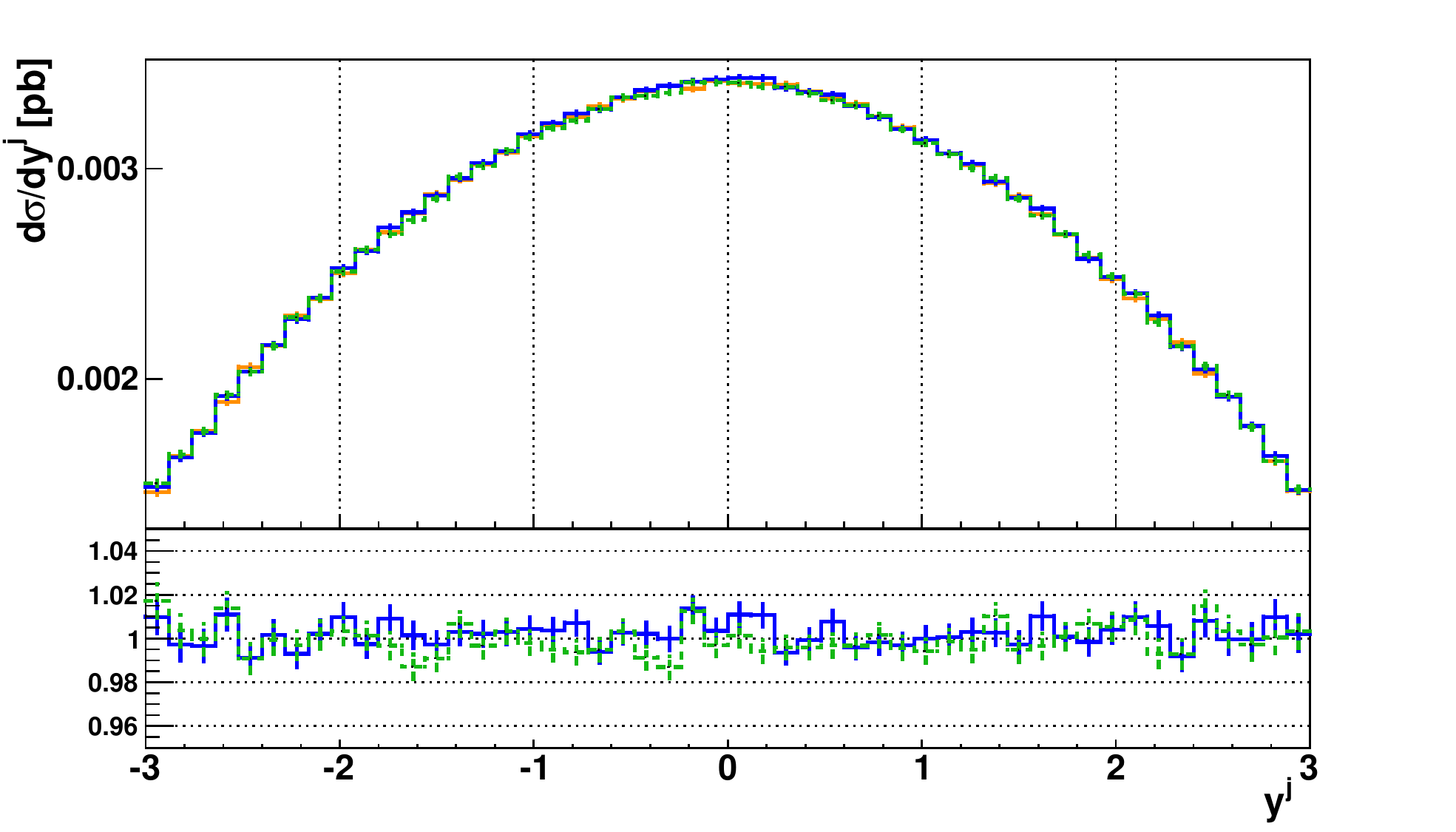}
\end{minipage}
\caption {Dependence of the \textsc{Powheg} results for the CMSSM point $10.4.5$ obtained with the DS$^*$-III method on the regularizing $\go$ width  $\Gamma_{\go}$. The lower part of each plot shows the ratio for the results with $\Gamma_{\go} = 0.1\,\text{GeV}\,\text{or}\,10\,\text{GeV}$ and $\Gamma_{\go} = 1\,\text{GeV}$.}
\label{fig:OSSDSgamma}
\efig

\subsection{Influence of Different Parton Showers}
\label{sec:diffPS}
To test the influence of different parton showers, we have interfaced the LHE files obtained from the \PB~ with different shower programs: \textsc{Pythia 6} (version 6.4.26) \cite{pythia6}, where we invoked the usage of the $p_T$-ordered shower as appropriate for the \textsc{Powheg} method by calling the routine {\tt PYEVNW}, and \textsc{Herwig++} (version 2.6.1) \cite{herwigpp,herwigpp26} both with the default shower and the Dipole shower\footnote{The \textsc{Herwig++} default shower is angular-ordered, hence even after applying a $p_T$-veto the thus obtained results are not complete. In principle one has to add a truncated shower, which adds soft, wide-angle radiation, see \cite{nason}, but as this option is not available in \textsc{Herwig++} our results do not contain this additional radiation. However, by comparing with the output of the Dipole Shower, which is $p_T$-ordered, an estimate of the importance of these left-out contributions is possible.} \cite{herwigdp1,herwigdp2}. When comparing 
the results obtained with the different showers, we will focus on observables related to the jet originating from the first emission created according to the \textsc{Powheg} method. These observables play {\it e.g.} an important role in the disentangling of $\sq$ and $\go$ production in case of scenarios where the shortest possible cascades are predominant, {\it i.e.}~the $\go$ decays into $\sq \bar{q}$ and the $\sq$ into $q \tilde{\chi}^0_1$. Further studies on parton shower effects for these processes applying merging techniques to combine matrix elements for $\sq\sq + 1\, \text{or}\, 2$ partons with \textsc{Pythia 6} can be found in the literature \cite{sqmerging,sqmerging2}.

As we are mainly interested in the effects of the parton showers, we switched off hadronisation and simulation of the underlying event in the used programs. The \textsc{Pythia} results have been obtained with the Perugia 0 tune \cite{perugia} (\textsc{MSTP(5)} = 320). A comparison with the Perugia 11 tune (\textsc{MSTP(5)} = 350) shows only small discrepancies (up to $-4\%$ in the jet observables if the $\sq$ decays are not included and up to $-8\%$ for the third hardest jet with the decays $\sq\rightarrow q\tilde{\chi}^0_1$ taken into account). 

To study solely the effects of the parton showers on the results at production level we consider in a first step again the case of undecayed $\sq$. However, interfacing the \textsc{Powheg} events to the \textsc{Herwig++} Dipole shower with undecayed scalar particles is not possible, as the splitting kernels which invoke the $\sq$ are not implemented in the current version. Therefore we compare in Fig.~\ref{fig:SH_undec} only the default shower of \textsc{Herwig++} with \textsc{Pythia} and the NLO results. Considering first the inclusive quantities $p_T^{\sq}$ and $m^{\sq\sq}$ in the upper row we note that both showers hardly affect the NLO prediction for these distributions, as is expected for final state particles with masses of $\matO(\text{TeV})$.  

\bfig[t]
  \begin{minipage}{0.5\textwidth}
 \includegraphics[width=\textwidth,height=50mm]{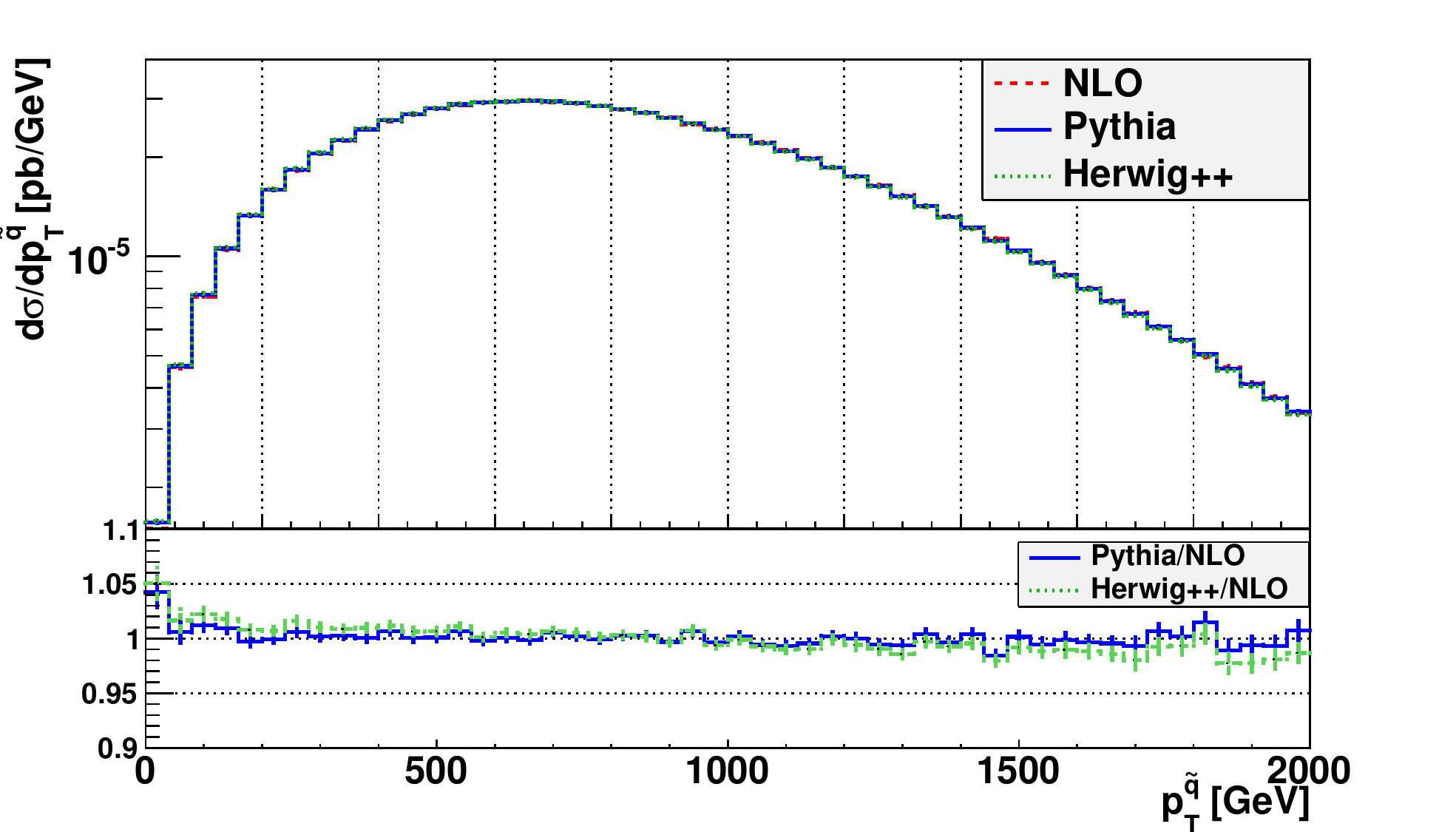}
 \newline
 \includegraphics[width=\textwidth,height=50mm]{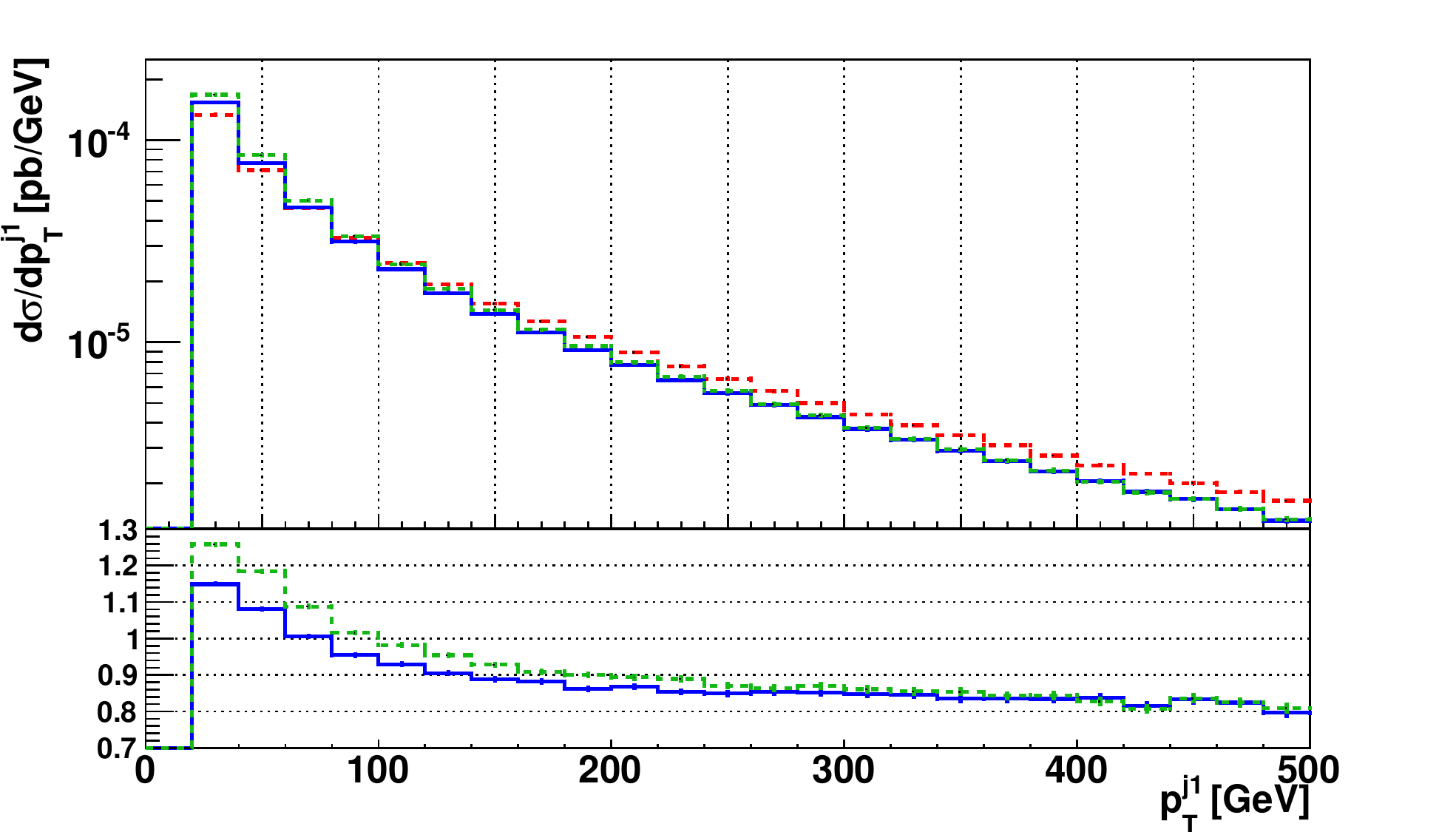}
\end{minipage}
\begin{minipage}{0.5\textwidth}
 \includegraphics[width=\textwidth,height=50mm]{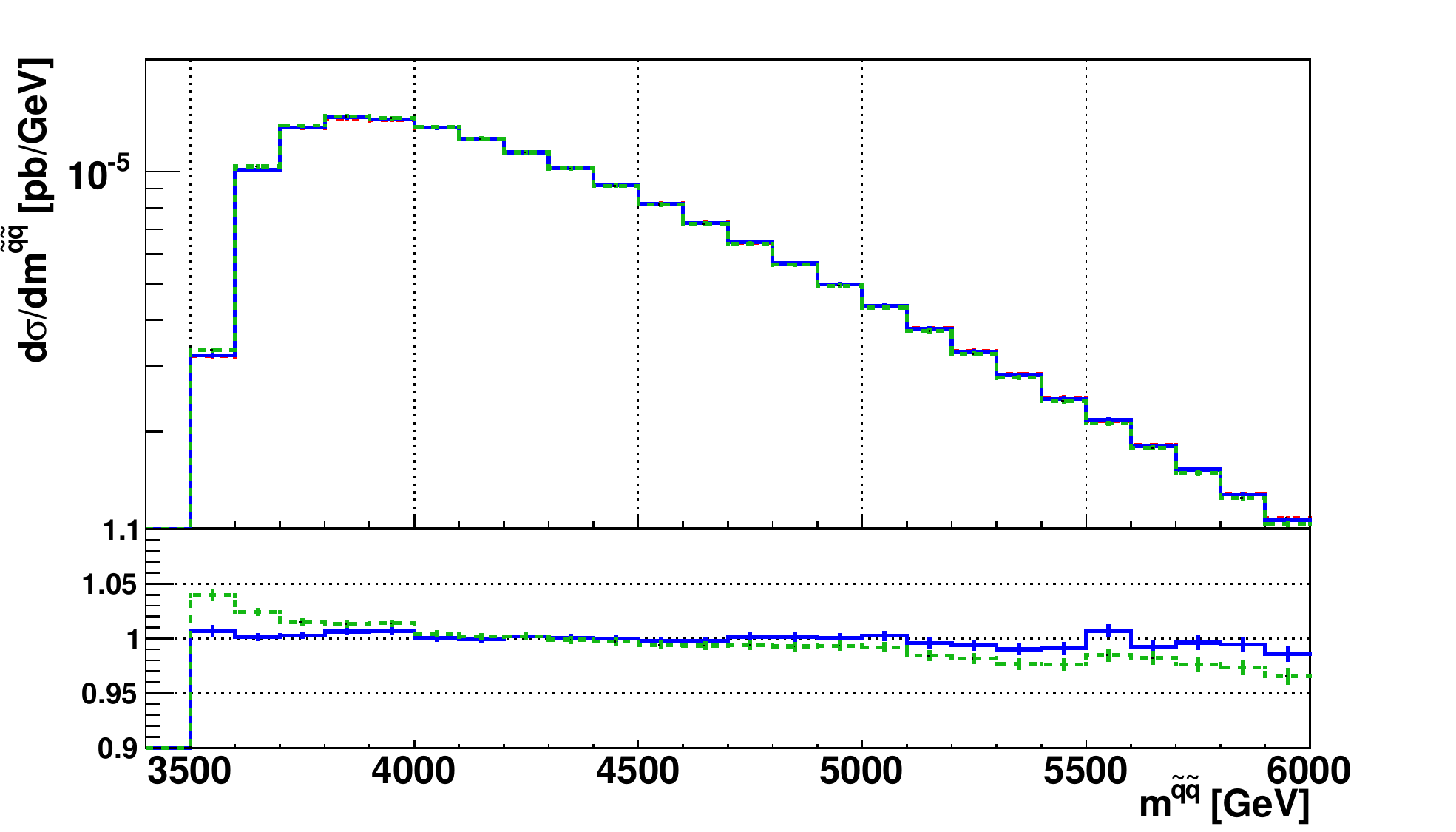}
 \newline
 \includegraphics[width=\textwidth,height=50mm]{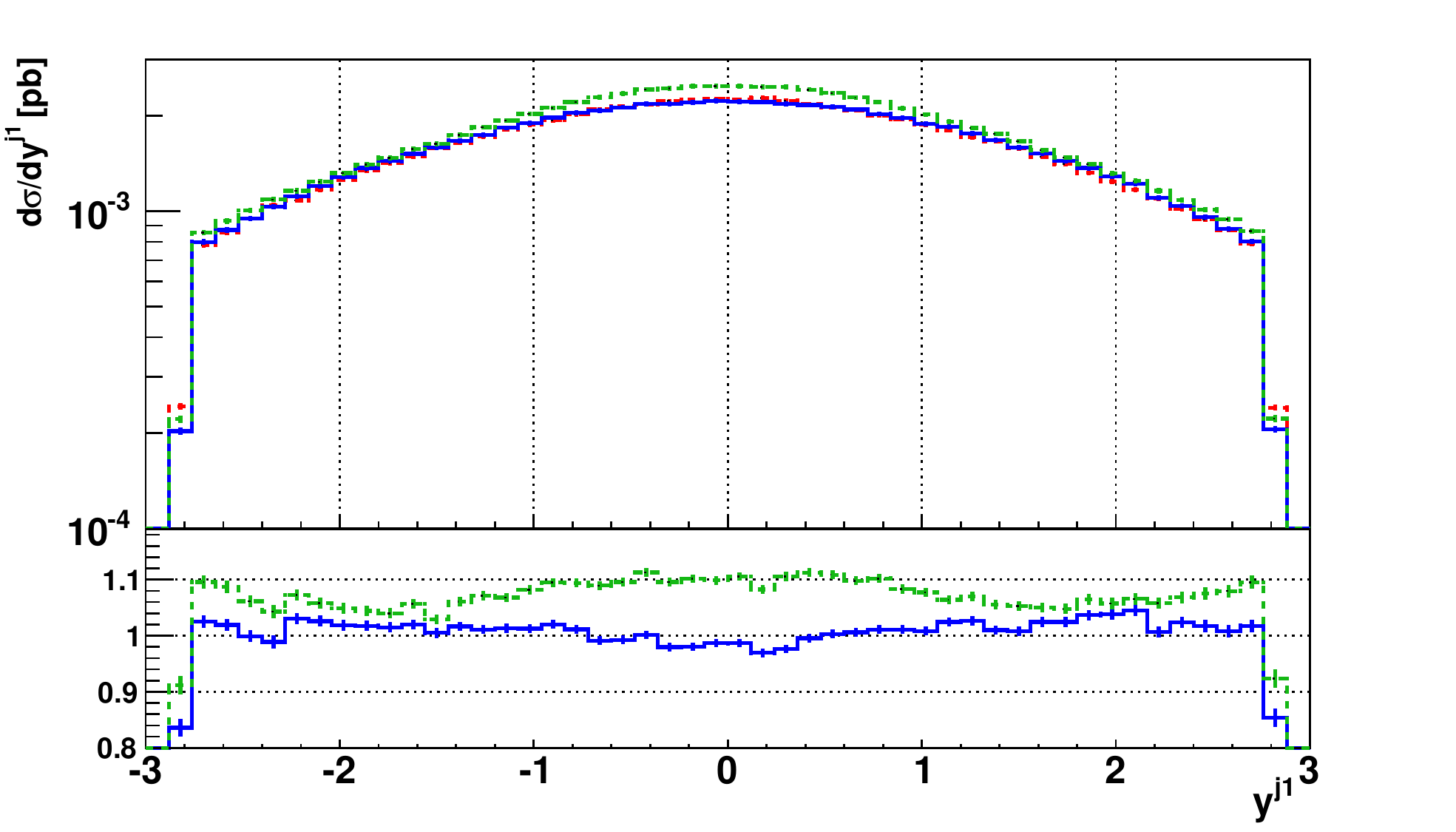}
\end{minipage}
\caption {Results after applying a parton shower obtained with \textsc{Pythia} and the \textsc{Herwig++} default shower compared to NLO predictions for the CMSSM point $10.3.6^*$. Shown are the results for $p_T^{\sq}$, $m^{\sq\sq}$  and $p_T^j$, $y^j$ (for the hardest jet). The lower part of each plot shows the ratio of the shower results and the NLO prediction.}
\label{fig:SH_undec}
\efig

In the lower row of Fig.~\ref{fig:SH_undec} the $p_T$ and rapidity of the hardest jet are shown. Considering first the predictions for $p_T^{j1}$ we notice that both showers agree for $p_T^{j1}\gtrsim 200\,\text{GeV}$ and predict lower rates than the NLO distribution for $p_T^{j1}\gtrsim 100\,\text{GeV}$. This behaviour is caused by additional radiation produced in the showering stage that may be too hard and/or develop too large angles to be clustered together with the original parton into the hardest jet. For smaller $p_T^{j1}$ values the \textsc{Herwig++} result is up to $10\%$ larger than the \textsc{Pythia} prediction. 

For the $y^{j1}$ distribution in Fig.~\ref{fig:SH_undec} we observe some discrepancies between the showers, too. While \textsc{Pythia} essentially reproduces the NLO result, the \textsc{Herwig++} default shower has a higher jet-rate especially in the central region. The observed difference in the shape of the curve is caused by relatively soft jets. Considering the same quantity for jets with $p_t^{j1}> 100\,\text{GeV}$ (instead of $p_t^{j1}> 20\,\text{GeV}$) the two shower predictions coincide around $y^{j1}=0$. This discrepancy can be traced back to initial state radiation (ISR): Comparing the same observable with ISR turned off, the two showers agree with each other. The \textsc{Herwig++} prediction without ISR stays more or less the same in the central region, whereas the \textsc{Pythia} result goes up by almost $10\%$ around $y^{j1}=0$. This observation can be attributed to the fact that \textsc{Pythia} is known to create more soft wide-angle radiation and therefore \lq pulls' the third jet away from 
the central region. A similar effect was 
described recently in a study on parton shower effects in vector boson fusion, see \cite{VBFPS}. 

While being of some interest for the understanding of the different parton showers, event samples with undecayed $\sq$ are obviously not very relevant for phenomenological studies. As a last step we therefore consider again the simplest possible decay channel $\sq\rightarrow q \tilde{\chi}^0_1$ and compare the output of \textsc{Pythia} and \textsc{Herwig++} (now with both the default and the Dipole shower). The decays are performed by the shower programs directly, but we use again the BRs from Tab.~\ref{tab:brs}. We have checked that the distributions after the decay without parton shower perfectly agree with the results obtained with our own decay routine. 

In Fig.~\ref{fig:SH_dec} we plot the missing transverse energy $E_T^{\text{miss}}$ carried away by the $\tilde{\chi}^0_1$, and the $p_T$-distributions of the three hardest jets as obtained with the three parton showers, compared to the NLO prediction. The $E_T^{\text{miss}}$ shape is barely affected by the showers. Only at the very end of the shown range (where the quarks from the original decay tend to be rather soft) both \textsc{Pythia} and the \textsc{Herwig++} default shower drop significantly below the NLO curve. Turning next to the jet distributions a large discrepancy between both \textsc{Herwig++} showers and the \textsc{Pythia} results for all $p_T^j$ distributions is obvious. While the two hardest jets are significantly softer in case of \textsc{Pythia}, the rate for a third jet is much higher than in the \textsc{Herwig++} results. 
\bfig[t]
  \begin{minipage}{0.5\textwidth}
 \includegraphics[width=\textwidth,height=50mm]{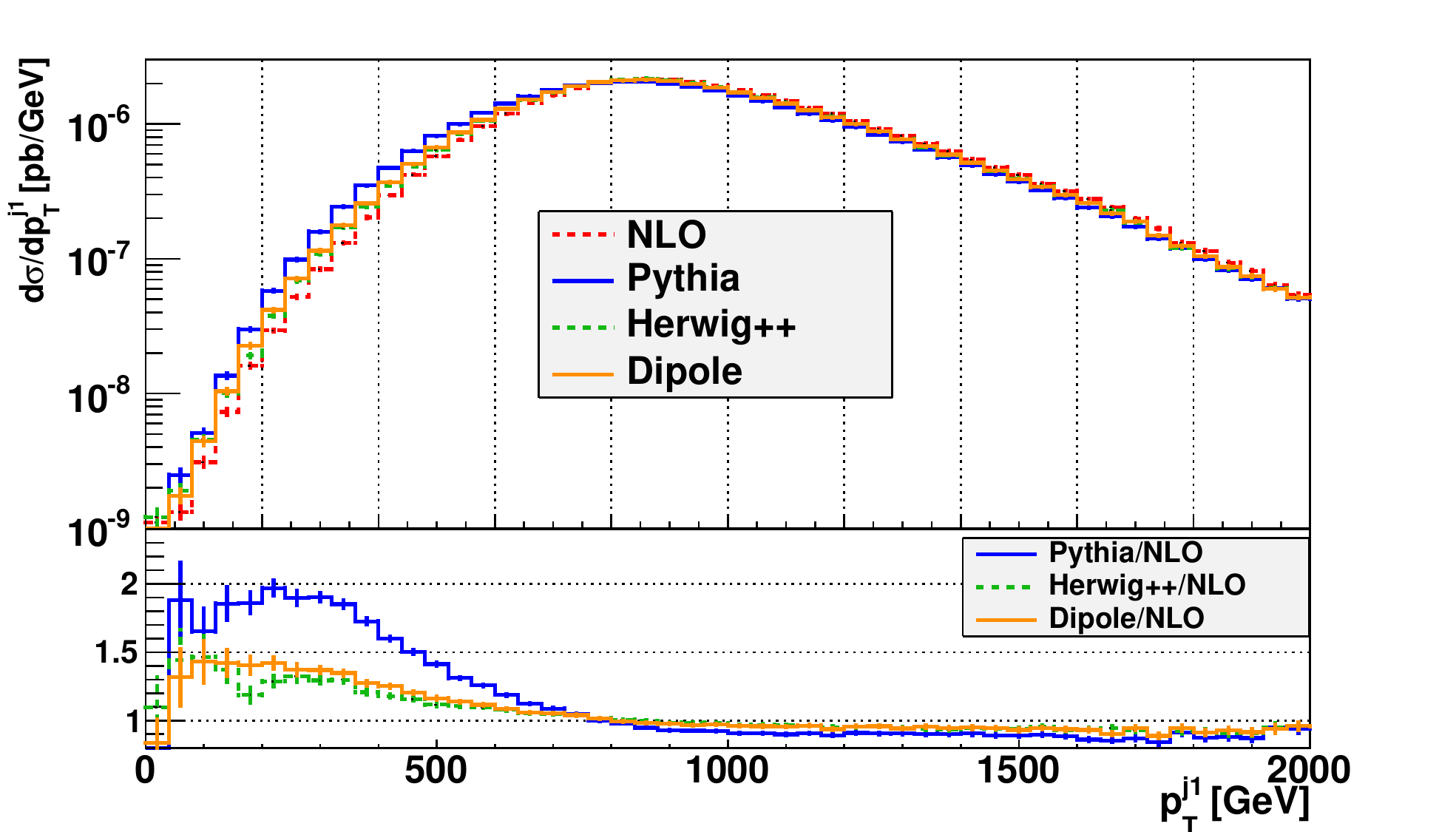}
 \newline
 \includegraphics[width=\textwidth,height=50mm]{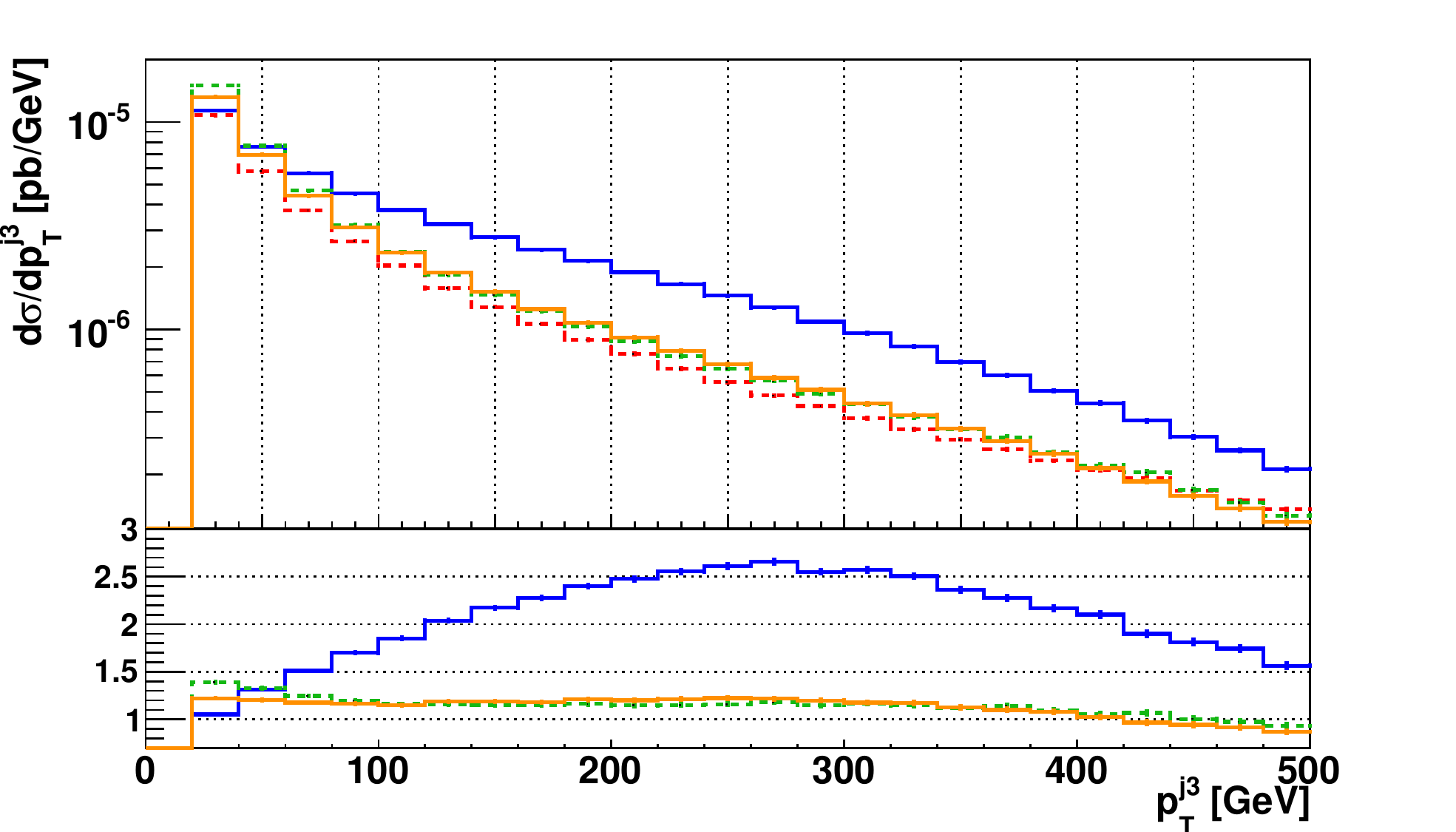}
\end{minipage}
\begin{minipage}{0.5\textwidth}
 \includegraphics[width=\textwidth,height=50mm]{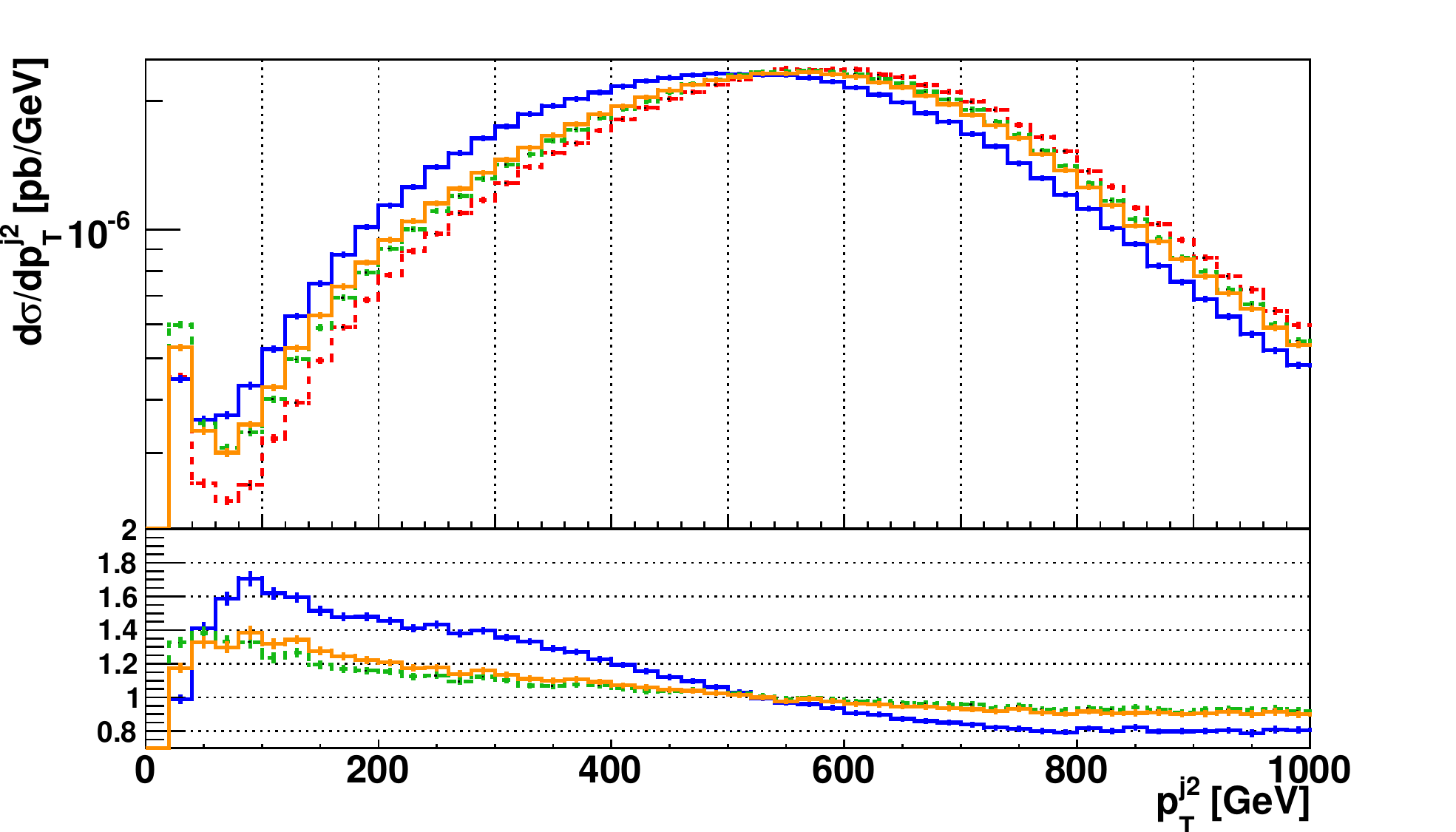}
 \newline
 \includegraphics[width=\textwidth,height=50mm]{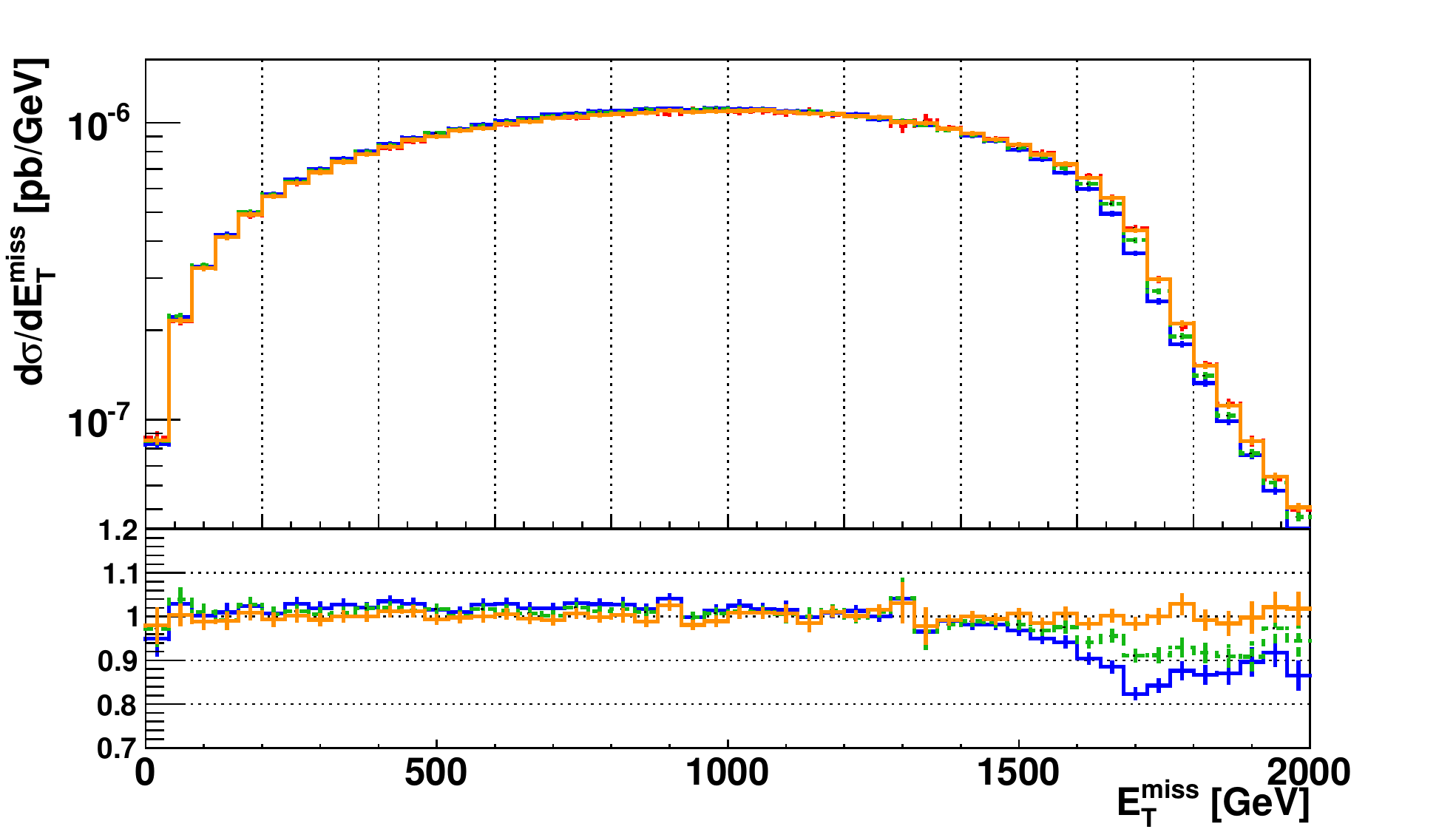}
\end{minipage}
\caption {Comparison of the differential cross sections for the $p_T$ of the three hardest jets and for the missing transverse energy $E_T^{\text{miss}}$ for the three parton showers and the NLO results. The CMSSM point $10.3.6^*$ was used and the lower part of each plot shows the ratio of the shower results and the NLO prediction.}
\label{fig:SH_dec}
\efig
The reason for these large effects is related to the way the decays of the $\sq$ are performed in the showers and the way the $p_T$-veto is applied in case of \textsc{Herwig++}. \textsc{Pythia} performs the decays during the showering stage and creates the additional radiation off the $\sq$-decay products independently of the radiation related directly to the production process ({\it i.e.}~ISR and radiation off the parton created in the \textsc{Powheg} simulation). The starting scale for the shower is related to the mass of the decaying particle. In contrast, the produced particles in \textsc{Herwig++} are decayed before the parton shower. The imposed $p_T$-veto with the veto-scale being determined by $p_T^{PWG}$, the $p_T$ of the first (\textsc{Powheg}) emission, is then applied for radiation related both to the production and the decay process. The starting scale for the final state showers from the $q$ produced in the $\sq$ decay is therefore much smaller than in the simulation with \textsc{Pythia}. To 
compensate this 
effect, we have modified \textsc{Pythia} such that $p_T^{PWG}$ determines the starting scale for all types of radiation.\footnote{The
  sole purpose of this change is to compare the parton showers on an equal footing. A realistic prediction for phenomenological discussions should instead treat production and decay consistently at the same order of perturbation theory. After matching the full process to a parton shower with the \textsc{Powheg} method, the starting scale for the shower is unambiguously related to the $p_T$ of the first (hardest) emission, regardless of its origin. A comparison of {\it e.g.}~event rates with experimental data makes only sense after this extension of our calculation. We leave this to future work.}

\bfig
  \begin{minipage}{0.5\textwidth}
 \includegraphics[width=\textwidth,height=50mm]{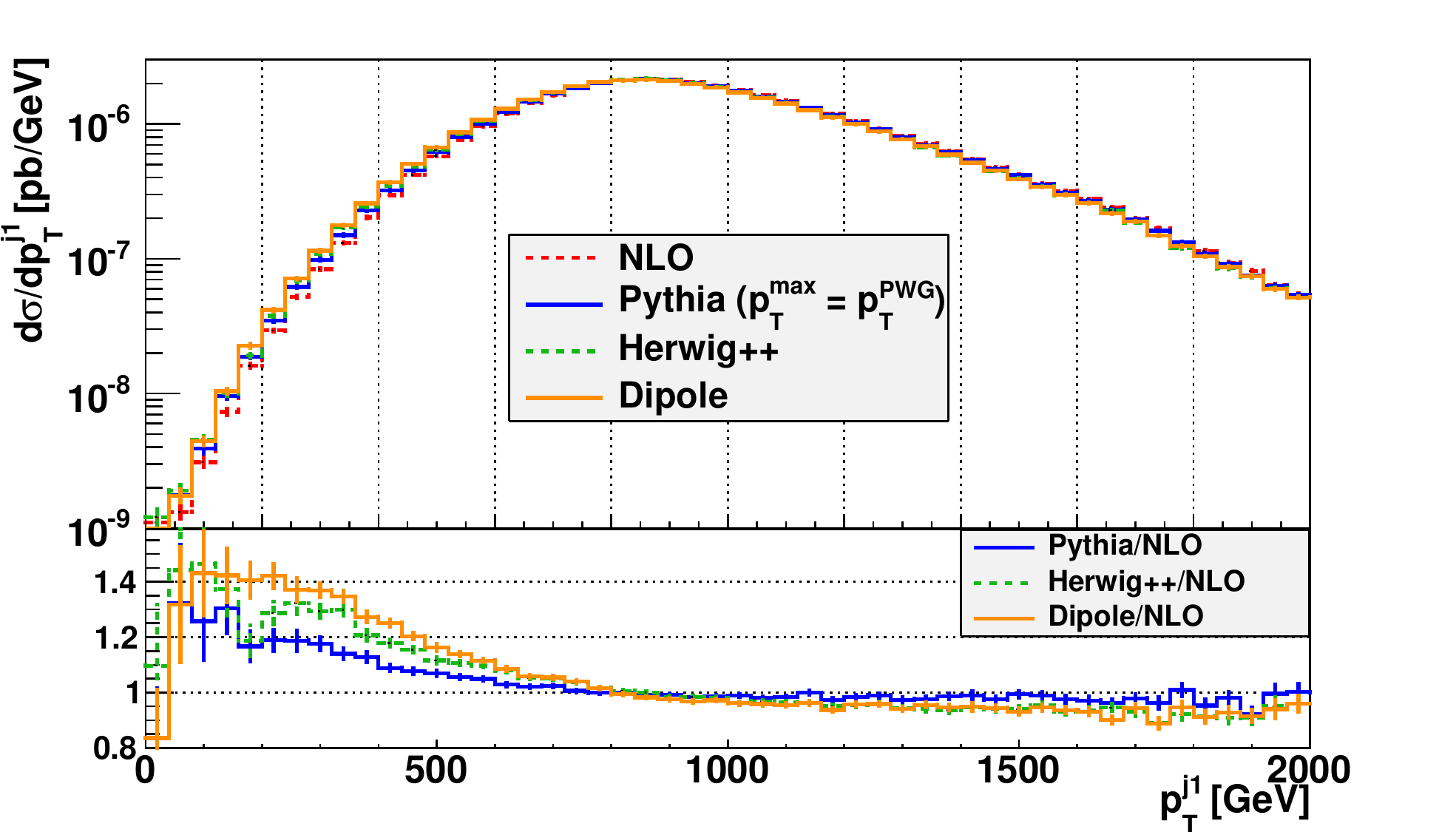}
 \newline
 \includegraphics[width=\textwidth,height=50mm]{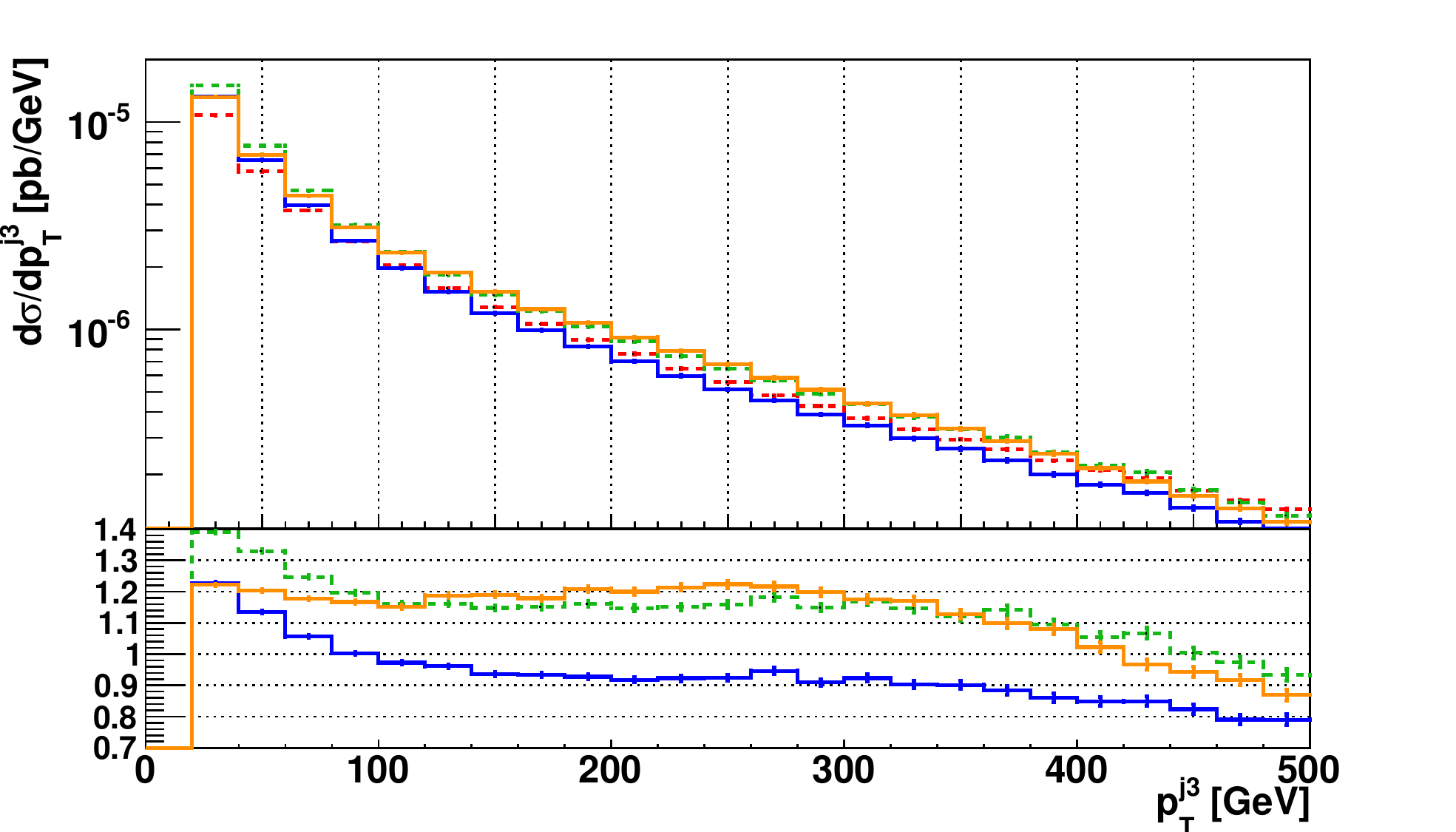}
\end{minipage}
\begin{minipage}{0.5\textwidth}
 \includegraphics[width=\textwidth,height=50mm]{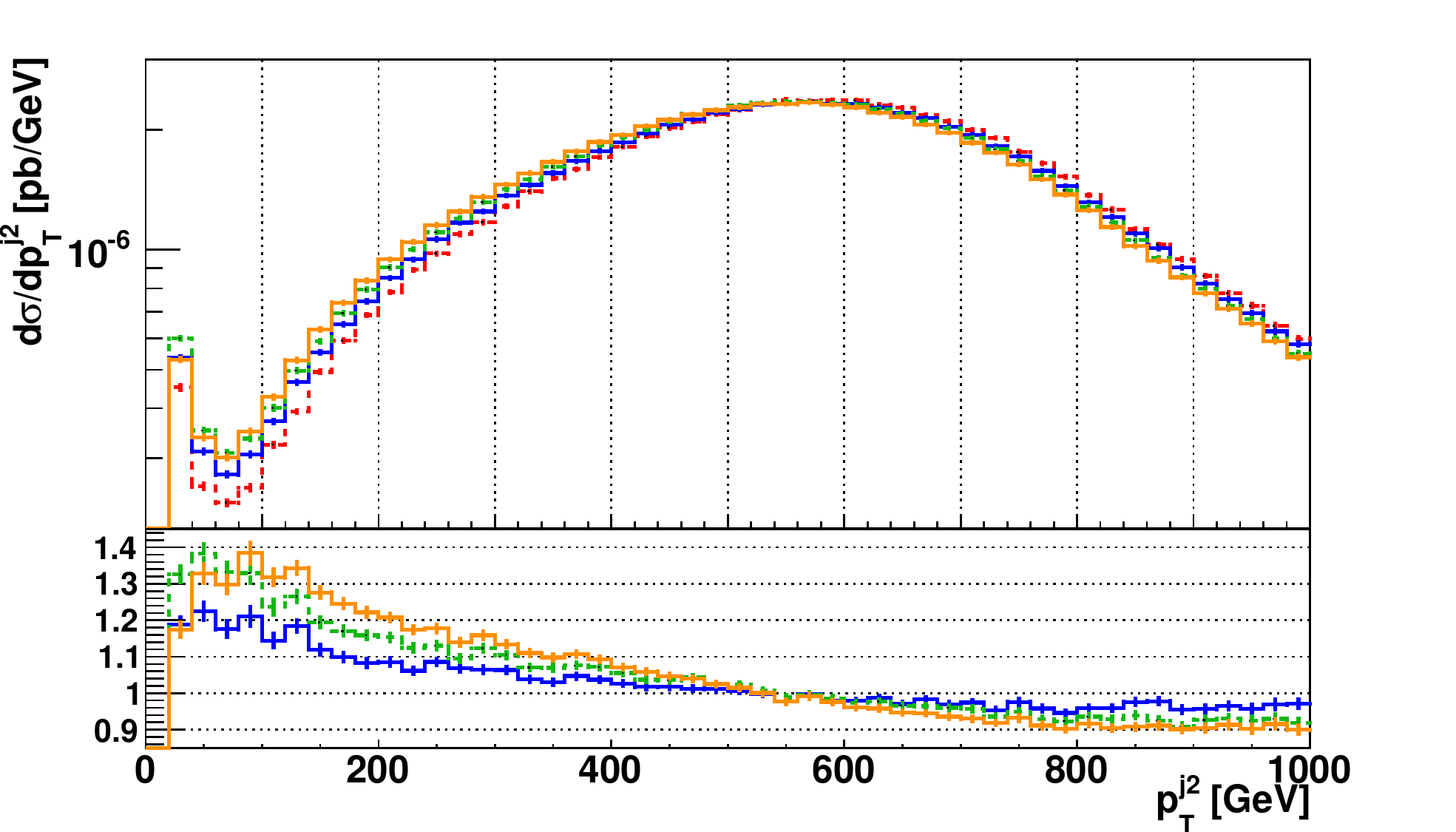}
 \newline
 \includegraphics[width=\textwidth,height=50mm]{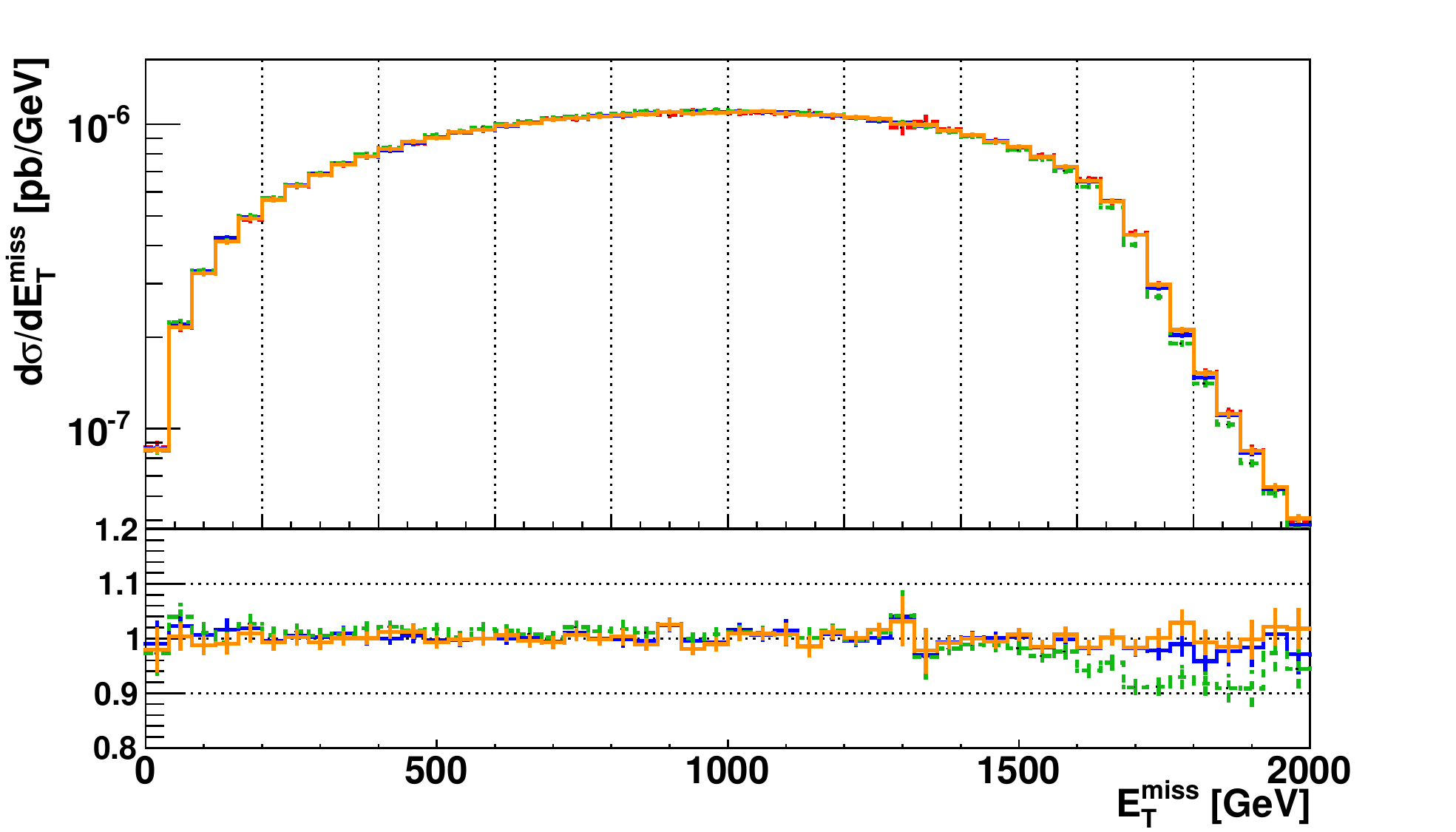}
\end{minipage}
\caption {Same as Fig.~\ref{fig:SH_dec}, but modified \textsc{Pythia} such that the starting scale for the shower is always set to $p_T^{PWG}$, see text.}
\label{fig:SH_dec_pythmod}
\efig

The resulting distributions are shown in Fig.~\ref{fig:SH_dec_pythmod}. Looking again first at the differential cross section for $E_T^{\text{miss}}$ we observe that the \textsc{Pythia} shape now shows no longer any deviation from the NLO result. Comparing the $p_T$-distributions of the three hardest jets it is obvious that the difference between \textsc{Pythia} and the \textsc{Herwig++} showers has shrunk considerably. The two hardest jets are slightly softer than in the NLO result, but match essentially the NLO curve in the hard region (up to deviations of $\matO(10\%)$). Comparing the three shower MCs we note that the most pronounced differences occur (as expected) in the low $p_T$ region, where \textsc{Pythia} predicts slightly lower rates. The Dipole shower and the \textsc{Herwig++} default shower agree rather well with each other in the whole range. The distribution for the third jet develops larger discrepancies: while both \textsc{Herwig++} showers predict higher rates than the NLO calculation up to 
$p_T^{j3}\approx 400\,\text{GeV}$ and agree quite well with each other, the \textsc{Pythia} result ranges slightly below the NLO curve for $p_T^{j3}\gtrsim 100\,\text{GeV}$ and deviates up to $30\%$ from the \textsc{Herwig++} shower results. \\

Considering the rapidity distributions of the second and the third hardest jet depicted in Fig.~\ref{fig:SH_dec_pythmod_y} we observe that all showers essentially reproduce the NLO result for the second jet (this also holds for the hardest jet). The results of the third jet show, however, rather large differences between the showers, again as in the case of undecayed $\sq$ in the central region of the detector. While \textsc{Pythia} ranges only slightly above the NLO prediction, the \textsc{Herwig++} showers (in particular the default shower) predict higher rates around $y^{j3}=0$.\\

These differences can again be attributed to a large extent to differences in the IS shower. Turning off ISR, the Dipole shower and \textsc{Pythia} predict (within $\matO(10\%)$) identical $y^{j3}$ distributions. The \textsc{Herwig++} default shower, however, still deviates by more than $20\%$ from this result. The $p_T^{j3}$ curves for the \textsc{Herwig++} showers are still nearly identical for $p_T^{j3}>100\,\text{GeV}$, while the difference to \textsc{Pythia} is reduced to $<10\%$. However, for soft jets the default shower deviates by up to $+15\%$ from the other two shower MCs. To clarify if these effects are caused solely by the missing truncated shower in \textsc{Herwig++} or if the differences in the shower algorithms (especially the size of the available phase space for radiation) are responsible for the observed discrepancies would require more detailed studies.\\

 \bfig
  \begin{minipage}{0.5\textwidth}
 \includegraphics[width=\textwidth,height=48mm]{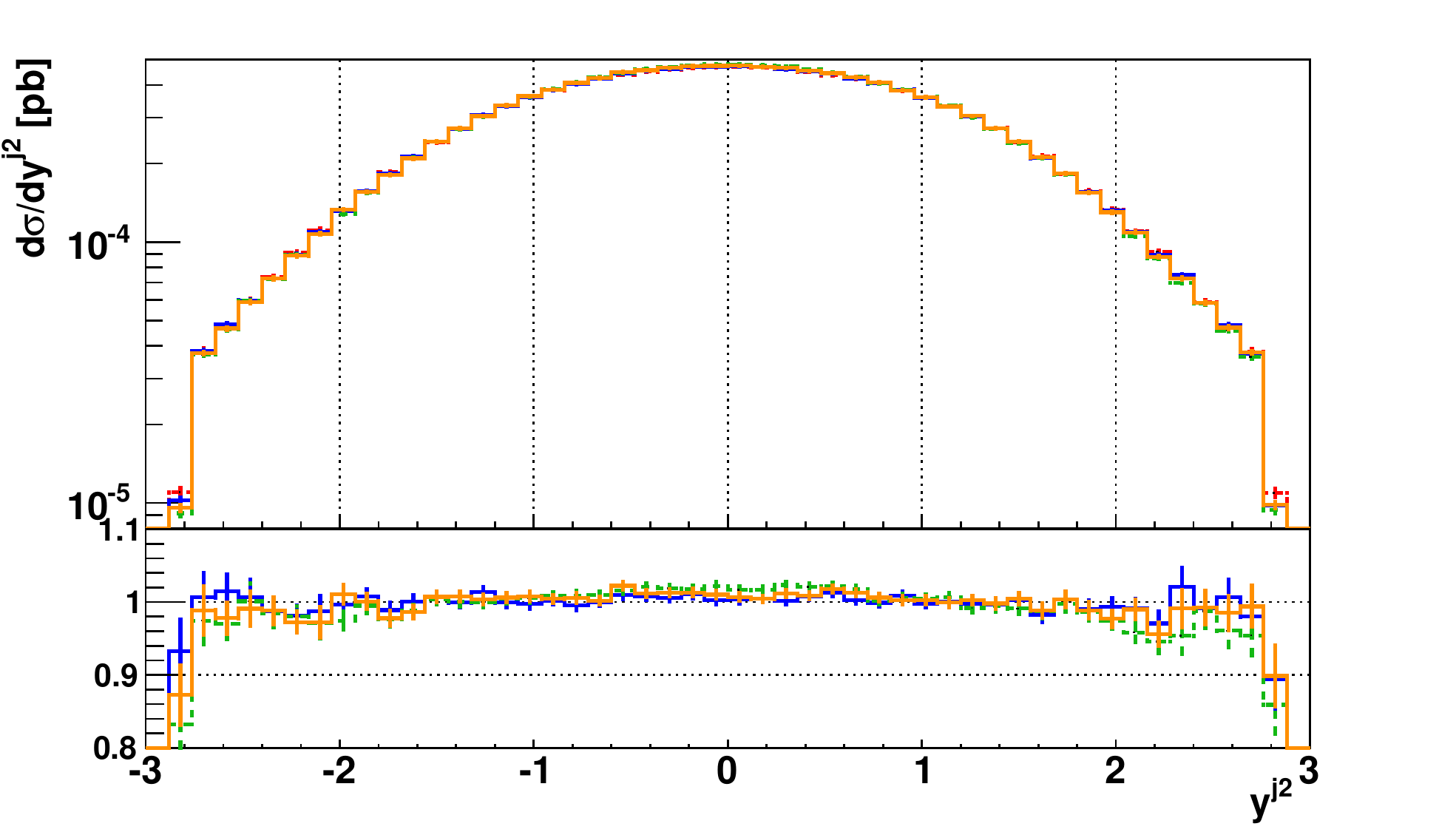}
\end{minipage}
\begin{minipage}{0.5\textwidth}
 \includegraphics[width=\textwidth,height=48mm]{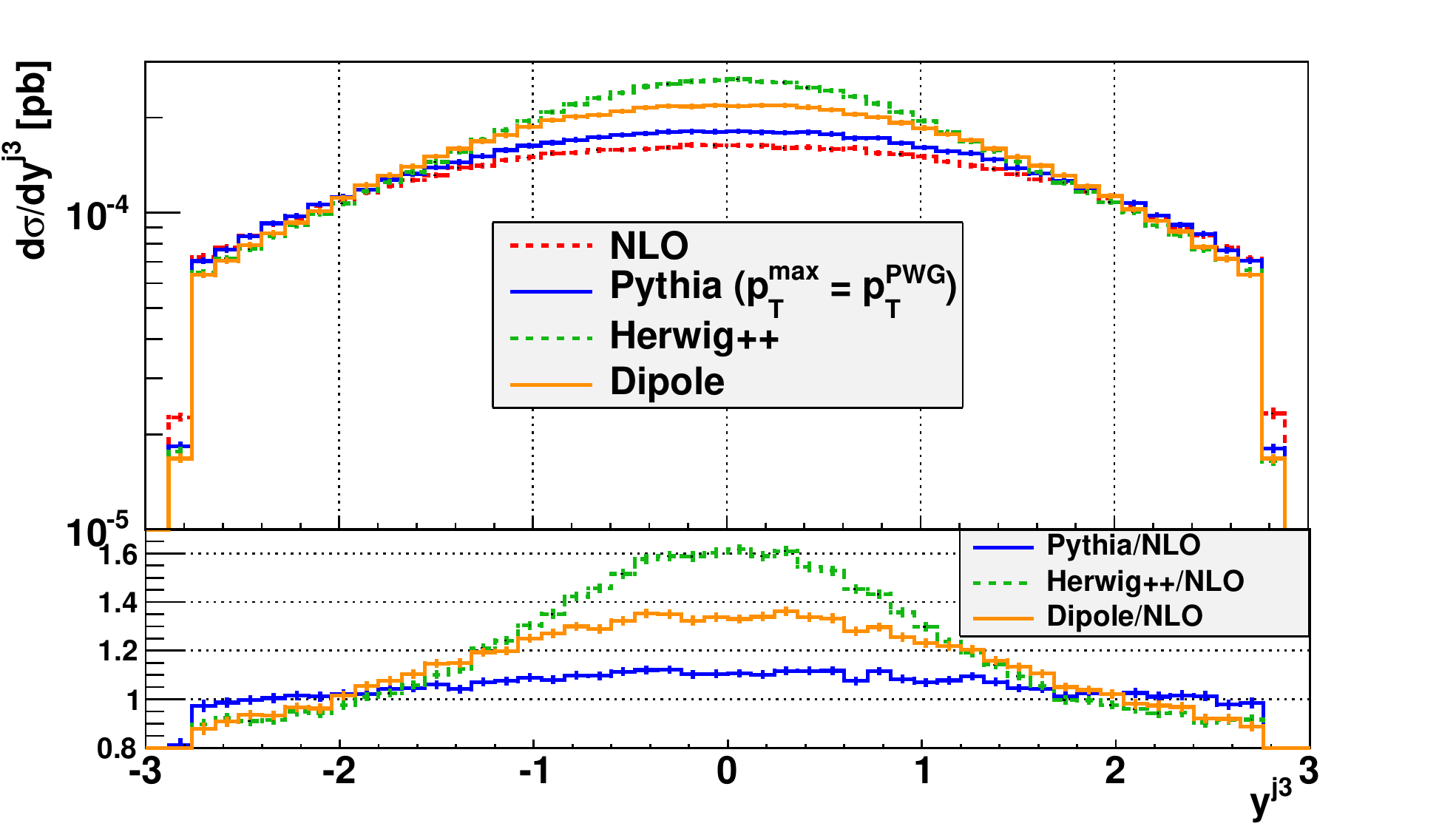}
\end{minipage}
\caption {Rapidity distributions for the second and third hardest jet. The \textsc{Pythia} results were obtained with the modified starting scale.}
\label{fig:SH_dec_pythmod_y}
\efig

A further interesting observable for the comparison of the jet structure of an event with a high multiplicity of partons in the final state is the shape $\rho(r)$ of the jets. We use a definition similar to \cite{atlasrhor} and define for the shape of the $i^{\text{th}}$ jet
\be
  \rho_{ji}(r)=\frac{1}{\Delta r} \frac{p_T^{ji}(r-\Delta r/2, r+\Delta r/2)}{p_T^{ji}(0,R)}, \quad \Delta r/2\leq r\leq R-\Delta r/2\ ,
 \label{rhor}
\ee
with the distance $r=\sqrt{\Delta y^2 + \Delta \phi^2}$ relative to the jet-axis. Here $p_T^{ji}(r_1, r_2)$ is the summed transverse momentum of all partons which are clustered into the jet under consideration and lie in an annulus with inner/outer radius $r_1$/$r_2$ around the jet axis, {\it i.e.}~have a distance $r_1\leq r\leq r_2$ to the jet axis. We used $\Delta r = r_2-r_1= 0.05$ for our analysis.\\

The result for the three hardest jets is presented in Fig.~\ref{fig:SH_dec_rhor} (where we used again our modified \textsc{Pythia} version). Comparing the obtained jet shapes we note that \textsc{Pythia} and the \textsc{Herwig++} default shower essentially predict the same shapes for the two hardest jets, while the Dipole shower is slightly broader. The third jet, in contrast, is much broader in the simulation with \textsc{Pythia} than with the \textsc{Herwig++} showers. 

This observation matches the observations made for the $y^{j3}$ distributions: \textsc{Pythia} seems to generate more soft, wide-angle partons and thus \lq dilutes' the structure of the original \lq parton-jet' for the rather soft third jet, while the hard jets are affected little due to their intrinsically high $p_T$.

\bfig[t]
  \begin{minipage}{0.32\textwidth}
 \includegraphics[width=\textwidth,height=35mm]{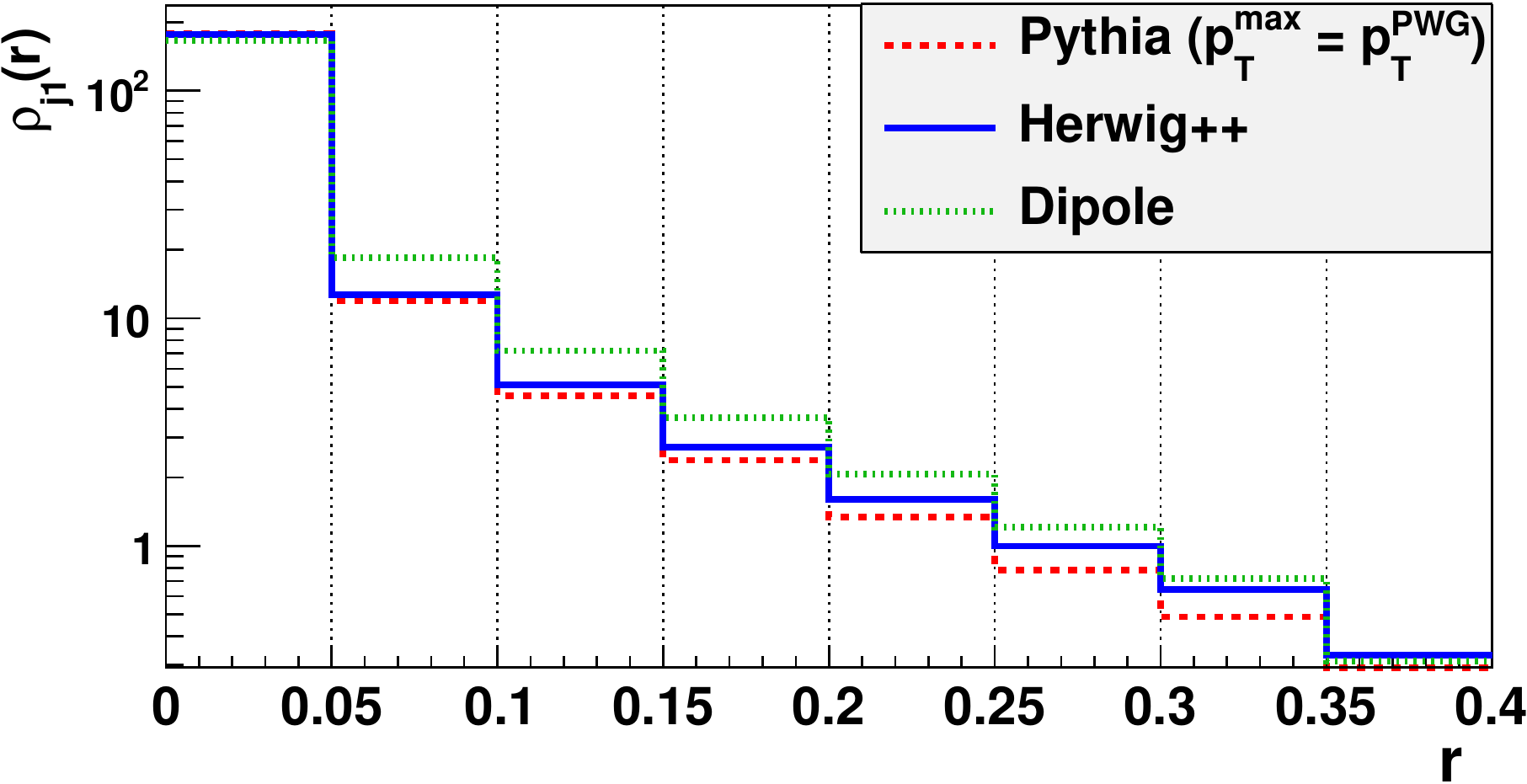}
\end{minipage}
  \begin{minipage}{0.32\textwidth}
 \includegraphics[width=\textwidth,height=35mm]{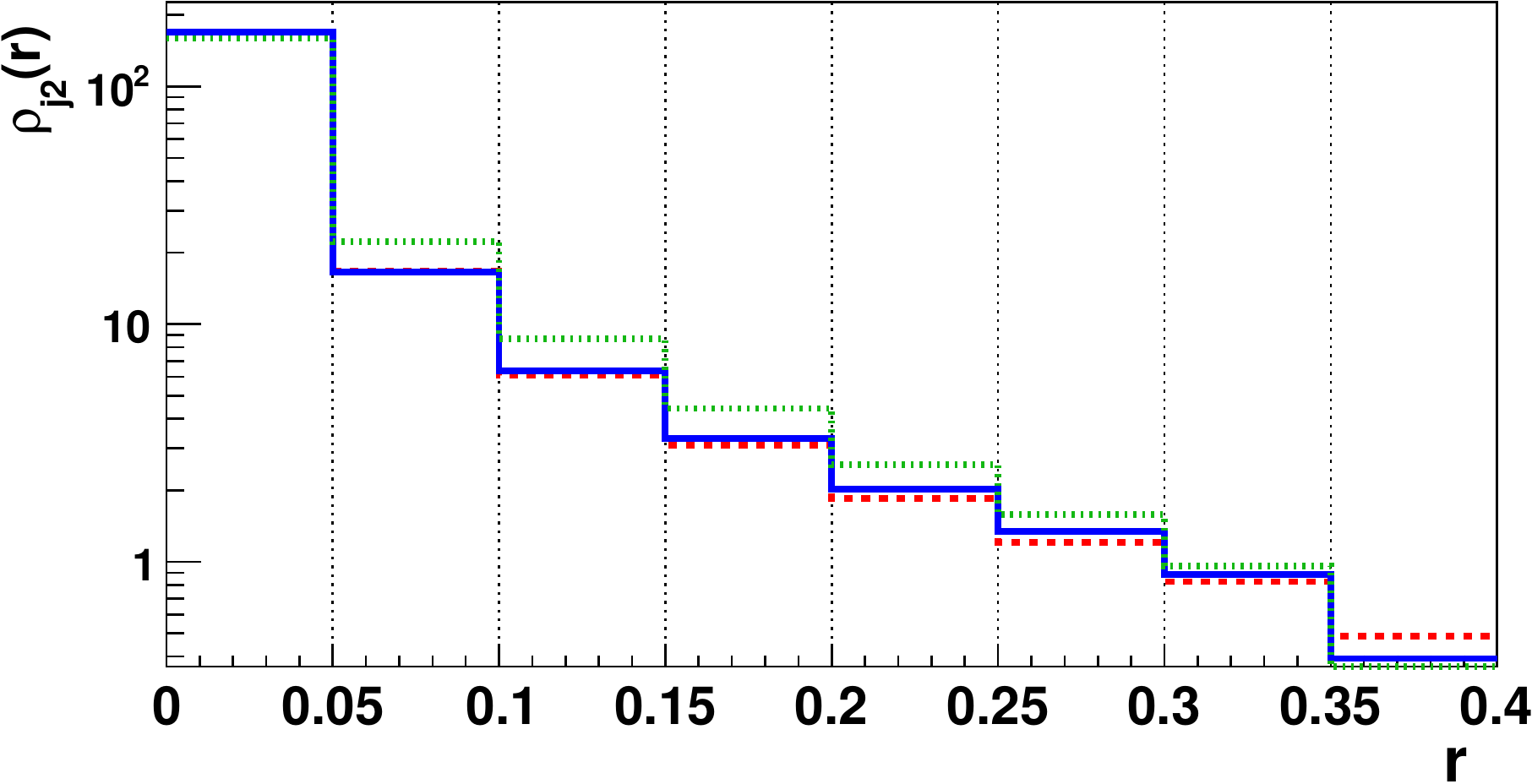}
\end{minipage}
  \begin{minipage}{0.32\textwidth}
 \includegraphics[width=\textwidth,height=35mm]{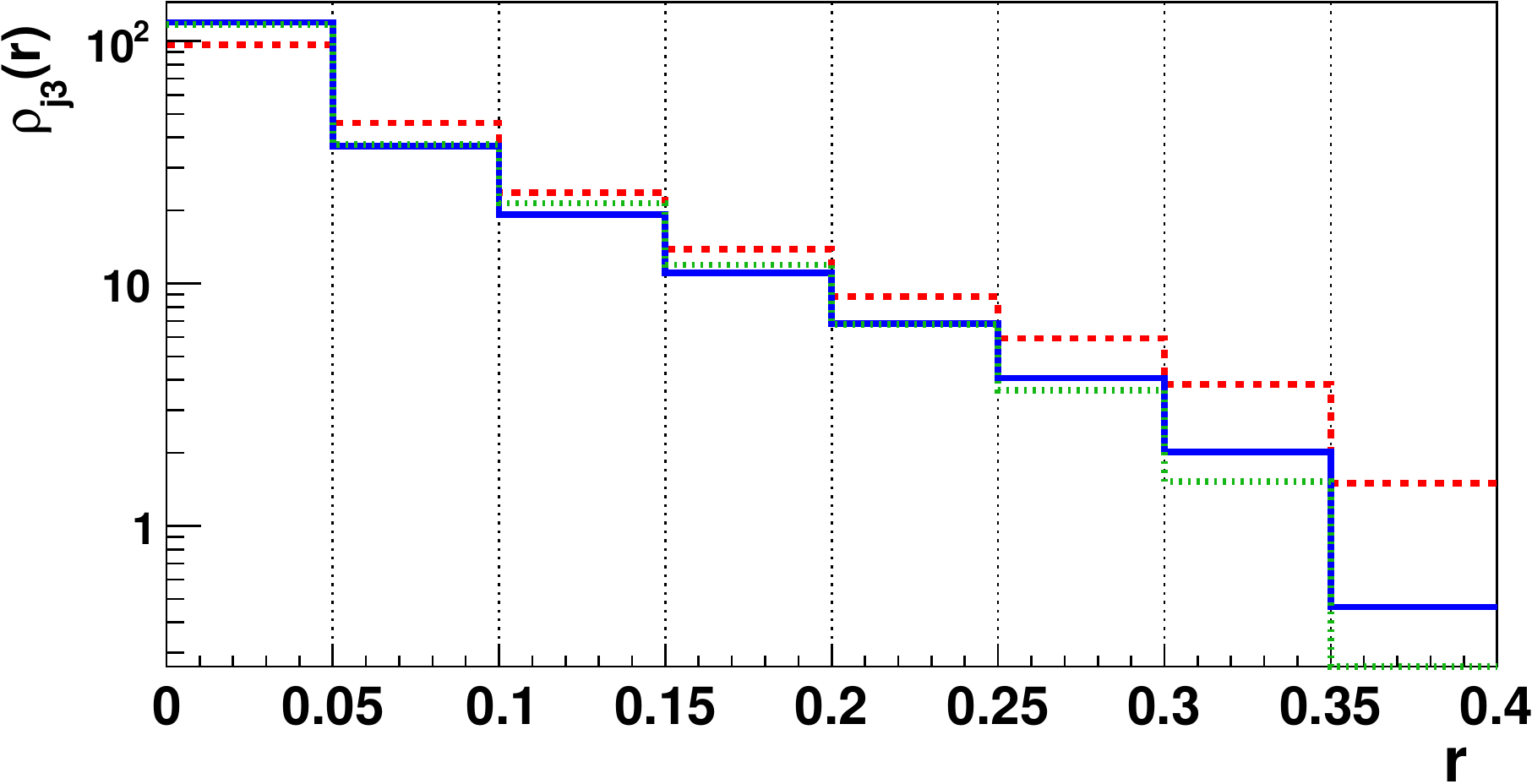}
\end{minipage}
\caption {The jet shapes $\rho(r)$ for the three hardest jets with the modified \textsc{Pythia}. The error bars are not shown.}
\label{fig:SH_dec_rhor}
\efig

\section{Conclusions}
\label{ch:conclusion}
A tremendous effort has been and will be devoted to the search for supersymmetric particles at the LHC. For the interpretation of the experimental data precise theoretical predictions are crucial. 
The work presented in this paper contributes to this effort by providing NLO corrections to the pair production of squarks of the first two generations in a flexible partonic Monte Carlo 
program. In contrast to previous calculations no assumptions regarding the squark masses have been made and the different subchannels have been treated independently. The subtraction of on-shell intermediate $\go$ has been performed with a new approach and compared to several methods proposed in the literature. The differences between these methods turned out to be negligible for total rates and quite small for distributions in general. While there have been published two independent calculations of the NLO corrections to squark pair production recently \cite{hollik,plehn}, these fixed order results have not been matched to parton showers so far. In the second part of this work we present the matching of our NLO calculation using the \textsc{Powheg} method. The \PB\ has been used as a framework, adapted to accommodate strongly interacting SUSY particles and the subtraction of on-shell intermediate gluinos. 

It has been found that the $K$-factors in the individual subchannels can vary by about $20\%$. Thus a proper NLO treatment of individual channels, rather than using an averaged $K$-factor, is mandatory to improve the accuracy of the cross section prediction. As expected, the higher-order corrections substantially reduce the scale dependence, from about $\pm40\%$ at LO to $\pm10\%$ at NLO. While the shape of semi-inclusive distributions like the squark transverse momentum is hardly affected by NLO corrections, more exclusive observables, including {\it e.g.} invariant mass distributions, are more sensitive to higher-order terms. $K$-factors have been found to vary up to $\pm20\%$ depending on the kinematics, both at the level of squark production and at the level of realistic final states from squark decays $\sq\rightarrow \tilde{\chi}^0_1 q$ supplemented by the clustering of partons to form jets. Simply scaling LO distributions with the global $K$-factor obtained from the total cross section will thus not 
provide an 
accurate prediction of exclusive observables, irrespective of whether fixed or dynamical scales are used. 
 
 Comparing the \textsc{Powheg} results to the NLO distributions the expected agreement for inclusive observables has been found. Some discrepancies in the results for observables which are sensitive to additional radiation have been observed and could be attributed to spurious higher-order terms, which have to be suppressed. 
 
Matched NLO plus parton shower results have been obtained for three different showers: the $p_T$-ordered shower of \textsc{Pythia 6}, the default shower and the Dipole shower of \textsc{Herwig++}. As expected, inclusive quantities as the transverse momentum distribution of the squarks are hardly affected by shower radiation. The distributions of the hardest QCD jet are modified by up to $20\%$ compared to NLO, with differences between different showers of $\matO(10\%)$. A consistent comparison of the different showers for final-states including the decays $\sq\rightarrow \tilde{\chi}^0_1 q$ required a modification of the starting scale for the radiation off the decay products in \textsc{Pythia}. Sizeable differences between the different showers were found for example in the distributions of the third-hardest jet. These differences could be traced back to the implementation of initial state radiation. We note that a complete description of the SUSY processes should include NLO plus parton shower 
corrections not only in the production but also in the decay processes. This is left for future work. 

Exploiting current and future LHC data to search for physics beyond the Standard Model requires an accurate theoretical prediction for exclusive observables, including distributions and cross sections with kinematic cuts. The fully differential calculation of the SUSY-QCD corrections to squark pair production matched with parton showers shows that LO predictions scaled with inclusive $K$-factors often fail to properly describe such exclusive observables. The results presented here provide a first step towards a fully differential description of SUSY particle production and decay at the LHC, and should form the theoretical basis for future experimental analyses. 

\section*{Acknowledgments}
We thank Bastian Feigl, Stefan Gieseke, Karol Kova\v{r}\'{i}k and Le Duc Ninh for useful discussions. Furthermore, we thank David L\'opez-Val and Tilman Plehn for the quick replies to our questions on the comparison with the {\tt MadGolem} results of \cite{plehn}.

This research has been supported in part by the German Research Foundation (DFG) via the Sonderforschungsbereich/Transregio SFB/TR-9 ``Computational Particle Physic''.
C.H. has been supported by the \lq Graduiertenkolleg 1694'. The work of R.G. is partially supported by the Swiss National Science  Foundation.


\bibliographystyle{utphys.bst}
\bibliography{paper}

\providecommand{\href}[2]{#2}\begingroup\raggedright\begin{thebibliography}{100}

\bibitem{Volkov}
D.~Volkov and V.~Akulov, {\em {Is the Neutrino a Goldstone Particle?}}
\href{http://dx.doi.org/10.1016/0370-2693(73)90490-5}{Phys.Lett. {\bf B46}
  (1973)  109--110}.

\bibitem{golfand}
Y.~Golfand and E.~Likhtman, {\em {Extension of the Algebra of Poincare Group
  Generators and Violation of p Invariance}}.
JETP Lett. {\bf 13} (1971)  323--326.

\bibitem{Wess}
J.~Wess and B.~Zumino, {\em {Supergauge Transformations in Four-Dimensions}}.
\href{http://dx.doi.org/10.1016/0550-3213(74)90355-1}{Nucl.Phys. {\bf B70}
  (1974)  39--50}.

\bibitem{Sohnius}
M.~Sohnius, {\em {Introducing Supersymmetry}}.
\href{http://dx.doi.org/10.1016/0370-1573(85)90023-7}{Phys.Rept. {\bf 128}
  (1985)  39--204}.

\bibitem{Nilles}
H.~P. Nilles, {\em {Supersymmetry, Supergravity and Particle Physics}}.
\href{http://dx.doi.org/10.1016/0370-1573(84)90008-5}{Phys.Rept. {\bf 110}
  (1984)  1--162}.

\bibitem{HaberKane}
H.~E. Haber and G.~L. Kane, {\em {The Search for Supersymmetry: Probing Physics
  Beyond the Standard Model}}.
\href{http://dx.doi.org/10.1016/0370-1573(85)90051-1}{Phys.Rept. {\bf 117}
  (1985)  75--263}.

\bibitem{Gunion1}
J.~Gunion and H.~E. Haber, {\em {Higgs Bosons in Supersymmetric Models. 1.}}
\href{http://dx.doi.org/10.1016/0550-3213(86)90340-8}{Nucl.Phys. {\bf B272}
  (1986)  1}.

\bibitem{Gunion2}
J.~Gunion and H.~E. Haber, {\em {Higgs Bosons in Supersymmetric Models. 2.
  Implications for Phenomenology}}.
\href{http://dx.doi.org/10.1016/0550-3213(86)90050-7}{Nucl.Phys. {\bf B278}
  (1986)  449}.

\bibitem{Gunion3}
J.~F. Gunion and H.~E. Haber, {\em {Higgs Bosons in Supersymmetric Models. 3.
  Decays into Neutralinos and Charginos}}.
\href{http://dx.doi.org/10.1016/0550-3213(88)90259-3}{Nucl.Phys. {\bf B307}
  (1988)  445}.

\bibitem{squarklo1}
G.~L. Kane and J.~Leveille, {\em {Experimental Constraints on Gluino Masses and
  Supersymmetric Theories}}.
\href{http://dx.doi.org/10.1016/0370-2693(82)90968-6}{Phys.Lett. {\bf B112}
  (1982)  227}.

\bibitem{squarklo2}
P.~Harrison and C.~Llewellyn~Smith, {\em {Hadroproduction of Supersymmetric
  Particles}}.
\href{http://dx.doi.org/10.1016/0550-3213(83)90510-2}{Nucl.Phys. {\bf B213}
  (1983)  223}.

\bibitem{squarklo3}
E.~Reya and D.~Roy, {\em {Supersymmetric Particle Production at p anti-p
  Collider Energies}}.
\href{http://dx.doi.org/10.1103/PhysRevD.32.645}{Phys.Rev. {\bf D32} (1985)
  645}.

\bibitem{squarklo4}
S.~Dawson, E.~Eichten, and C.~Quigg, {\em {Search for Supersymmetric Particles
  in Hadron - Hadron Collisions}}.
\href{http://dx.doi.org/10.1103/PhysRevD.31.1581}{Phys.Rev. {\bf D31} (1985)
  1581}.

\bibitem{squarknlo1}
W.~Beenakker, R.~Hopker, M.~Spira, and P.~Zerwas, {\em {Squark production at
  the Tevatron}}.
  \href{http://dx.doi.org/10.1103/PhysRevLett.74.2905}{Phys.Rev.Lett. {\bf 74}
  (1995)  2905--2908},
\href{http://arxiv.org/abs/hep-ph/9412272}{{\tt arXiv:hep-ph/9412272}}.

\bibitem{squarknlo2}
W.~Beenakker, R.~Hopker, M.~Spira, and P.~Zerwas, {\em {Gluino pair production
  at the Tevatron}}. \href{http://dx.doi.org/10.1007/s002880050016}{Z.Phys.
  {\bf C69} (1995)  163--166},
\href{http://arxiv.org/abs/hep-ph/9505416}{{\tt arXiv:hep-ph/9505416}}.

\bibitem{prospino}
W.~Beenakker, R.~Hopker, M.~Spira, and P.~Zerwas, {\em {Squark and gluino
  production at hadron colliders}}.
  \href{http://dx.doi.org/10.1016/S0550-3213(97)00084-9}{Nucl.Phys. {\bf B492}
  (1997)  51--103},
\href{http://arxiv.org/abs/hep-ph/9610490}{{\tt arXiv:hep-ph/9610490}}.

\bibitem{squarknlo3}
W.~Beenakker, M.~Kramer, T.~Plehn, M.~Spira, and P.~Zerwas, {\em {Stop
  production at hadron colliders}}.
  \href{http://dx.doi.org/10.1016/S0550-3213(98)00014-5}{Nucl.Phys. {\bf B515}
  (1998)  3--14},
\href{http://arxiv.org/abs/hep-ph/9710451}{{\tt arXiv:hep-ph/9710451}}.

\bibitem{hollik}
W.~Hollik, J.~M. Lindert, and D.~Pagani, {\em {NLO corrections to squark-squark
  production and decay at the LHC}}.
\href{http://arxiv.org/abs/hep-ph/1207.1071}{{\tt arXiv:hep-ph/1207.1071}}.

\bibitem{hollik2}
W.~Hollik, J.~M. Lindert, and D.~Pagani, {\em {On cascade decays of squarks at
  the LHC in NLO QCD}}.
\href{http://arxiv.org/abs/hep-ph/1303.0186}{{\tt arXiv:hep-ph/1303.0186}}.

\bibitem{plehn}
D.~Goncalves-Netto, D.~Lopez-Val, K.~Mawatari, T.~Plehn, and I.~Wigmore, {\em
  {Automated Squark and Gluino Production to Next-to-Leading Order}}.
\href{http://arxiv.org/abs/hep-ph/1211.0286}{{\tt arXiv:hep-ph/1211.0286}}.

\bibitem{sqbnlo1}
U.~Langenfeld and S.-O. Moch, {\em {Higher-order soft corrections to squark
  hadro-production}}.
  \href{http://dx.doi.org/10.1016/j.physletb.2009.04.002}{Phys.Lett. {\bf B675}
  (2009)  210--221},
\href{http://arxiv.org/abs/hep-ph/0901.0802}{{\tt arXiv:hep-ph/0901.0802}}.

\bibitem{sqbnlo2}
A.~Kulesza and L.~Motyka, {\em {Threshold resummation for squark-antisquark and
  gluino-pair production at the LHC}}.
  \href{http://dx.doi.org/10.1103/PhysRevLett.102.111802}{Phys.Rev.Lett. {\bf
  102} (2009)  111802},
\href{http://arxiv.org/abs/hep-ph/0807.2405}{{\tt arXiv:hep-ph/0807.2405}}.

\bibitem{sqbnlo3}
A.~Kulesza and L.~Motyka, {\em {Soft gluon resummation for the production of
  gluino-gluino and squark-antisquark pairs at the LHC}}.
  \href{http://dx.doi.org/10.1103/PhysRevD.80.095004}{Phys.Rev. {\bf D80}
  (2009)  095004},
\href{http://arxiv.org/abs/hep-ph/0905.4749}{{\tt arXiv:hep-ph/0905.4749}}.

\bibitem{sqbnlo4}
W.~Beenakker, S.~Brensing, M.~Kramer, A.~Kulesza, E.~Laenen, and I.~Niessen,
  {\em {Soft-gluon resummation for squark and gluino hadroproduction}}.
  \href{http://dx.doi.org/10.1088/1126-6708/2009/12/041}{JHEP {\bf 0912} (2009)
   041},
\href{http://arxiv.org/abs/hep-ph/0909.4418}{{\tt arXiv:hep-ph/0909.4418}}.

\bibitem{sqbnlo5}
W.~Beenakker, S.~Brensing, M.~Kramer, A.~Kulesza, E.~Laenen, and I.~Niessen,
  {\em {Supersymmetric top and bottom squark production at hadron colliders}}.
  \href{http://dx.doi.org/10.1007/JHEP08(2010)098}{JHEP {\bf 1008} (2010)
  098},
\href{http://arxiv.org/abs/hep-ph/1006.4771}{{\tt arXiv:hep-ph/1006.4771}}.

\bibitem{sqbnlo6}
W.~Beenakker, S.~Brensing, M.~Kramer, A.~Kulesza, E.~Laenen, and I.~Niessen,
  {\em {NNLL resummation for squark-antisquark pair production at the LHC}}.
  \href{http://dx.doi.org/10.1007/JHEP01(2012)076}{JHEP {\bf 1201} (2012)
  076},
\href{http://arxiv.org/abs/hep-ph/1110.2446}{{\tt arXiv:hep-ph/1110.2446}}.

\bibitem{sqbnlo7}
M.~Beneke, P.~Falgari, and C.~Schwinn, {\em {Threshold resummation for pair
  production of coloured heavy (s)particles at hadron colliders}}.
  \href{http://dx.doi.org/10.1016/j.nuclphysb.2010.09.009}{Nucl.Phys. {\bf
  B842} (2011)  414--474},
\href{http://arxiv.org/abs/hep-ph/1007.5414}{{\tt arXiv:hep-ph/1007.5414}}.

\bibitem{sqbnlo9}
W.~Beenakker, T.~Janssen, S.~Lepoeter, M.~Kramer, A.~Kulesza, E.~Laenen,
  I.~Niessen, S.~Thewes, and T.~Van~Daal, {\em {Towards NNLL resummation: hard
  matching coefficients for squark and gluino hadroproduction}}.
\href{http://arxiv.org/abs/hep-ph/1304.6354}{{\tt arXiv:hep-ph/1304.6354}}.

\bibitem{sqbnlo10}
A.~Broggio, A.~Ferroglia, M.~Neubert, L.~Vernazza, and L.~L. Yang, {\em
  {Approximate NNLO Predictions for the Stop-Pair Production Cross Section at
  the LHC}}.
\href{http://arxiv.org/abs/1304.2411}{{\tt arXiv:1304.2411 [hep-ph]}}.

\bibitem{sqthresh1}
M.~R. Kauth, A.~Kress, and J.~H. Kuhn, {\em {Gluino-Squark Production at the
  LHC: The Threshold}}. \href{http://dx.doi.org/10.1007/JHEP12(2011)104}{JHEP
  {\bf 1112} (2011)  104},
\href{http://arxiv.org/abs/hep-ph/1108.0542}{{\tt arXiv:hep-ph/1108.0542}}.

\bibitem{sqthresh2}
M.~R. Kauth, J.~H. Kuhn, P.~Marquard, and M.~Steinhauser, {\em {Gluino Pair
  Production at the LHC: The Threshold}}.
  \href{http://dx.doi.org/10.1016/j.nuclphysb.2011.11.024}{Nucl.Phys. {\bf
  B857} (2012)  28--64},
\href{http://arxiv.org/abs/hep-ph/1108.0361}{{\tt arXiv:hep-ph/1108.0361}}.

\bibitem{sqthresh3}
K.~Hagiwara and H.~Yokoya, {\em {Bound-state effects on gluino-pair production
  at hadron colliders}}.
  \href{http://dx.doi.org/10.1088/1126-6708/2009/10/049}{JHEP {\bf 0910} (2009)
   049},
\href{http://arxiv.org/abs/hep-ph/0909.3204}{{\tt arXiv:hep-ph/0909.3204}}.

\bibitem{ewlo1}
S.~Bornhauser, M.~Drees, H.~K. Dreiner, and J.~S. Kim, {\em {Electroweak
  contributions to squark pair production at the LHC}}.
  \href{http://dx.doi.org/10.1103/PhysRevD.76.095020}{Phys.Rev. {\bf D76}
  (2007)  095020},
\href{http://arxiv.org/abs/hep-ph/0709.2544}{{\tt arXiv:hep-ph/0709.2544}}.

\bibitem{ewlo2}
A.~Arhrib, R.~Benbrik, K.~Cheung, and T.-C. Yuan, {\em {Higgs boson enhancement
  effects on squark-pair production at the LHC}}.
  \href{http://dx.doi.org/10.1007/JHEP02(2010)048}{JHEP {\bf 1002} (2010)
  048},
\href{http://arxiv.org/abs/hep-ph/0911.1820}{{\tt arXiv:hep-ph/0911.1820}}.

\bibitem{ewnlo1}
W.~Hollik, M.~Kollar, and M.~K. Trenkel, {\em {Hadronic production of
  top-squark pairs with electroweak NLO contributions}}.
  \href{http://dx.doi.org/10.1088/1126-6708/2008/02/018}{JHEP {\bf 0802} (2008)
   018},
\href{http://arxiv.org/abs/hep-ph/0712.0287}{{\tt arXiv:hep-ph/0712.0287}}.

\bibitem{ewnlo2}
M.~Beccaria, G.~Macorini, L.~Panizzi, F.~Renard, and C.~Verzegnassi, {\em
  {Stop-antistop and sbottom-antisbottom production at LHC: A One-loop search
  for model parameters dependence}}.
  \href{http://dx.doi.org/10.1142/S0217751X08041694}{Int.J.Mod.Phys. {\bf A23}
  (2008)  4779--4810},
\href{http://arxiv.org/abs/hep-ph/0804.1252}{{\tt arXiv:hep-ph/0804.1252}}.

\bibitem{ewnlo3}
W.~Hollik and E.~Mirabella, {\em {Squark anti-squark pair production at the
  LHC: The Electroweak contribution}}.
  \href{http://dx.doi.org/10.1088/1126-6708/2008/12/087}{JHEP {\bf 0812} (2008)
   087},
\href{http://arxiv.org/abs/hep-ph/0806.1433}{{\tt arXiv:hep-ph/0806.1433}}.

\bibitem{ewnlo4}
W.~Hollik, E.~Mirabella, and M.~K. Trenkel, {\em {Electroweak contributions to
  squark-gluino production at the LHC}}.
  \href{http://dx.doi.org/10.1088/1126-6708/2009/02/002}{JHEP {\bf 0902} (2009)
   002},
\href{http://arxiv.org/abs/hep-ph/0810.1044}{{\tt arXiv:hep-ph/0810.1044}}.

\bibitem{ewnlo5}
E.~Mirabella, {\em {NLO electroweak contributions to gluino pair production at
  hadron colliders}}.
  \href{http://dx.doi.org/10.1088/1126-6708/2009/12/012}{JHEP {\bf 0912} (2009)
   012},
\href{http://arxiv.org/abs/hep-ph/0908.3318}{{\tt arXiv:hep-ph/0908.3318}}.

\bibitem{ewnlo6}
J.~Germer, W.~Hollik, E.~Mirabella, and M.~K. Trenkel, {\em {Hadronic
  production of squark-squark pairs: The electroweak contributions}}.
  \href{http://dx.doi.org/10.1007/JHEP08(2010)023}{JHEP {\bf 1008} (2010)
  023},
\href{http://arxiv.org/abs/hep-ph/1004.2621}{{\tt arXiv:hep-ph/1004.2621}}.

\bibitem{ewnlo7}
J.~Germer, W.~Hollik, and E.~Mirabella, {\em {Hadronic production of
  bottom-squark pairs with electroweak contributions}}.
  \href{http://dx.doi.org/10.1007/JHEP05(2011)068}{JHEP {\bf 1105} (2011)
  068},
\href{http://arxiv.org/abs/hep-ph/1103.1258}{{\tt arXiv:hep-ph/1103.1258}}.

\bibitem{prospino_manual}
W.~Beenakker, R.~Hopker, and M.~Spira, {\em {PROSPINO: A Program for the
  production of supersymmetric particles in next-to-leading order QCD}}.
\href{http://arxiv.org/abs/hep-ph/9611232}{{\tt arXiv:hep-ph/9611232}}.

\bibitem{mythesis}
E.~Popenda, {\em {Higher Order Corrections to Supersymmetric Production and
  Decay Processes at the LHC}}. PhD Thesis (2012)  Karlsruhe Institute of
  Technology.

\bibitem{sqbnlo8}
P.~Falgari, C.~Schwinn, and C.~Wever, {\em {NLL soft and Coulomb resummation
  for squark and gluino production at the LHC}}.
  \href{http://dx.doi.org/10.1007/JHEP06(2012)052}{JHEP {\bf 1206} (2012)
  052},
\href{http://arxiv.org/abs/hep-ph/1202.2260}{{\tt arXiv:hep-ph/1202.2260}}.

\bibitem{sqbnlo11}
W.~Beenakker, S.~Brensing, M.~Kramer, A.~Kulesza, E.~Laenen, {\em et al.}, {\em
  {Squark and Gluino Hadroproduction}}.
  \href{http://dx.doi.org/10.1142/S0217751X11053560}{Int.J.Mod.Phys. {\bf A26}
  (2011)  2637--2664},
\href{http://arxiv.org/abs/1105.1110}{{\tt arXiv:1105.1110 [hep-ph]}}.

\bibitem{skandsQCDforcolliders}
P.~Z. Skands, {\em {QCD for Collider Physics}}.
\href{http://arxiv.org/abs/hep-ph/1104.2863}{{\tt arXiv:hep-ph/1104.2863}}.

\bibitem{mcatnlo}
S.~Frixione and B.~R. Webber, {\em {Matching NLO QCD computations and parton
  shower simulations}}. JHEP {\bf 0206} (2002)  029,
\href{http://arxiv.org/abs/hep-ph/0204244}{{\tt arXiv:hep-ph/0204244}}.

\bibitem{nason}
P.~Nason, {\em {A New method for combining NLO QCD with shower Monte Carlo
  algorithms}}. \href{http://dx.doi.org/10.1088/1126-6708/2004/11/040}{JHEP
  {\bf 0411} (2004)  040},
\href{http://arxiv.org/abs/hep-ph/0409146}{{\tt arXiv:hep-ph/0409146}}.

\bibitem{powheg}
S.~Frixione, P.~Nason, and C.~Oleari, {\em {Matching NLO QCD computations with
  Parton Shower simulations: the POWHEG method}}.
  \href{http://dx.doi.org/10.1088/1126-6708/2007/11/070}{JHEP {\bf 0711} (2007)
   070},
\href{http://arxiv.org/abs/hep-ph/0709.2092}{{\tt arXiv:hep-ph/0709.2092}}.

\bibitem{powhegbox}
S.~Alioli, P.~Nason, C.~Oleari, and E.~Re, {\em {A general framework for
  implementing NLO calculations in shower Monte Carlo programs: the POWHEG
  BOX}}. \href{http://dx.doi.org/10.1007/JHEP06(2010)043}{JHEP {\bf 1006}
  (2010)  043},
\href{http://arxiv.org/abs/hep-ph/1002.2581}{{\tt arXiv:hep-ph/1002.2581}}.

\bibitem{pythia6}
T.~Sjostrand, S.~Mrenna, and P.~Z. Skands, {\em {PYTHIA 6.4 Physics and
  Manual}}. \href{http://dx.doi.org/10.1088/1126-6708/2006/05/026}{JHEP {\bf
  0605} (2006)  026},
\href{http://arxiv.org/abs/hep-ph/0603175}{{\tt arXiv:hep-ph/0603175}}.

\bibitem{herwigpp}
M.~Bahr, S.~Gieseke, M.~Gigg, D.~Grellscheid, K.~Hamilton, {\em et al.}, {\em
  {Herwig++ Physics and Manual}}.
  \href{http://dx.doi.org/10.1140/epjc/s10052-008-0798-9}{Eur.Phys.J. {\bf C58}
  (2008)  639--707},
\href{http://arxiv.org/abs/hep-ph/0803.0883}{{\tt arXiv:hep-ph/0803.0883}}.

\bibitem{herwigpp26}
K.~Arnold, L.~d'Errico, S.~Gieseke, D.~Grellscheid, K.~Hamilton, {\em et al.},
  {\em {Herwig++ 2.6 Release Note}}.
\href{http://arxiv.org/abs/1205.4902}{{\tt hep-ph/arXiv:1205.4902}}.

\bibitem{herwigdp1}
S.~Platzer and S.~Gieseke, {\em {Coherent Parton Showers with Local Recoils}}.
  \href{http://dx.doi.org/10.1007/JHEP01(2011)024}{JHEP {\bf 1101} (2011)
  024},
\href{http://arxiv.org/abs/hep-ph/0909.5593}{{\tt arXiv:hep-ph/0909.5593}}.

\bibitem{herwigdp2}
S.~Platzer and S.~Gieseke, {\em {Dipole Showers and Automated NLO Matching in
  Herwig++}}.
  \href{http://dx.doi.org/10.1140/epjc/s10052-012-2187-7}{Eur.Phys.J. {\bf C72}
  (2012)  2187},
\href{http://arxiv.org/abs/hep-ph/1109.6256}{{\tt arXiv:hep-ph/1109.6256}}.

\bibitem{dimreg}
G.~'t~Hooft and M.~Veltman, {\em {Regularization and Renormalization of Gauge
  Fields}}.
\href{http://dx.doi.org/10.1016/0550-3213(72)90279-9}{Nucl.Phys. {\bf B44}
  (1972)  189--213}.

\bibitem{martinvaughn}
S.~P. Martin and M.~T. Vaughn, {\em {Regularization dependence of running
  couplings in softly broken supersymmetry}}.
  \href{http://dx.doi.org/10.1016/0370-2693(93)90136-6}{Phys.Lett. {\bf B318}
  (1993)  331--337},
\href{http://arxiv.org/abs/hep-ph/9308222}{{\tt arXiv:hep-ph/9308222}}.

\bibitem{msbar}
W.~A. Bardeen, A.~Buras, D.~Duke, and T.~Muta, {\em {Deep Inelastic Scattering
  Beyond the Leading Order in Asymptotically Free Gauge Theories}}.
\href{http://dx.doi.org/10.1103/PhysRevD.18.3998}{Phys.Rev. {\bf D18} (1978)
  3998}.

\bibitem{pdg}
{\bf Particle Data Group} Collaboration, J.~Beringer {\em et al.}, {\em {Review
  of Particle Physics (RPP)}}.
\href{http://dx.doi.org/10.1103/PhysRevD.86.010001}{Phys.Rev. {\bf D86} (2012)
  010001}.

\bibitem{Collins}
J.~C. Collins, F.~Wilczek, and A.~Zee, {\em {Low-Energy Manifestations of Heavy
  Particles: Application to the Neutral Current}}.
\href{http://dx.doi.org/10.1103/PhysRevD.18.242}{Phys.Rev. {\bf D18} (1978)
  242}.

\bibitem{feynarts}
T.~Hahn, {\em {Generating Feynman diagrams and amplitudes with FeynArts 3}}.
  \href{http://dx.doi.org/10.1016/S0010-4655(01)00290-9}{Comput.Phys.Commun.
  {\bf 140} (2001)  418--431},
\href{http://arxiv.org/abs/hep-ph/0012260}{{\tt arXiv:hep-ph/0012260}}.

\bibitem{feynartsmssm}
T.~Hahn and C.~Schappacher, {\em {The Implementation of the minimal
  supersymmetric standard model in FeynArts and FormCalc}}.
  \href{http://dx.doi.org/10.1016/S0010-4655(01)00436-2}{Comput.Phys.Commun.
  {\bf 143} (2002)  54--68},
\href{http://arxiv.org/abs/hep-ph/0105349}{{\tt arXiv:hep-ph/0105349}}.

\bibitem{formcalc1}
T.~Hahn and M.~Perez-Victoria, {\em {Automatized one loop calculations in
  four-dimensions and D-dimensions}}.
  \href{http://dx.doi.org/10.1016/S0010-4655(98)00173-8}{Comput.Phys.Commun.
  {\bf 118} (1999)  153--165},
\href{http://arxiv.org/abs/hep-ph/9807565}{{\tt arXiv:hep-ph/9807565}}.

\bibitem{formcalc2}
T.~Hahn, {\em {A Mathematica interface for FormCalc-generated code}}.
  \href{http://dx.doi.org/10.1016/j.cpc.2007.09.004}{Comput.Phys.Commun. {\bf
  178} (2008)  217--221},
\href{http://arxiv.org/abs/hep-ph/0611273}{{\tt arXiv:hep-ph/0611273}}.

\bibitem{cs}
S.~Catani and M.~Seymour, {\em {A General algorithm for calculating jet
  cross-sections in NLO QCD}}.
  \href{http://dx.doi.org/10.1016/S0550-3213(96)00589-5}{Nucl.Phys. {\bf B485}
  (1997)  291--419},
\href{http://arxiv.org/abs/hep-ph/9605323}{{\tt arXiv:hep-ph/9605323}}.

\bibitem{cdst}
S.~Catani, S.~Dittmaier, M.~H. Seymour, and Z.~Trocsanyi, {\em {The Dipole
  formalism for next-to-leading order QCD calculations with massive partons}}.
  \href{http://dx.doi.org/10.1016/S0550-3213(02)00098-6}{Nucl.Phys. {\bf B627}
  (2002)  189--265},
\href{http://arxiv.org/abs/hep-ph/0201036}{{\tt arXiv:hep-ph/0201036}}.

\bibitem{autodipole}
K.~Hasegawa, S.~Moch, and P.~Uwer, {\em {AutoDipole: Automated generation of
  dipole subtraction terms}}.
  \href{http://dx.doi.org/10.1016/j.cpc.2010.06.044}{Comput.Phys.Commun. {\bf
  181} (2010)  1802--1817},
\href{http://arxiv.org/abs/hep-ph/0911.4371}{{\tt arXiv:hep-ph/0911.4371}}.

\bibitem{superautodipole}
K.~Hasegawa, {\em {Super AutoDipole}}.
  \href{http://dx.doi.org/10.1140/epjc/s10052-010-1452-x}{Eur.Phys.J. {\bf C70}
  (2010)  285--293},
\href{http://arxiv.org/abs/hep-ph/1007.1585}{{\tt arXiv:hep-ph/1007.1585}}.

\bibitem{madgraph1}
J.~Alwall, P.~Demin, S.~de~Visscher, R.~Frederix, M.~Herquet, {\em et al.},
  {\em {MadGraph/MadEvent v4: The New Web Generation}}.
  \href{http://dx.doi.org/10.1088/1126-6708/2007/09/028}{JHEP {\bf 0709} (2007)
   028},
\href{http://arxiv.org/abs/hep-ph/0706.2334}{{\tt arXiv:hep-ph/0706.2334}}.

\bibitem{madgraph2}
T.~Stelzer and W.~Long, {\em {Automatic generation of tree level helicity
  amplitudes}}.
  \href{http://dx.doi.org/10.1016/0010-4655(94)90084-1}{Comput.Phys.Commun.
  {\bf 81} (1994)  357--371},
\href{http://arxiv.org/abs/hep-ph/9401258}{{\tt arXiv:hep-ph/9401258}}.

\bibitem{helas}
H.~Murayama, I.~Watanabe, and K.~Hagiwara, {\em HELAS: HELicity Amplitude
  Subroutines for Feynman Diagram Evaluations} Tech. Rep. KEK-91-11, KEK, 1992.

\bibitem{dieter&co}
T.~Figy, C.~Oleari, and D.~Zeppenfeld, {\em {Next-to-leading order jet
  distributions for Higgs boson production via weak boson fusion}}.
  \href{http://dx.doi.org/10.1103/PhysRevD.68.073005}{Phys.Rev. {\bf D68}
  (2003)  073005},
\href{http://arxiv.org/abs/hep-ph/0306109}{{\tt arXiv:hep-ph/0306109}}.

\bibitem{twmcatnlo}
S.~Frixione, E.~Laenen, P.~Motylinski, B.~R. Webber, and C.~D. White, {\em
  {Single-top hadroproduction in association with a W boson}}.
  \href{http://dx.doi.org/10.1088/1126-6708/2008/07/029}{JHEP {\bf 0807} (2008)
   029},
\href{http://arxiv.org/abs/hep-ph/0805.3067}{{\tt arXiv:hep-ph/0805.3067}}.

\bibitem{feyncalc}
R.~Mertig, M.~Bohm, and A.~Denner, {\em {FEYN CALC: Computer algebraic
  calculation of Feynman amplitudes}}.
\href{http://dx.doi.org/10.1016/0010-4655(91)90130-D}{Comput.Phys.Commun. {\bf
  64} (1991)  345--359}.

\bibitem{byckling}
E.~Byckling and K.~Kajantie, {\em Particle kinematics}.
\newblock A Wiley-Interscience Publication. Wiley, London [u.a.], 1973.

\bibitem{vegas}
G.~P. Lepage, {\em A new algorithm for adaptive multidimensional integration}.
  \href{http://dx.doi.org/DOI:10.1016/0021-9991(78)90004-9}{Journal of
  Computational Physics {\bf 27} (1978) no.~2, 192 -- 203}.

\bibitem{vbfnlo1}
K.~Arnold, J.~Bellm, G.~Bozzi, F.~Campanario, C.~Englert, {\em et al.}, {\em
  {Release Note -- Vbfnlo-2.6.0}}.
\href{http://arxiv.org/abs/hep-ph/1207.4975}{{\tt arXiv:hep-ph/1207.4975}}.

\bibitem{vbfnlo2}
K.~Arnold, J.~Bellm, G.~Bozzi, M.~Brieg, F.~Campanario, {\em et al.}, {\em
  {VBFNLO: A Parton Level Monte Carlo for Processes with Electroweak Bosons --
  Manual for Version 2.5.0}}.
\href{http://arxiv.org/abs/hep-ph/1107.4038}{{\tt arXiv:hep-ph/1107.4038}}.

\bibitem{vbfnlo3}
K.~Arnold, M.~Bahr, G.~Bozzi, F.~Campanario, C.~Englert, {\em et al.}, {\em
  {VBFNLO: A Parton level Monte Carlo for processes with electroweak bosons}}.
  \href{http://dx.doi.org/10.1016/j.cpc.2009.03.006}{Comput.Phys.Commun. {\bf
  180} (2009)  1661--1670},
\href{http://arxiv.org/abs/hep-ph/0811.4559}{{\tt arXiv:hep-ph/0811.4559}}.

\bibitem{susylha}
P.~Z. Skands, B.~Allanach, H.~Baer, C.~Balazs, G.~Belanger, {\em et al.}, {\em
  {SUSY Les Houches accord: Interfacing SUSY spectrum calculators, decay
  packages, and event generators}}.
  \href{http://dx.doi.org/10.1088/1126-6708/2004/07/036}{JHEP {\bf 0407} (2004)
   036},
\href{http://arxiv.org/abs/hep-ph/0311123}{{\tt arXiv:hep-ph/0311123}}.

\bibitem{susybench}
S.~AbdusSalam, B.~Allanach, H.~Dreiner, J.~Ellis, U.~Ellwanger, {\em et al.},
  {\em {Benchmark Models, Planes, Lines and Points for Future SUSY Searches at
  the LHC}}.
  \href{http://dx.doi.org/10.1140/epjc/s10052-011-1835-7}{Eur.Phys.J. {\bf C71}
  (2011)  1835},
\href{http://arxiv.org/abs/hep-ph/1109.3859}{{\tt arXiv:hep-ph/1109.3859}}.

\bibitem{dislepton}
B.~Jager, A.~von Manteuffel, and S.~Thier, {\em {Slepton pair production in the
  POWHEG BOX}}. \href{http://dx.doi.org/10.1007/JHEP10(2012)130}{JHEP {\bf
  1210} (2012)  130},
\href{http://arxiv.org/abs/hep-ph/1208.2953}{{\tt arXiv:hep-ph/1208.2953}}.

\bibitem{toph}
M.~Klasen, K.~Kovarik, P.~Nason, and C.~Weydert, {\em {Associated production of
  charged Higgs bosons and top quarks with POWHEG}}.
  \href{http://dx.doi.org/10.1140/epjc/s10052-012-2088-9}{Eur.Phys.J. {\bf C72}
  (2012)  2088},
\href{http://arxiv.org/abs/hep-ph/1203.1341}{{\tt arXiv:hep-ph/1203.1341}}.

\bibitem{fks}
S.~Frixione, Z.~Kunszt, and A.~Signer, {\em {Three jet cross-sections to
  next-to-leading order}}.
  \href{http://dx.doi.org/10.1016/0550-3213(96)00110-1}{Nucl.Phys. {\bf B467}
  (1996)  399--442},
\href{http://arxiv.org/abs/hep-ph/9512328}{{\tt arXiv:hep-ph/9512328}}.

\bibitem{twpwg}
E.~Re, {\em {Single-top Wt-channel production matched with parton showers using
  the POWHEG method}}.
  \href{http://dx.doi.org/10.1140/epjc/s10052-011-1547-z}{Eur.Phys.J. {\bf C71}
  (2011)  1547},
\href{http://arxiv.org/abs/hep-ph/1009.2450}{{\tt arXiv:hep-ph/1009.2450}}.

\bibitem{atlasexcl}
{\bf ATLAS Collaboration} Collaboration, {\em {Search for squarks and gluinos
  with the ATLAS detector using final states with jets and missing transverse
  momentum and 5.8 fb$^{-1}$ of $\sqrt{s}$=8 TeV proton-proton collision
  data}}.
\href{http://arxiv.org/abs/ATLAS-CONF-2012-109, ATLAS-COM-CONF-2012-140}{{\tt
  ATLAS-CONF-2012-109, ATLAS-COM-CONF-2012-140}}.

\bibitem{cmssusy}
{\bf CMS Collaboration} Collaboration, S.~Chatrchyan {\em et al.}, {\em {Search
  for gluino mediated bottom- and top-squark production in multijet final
  states in pp collisions at 8 TeV}}.
\href{http://arxiv.org/abs/1305.2390}{{\tt arXiv:1305.2390 [hep-ex]}}.

\bibitem{softsusy}
B.~Allanach, {\em {SOFTSUSY: a program for calculating supersymmetric
  spectra}}.
  \href{http://dx.doi.org/10.1016/S0010-4655(01)00460-X}{Comput.Phys.Commun.
  {\bf 143} (2002)  305--331},
\href{http://arxiv.org/abs/hep-ph/0104145}{{\tt arXiv:hep-ph/0104145}}.

\bibitem{sdecay}
M.~Muhlleitner, A.~Djouadi, and Y.~Mambrini, {\em {SDECAY: A Fortran code for
  the decays of the supersymmetric particles in the MSSM}}.
  \href{http://dx.doi.org/10.1016/j.cpc.2005.01.012}{Comput.Phys.Commun. {\bf
  168} (2005)  46--70},
\href{http://arxiv.org/abs/hep-ph/0311167}{{\tt arXiv:hep-ph/0311167}}.

\bibitem{cteq}
H.-L. Lai, M.~Guzzi, J.~Huston, Z.~Li, P.~M. Nadolsky, {\em et al.}, {\em {New
  parton distributions for collider physics}}.
  \href{http://dx.doi.org/10.1103/PhysRevD.82.074024}{Phys.Rev. {\bf D82}
  (2010)  074024},
\href{http://arxiv.org/abs/hep-ph/1007.2241}{{\tt arXiv:hep-ph/1007.2241}}.

\bibitem{lhapdf}
M.~Whalley, D.~Bourilkov, and R.~Group, {\em {The Les Houches accord PDFs
  (LHAPDF) and LHAGLUE}}.
\href{http://arxiv.org/abs/hep-ph/0508110}{{\tt arXiv:hep-ph/0508110}}.

\bibitem{fastjet}
M.~Cacciari and G.~P. Salam, {\em {Dispelling the $N^{3}$ myth for the $k_t$
  jet-finder}}.
  \href{http://dx.doi.org/10.1016/j.physletb.2006.08.037}{Phys.Lett. {\bf B641}
  (2006)  57--61},
\href{http://arxiv.org/abs/hep-ph/0512210}{{\tt arXiv:hep-ph/0512210}}.

\bibitem{antikt}
M.~Cacciari, G.~P. Salam, and G.~Soyez, {\em {The Anti-k(t) jet clustering
  algorithm}}. \href{http://dx.doi.org/10.1088/1126-6708/2008/04/063}{JHEP {\bf
  0804} (2008)  063},
\href{http://arxiv.org/abs/hep-ph/0802.1189}{{\tt arXiv:hep-ph/0802.1189}}.

\bibitem{ggH}
S.~Alioli, P.~Nason, C.~Oleari, and E.~Re, {\em {NLO Higgs boson production via
  gluon fusion matched with shower in POWHEG}}.
  \href{http://dx.doi.org/10.1088/1126-6708/2009/04/002}{JHEP {\bf 0904} (2009)
   002},
\href{http://arxiv.org/abs/hep-ph/0812.0578}{{\tt arXiv:hep-ph/0812.0578}}.

\bibitem{VV}
T.~Melia, P.~Nason, R.~Rontsch, and G.~Zanderighi, {\em {W+W-, WZ and ZZ
  production in the POWHEG BOX}}.
  \href{http://dx.doi.org/10.1007/JHEP11(2011)078}{JHEP {\bf 1111} (2011)
  078},
\href{http://arxiv.org/abs/hep-ph/1107.5051}{{\tt arXiv:hep-ph/1107.5051}}.

\bibitem{sqmerging}
T.~Plehn, D.~Rainwater, and P.~Z. Skands, {\em {Squark and gluino production
  with jets}}.
  \href{http://dx.doi.org/10.1016/j.physletb.2006.12.009}{Phys.Lett. {\bf B645}
  (2007)  217--221},
\href{http://arxiv.org/abs/hep-ph/0510144}{{\tt arXiv:hep-ph/0510144}}.

\bibitem{sqmerging2}
J.~Alwall, S.~de~Visscher, and F.~Maltoni, {\em {QCD radiation in the
  production of heavy colored particles at the LHC}}.
  \href{http://dx.doi.org/10.1088/1126-6708/2009/02/017}{JHEP {\bf 0902} (2009)
   017},
\href{http://arxiv.org/abs/hep-ph/0810.5350}{{\tt arXiv:hep-ph/0810.5350}}.

\bibitem{perugia}
P.~Z. Skands, {\em {Tuning Monte Carlo Generators: The Perugia Tunes}}.
  \href{http://dx.doi.org/10.1103/PhysRevD.82.074018}{Phys.Rev. {\bf D82}
  (2010)  074018},
\href{http://arxiv.org/abs/hep-ph/1005.3457}{{\tt arXiv:hep-ph/1005.3457}}.

\bibitem{VBFPS}
F.~Schissler and D.~Zeppenfeld, {\em {Parton Shower Effects on W and Z
  Production via Vector Boson Fusion at NLO QCD}}.
  \href{http://dx.doi.org/10.1007/JHEP04(2013)057}{JHEP {\bf 04} (2013)  057},
\href{http://arxiv.org/abs/hep-ph/1302.2884}{{\tt arXiv:hep-ph/1302.2884}}.

\bibitem{atlasrhor}
{\bf Atlas Collaboration} Collaboration, G.~Aad {\em et al.}, {\em {Study of
  Jet Shapes in Inclusive Jet Production in $pp$ Collisions at $\sqrt{s}=7$ TeV
  using the ATLAS Detector}}.
  \href{http://dx.doi.org/10.1103/PhysRevD.83.052003}{Phys.Rev. {\bf D83}
  (2011)  052003},
\href{http://arxiv.org/abs/hep-ex/1101.0070}{{\tt arXiv:hep-ex/1101.0070}}.

\end{thebibliography}\endgroup
\end{document}